\documentclass{aa}

\usepackage{natbib}
\bibpunct{(}{)}{;}{a}{}{,} 
\usepackage{graphicx}

\usepackage{txfonts}

\newcommand{\mri}{magneto-rotational instability}
\newcommand{\pns}{proto-neutron star}
\newcommand{\halb}{\frac{1}{2}}
\newcommand{\eqref}[1]{Eq.\,(\ref{#1})}
\newcommand{\eqsref}[1]{Eqs.\,(\ref{#1})}
\newcommand{\figref}[1]{Fig.\,\ref{#1}}
\newcommand{\tabref}[1]{Tab.\,\ref{#1}}
\newcommand{\secref}[1]{Sect.\,\ref{#1}}

\begin{document}

\title{
  Semi-global simulations of the magneto-rotational instability in
  core collapse supernovae
}
\titlerunning{MRI in core collapse supernovae}
\author{
  M.\,Obergaulinger\inst{1}
  \and
  P.\,Cerd\'a-Dur\'an\inst{1} 
  \and
  E.\,M{\"u}ller\inst{1} 
  \and
  M.A.\,Aloy\inst{2} 
}
\institute{
  Max-Planck-Institut f{\"u}r Astrophysik, Garching bei M{\"u}nchen 
  \and
  Departamento de Astronom\'{\i}a y Astrof\'{\i}sica, 
  Universidad de Valencia, 46100 Burjassot, Spain 
}
\date{Received day month year / Accepted day month year}

\abstract 
{
  Possible effects of magnetic fields in core collapse supernovae rely
  on an efficient amplification of the weak pre-collapse fields.  It
  has been suggested that the magneto-rotational instability (MRI)
  leads to a rapid growth for these weak seed fields.  Although plenty
  of MRI studies exist for accretion disks, the application of their
  results to core collapse supernovae is inhibited as the physics of
  supernova cores is substantially different from that of accretion
  discs.  }
{
  We address the problem of growth and saturation of the MRI in core
  collapse supernovae by studying its evolution by means of
  semi-global simulations, which combine elements of global and local
  simulations by taking the presence of global background gradients
  into account and using a local computational grid.  We investigate,
  in particular, the termination of the growth of the MRI and the
  properties of the turbulence in the saturated state.}
{
  We analyze the dispersion relation of the MRI to identify different
  regimes of the instability.  This analysis is complemented by
  semi-global ideal MHD simulations, where we consider core matter in
  a local computational box (size $\sim 1\,$km) rotating at
  sub-Keplerian velocity and where we allow for the presence of a
  radial entropy gradient, but neglect neutrino radiation.
}
{
  We identify six regimes of the MRI depending on the ratio of the
  entropy and angular velocity gradient.  Our numerical models
  confirm the instability criteria and growth rates for all regimes
  relevant to core-collapse supernovae.  The MRI grows exponentially
  on time scales of milliseconds, the flow and magnetic field
  geometries being dominated by channel flows. We find MHD
  turbulence and efficient transport of angular momentum. The MRI
  growth ceases once the channels are disrupted by resistive
  instabilities (stemming from to the finite conductivity of the
  numerical code), and MHD turbulence sets in.  From an analysis of
  the growth rates of the resistive instabilities, we deduce scaling
  laws for the termination amplitude of the MRI, which agree well
  with our numerical models.  We determine the dependence of the
  development of large-scale coherent flow structures in the
  saturated state on the aspect ratio of the simulation boxes.  }
{
  The MRI can grow rapidly under the conditions considered here, i.e.,
  a rapidly rotating core in hydrostatic equilibrium, possibly endowed
  with a nonvanishing entropy gradient, leading to fields exceeding
  $10^{15}~\mathrm{G}$.  More investigations are required to cover the
  parameter space more comprehensively and to include more physical
  effects. }
\keywords{MHD - Instabilities - Supernovae: general - Stars:
  magnetic fields} 

\maketitle


\section{Introduction}
\label{Sec:Intro}
The \mri{} (MRI) \citep{Balbus_Hawley__1991__ApJ__MRI} is a local
linear instability of weakly magnetized differentially rotating
fluids. A large number of analytic, as well as numerical studies
support the assertion that the MRI is the main agent for exciting
turbulence in Keplerian accretion disks \citep[for a review, see,
e.g.,][]{Balbus_Hawley__1998__RMP__MRI}.  The MRI amplifies seed
perturbations exponentially with time until turbulence sets in.  In
the turbulent state, the magnetic field, $\vec b$, gives rise to a
non-vanishing (spatial and temporal) mean Maxwell stress tensor
$\tens{M}_{ij} = b_i b_j$.  Simulations of accretion disks show a high
negative mean value of the component $\tens{M}_{\varpi \phi}$ (where
$\varpi$ and $\phi$ are the radial and azimuthal coordinate of a
cylindrical coordinate system), which gives rise to an efficient
outward transport of angular momentum.

\cite{Akiyama_etal__2003__ApJ__MRI_SN} pointed out that the layers
surrounding the nascent \pns{} quite generically fulfill the MRI
instability criteria.  Consequently, any (weak) seed magnetic field
will be amplified exponentially. In the saturated state of the MRI
instability, sustained magneto-hydrodynamic turbulence might then
provide an efficient means for an angular momentum redistribution and
for the conversion of rotational energy into thermal energy of the
gas.  Imparting additional thermal energy into the post-shock stellar
matter the MRI might thus be important for the currently favored
neutrino-driven core collapse supernova explosion mechanism
\citep[e.g.][]{Thompson_Quataert_Burrows_ApJ_2004__Vis_Rot_SN,
  Janka_etal__2007__PRD__SN_theory}, although possibly only for
rapidly and strongly differentially rotating progenitors.
Furthermore, the growth of the magnetic field resulting from the MRI
may provide the adequate physical conditions in the collapsed core to
launch bipolar outflows, which result in gamma-ray bursts
\citep{Aloy_Obergaulinger__2007__Rev_Mex__GRB-outflows}.  As the
physical conditions in accretion disks and stars differ significantly,
and as only a few analytic studies of the MRI in stars exist
\citep[e.g.,][]{Acheson__1978__RSTA__MHD-inst-stars}, it remains
unclear whether existing results on the MRI in disks apply to stars,
and particularly to supernovae, as well.

Numerical simulations of the MRI face a severe problem, since the
growth rate of MRI-unstable modes depends on the product of the
initial field strength and the wave number of the mode.  For a weak
field, only fairly short modes grow rapidly.  Simulations of
astrophysical flows, on the other hand, often fail to resolve just
those modes, as it would require prohibitively high computational
costs to cover spatial scales ranging from the global extent of the
astrophysical system (which may be much larger than the MRI-unstable
region) down to the wavelengths of the fastest growing MRI modes.
This dependence of the growth rate on the wavelength of the mode
suggests a twofold approximate numerical approach: one either performs
simulations which properly cover the global scales of the
astrophysical system foregoing to resolve the small scales set by the
wavelengths of the fastest growing MRI modes (\emph{global
  simulations}), or vice versa (\emph{local simulations}).

Local simulations evolve only a small part of the entire MRI-unstable
system, known as the \emph{shearing box}.  However, information on the
scales exceeding the size of the computational grid has to be provided
using suitably chosen boundary conditions.  No unique recipe exists
for this procedure, but the use of reflecting and periodic boundaries
is a common practice.  In most studies of accretion disks, the
boundary conditions of \cite{Balbus_Hawley__1992__ApJ__MRI_3} are used
in radial direction, which are essentially periodic boundary
conditions but account also for the relative shear between the inner
and the outer radial edge of the grid.  They are often combined with a
Galilei transformation into a frame of reference co-rotating at the
mean angular velocity of the shearing box, and a linearizion of the
angular velocity within the box.  Local (shearing box) simulations
using this kind of boundary treatment are commonly called
\emph{shearing-sheet} simulations.

The general drawback of local simulations obviously lies in their
inability to account accurately for large-scale phenomena. In
addition, there is only a limited possibility to model global
gradients other than differential rotation in shearing-sheet
simulations.  Independent of the boundary treatment only modes with a
wavelength less than the size of the grid can be excited, i.e.  modes
with a wavelength comparable to the dimensions of the whole system
cannot develop.  Consequently, MRI-driven turbulence may saturate at a
level determined (at least partially) by numerical rather than (only)
by physical parameters. A careful analysis is necessary to disentangle
the respective influence \citep[see
  e.g.,][]{Pessah_etal__2007__ApJL__MRI-scaling-laws,
  Fromang_Papaloizou__2007__AA__3d-local-MRI-disc-zero-net-flux_1,
  Regev_Umurhan__2008__AA__Viability_of_shearing_box}.

Global simulations, on the other hand, follow the evolution of the
entire system, albeit with a much coarser resolution than local ones.
Thus, they can account for the large-scale structure of stars and
disks, for the back-reaction of the MRI instigated turbulence on the
global flow, and allow one to draw conclusions on how the saturated
state depends on global properties of the system, e.g., the density or
pressure stratification.  However, foregoing the ability to resolve
short-wavelength modes, the growth of the MRI will be underestimated
or suppressed even entirely.  In many applications of numerical
analysis, it is possible to use suitable models for the unresolved
physics on sub-grid scales, e.g., sub-grid diffusivity.  This requires
a good knowledge of the physics on these scales, and is facilitated
greatly if processes at the unresolved scales act merely as a sink for
kinetic or magnetic energy cascading down from the integral scale.
During the growth of the MRI, however, the power shifts gradually from
short to long modes.  Thus, sub-grid models for global MRI simulations
tend to be complex, and are not used widely.

As a remedy for this problem, global simulations may be performed
using unrealistically strong initial fields to guarantee that the
fastest growing MRI modes are resolved numerically.  This approach
presumes that the unresolved MRI modes are able to amplify the much
weaker actual initial fields to the field strengths used as initial
value in global simulations.  This assumption can be justified, if the
MRI acting on the unresolved scales saturates at the initial field
strengths imposed in global simulations, i.e., if rapid amplification
by the MRI takes place over many orders of magnitude.  However, this
can only be proven by high-resolution local simulations.  Enhancing
the initial magnetic field by a constant factor throughout the
computational domain, as it is often done in global simulations, is
problematic as the MRI is a local instability, i.e. it is not expected
to cause a constant amplification of the field everywhere.  The
ambiguities regarding differences between the topology of this
artificially enhanced field and that of a field amplified locally by
the MRI add to the uncertainties clouding the influence of magnetic
fields on the overall dynamics.

Both the local and the global numerical approach has been used for
studying the MRI in accretion disks, and this combined effort has led to
the rapid development of the field.  Simulations of the MRI in core
collapse supernovae, on the other hand, have not yet reached this
advanced stage, mainly because of the weakness of the initial field of
the progenitors.  According to current stellar evolution models
\citep{Heger_Woosley_Spruit_2005}, the canonical pre-collapse magnetic
fields are so weak that they are unable to affect the dynamics of the
explosion unless they are amplified strongly.  Correspondingly, the
wavelengths of the fastest growing MRI modes are approximately a few
meters at most
.\footnote{Note, however, that these predictions still involve
  uncertainties, and hence rare, but much more strongly magnetized
  progenitors cannot be excluded presently.}
Thus, the possible importance of MHD effects in core collapse
supernovae depends on the existence of mechanisms which can amplify
the field efficiently during core collapse and the post-bounce phase.
The timescale available for the growth of the magnetic field is set by
the time required to turn the accretion of matter onto the \pns~into
an explosion, i.e., a few hundreds of milliseconds.  As already
mentioned above \cite{Akiyama_etal__2003__ApJ__MRI_SN} suggested that
the MRI might provide this mechanism. They estimated the saturation
field strength to be $10^{15}$ -- $10^{16}\,\mathrm{G}$, i.e.,
clearly in excess of the artificially enhanced initial field strengths
used in global simulations.

Up to now, there exist only global simulations of MHD core collapse
supernovae, which evolve the entire core of a massive star through
gravitational collapse, bounce, and explosion
\citep[e.g.,][]{Kotake_etal__2004__PRD__MagCollapse_GW,
  Yamada_Sawai__2004__ApJ__MHDCollapse,
  Takiwaki_etal__2004__ApJ__MHD_SN,
  Obergaulinger_Aloy_Mueller__2006__AA__MR_collapse,
  Obergaulinger_et_al__2006__AA__MR_collapse_TOV,
  Burrows_etal__2007__ApJ__MHD-SN}.  These global simulations fail to
find the MRI unless they employ drastically stronger initial fields.
\cite{Obergaulinger_Aloy_Mueller__2006__AA__MR_collapse,
  Obergaulinger_et_al__2006__AA__MR_collapse_TOV}, e.g., require a
pre-collapse field strength exceeding $10^{12}\,\mathrm{G}$ to
resolve the MRI in the post-bounce state.  The rationale behind the
artificially increased initial field strengths is that, once triggered by
the differential rotation in the \pns, the MRI will exponentially
amplify a much weaker seed field up to the values used in the
simulations.

Due to the lack of local simulations, the importance of the MRI in MHD
core collapse models remains unclear.  As a first step to resolve this
issue, we have performed high-resolution simulations of small parts of
simplified post-bounce, rotating, magnetized cores.  We have used a
recently developed high-resolution MHD code, and employed
\emph{shearing-disk} boundary conditions
\citep{Klahr_Bodenheimer__2003__ApJ__Global-baroclinic-inst-disc}.
These boundary conditions derive from the shearing-sheet boundary
conditions of \cite{Balbus_Hawley__1992__ApJ__MRI_3}, but allow one to
consider global gradients of, e.g., density or entropy.  Combining
elements of global and local simulations, viz.~the presence of global
background gradients, and a high-resolution local grid, we find it
justified to call our approach \emph{semi-global} (for more details
see \secref{sSec:BC}).

Differences in the physical conditions in disks and stars impede the
direct application of the MRI results from accretion disks to
supernovae.  Most obviously, the geometry of both systems differs
strongly.  Furthermore, while accretion discs are stabilized against
gravity by (Keplerian) rotation, stars are supported mainly by pressure
gradients, with only a minor contribution from rotation, i.e.  thermal
stratification is much more important in stars than in disks.  Thus,
entropy gradients can stabilize an MRI-unstable region or modify the
instability in convectively unstable regions.  Consequently, the problem
of the MRI in core collapse supernovae has to be addressed by
simulations accounting for their specific properties, which is the goal
of this study. We investigate the growth of the MRI from initial fields
comparable to the ones expected from realistic stellar evolution
modeling, and we seek to probe the possibility of MRI-driven field
amplification under typical conditions of supernova cores and on
timescales similar to the dynamic times of the system, (i.e., a few tens
of milliseconds).  Apart from the restrictions inherent to local and
semi-global simulations, several simplifications limit our approach: we
use simplified initial equilibrium models, a simplified equation of
state, and neglect neutrino heating and cooling.  The main physical
questions that we try to address are: (i) does the MRI grow on
sufficiently short time scales to influence the explosion, i.e., within
at most 100 msec, given typical post-bounce rotation profiles and
magnetic fields? (ii) How does an entropy gradient affect the growth of
the instability?  (iii) How does the saturated state of MRI-driven
turbulence depend on these factors?  In particular, is the saturation
field strength estimated by \cite{Akiyama_etal__2003__ApJ__MRI_SN},
i.e., the conversion of most of the rotational energy into magnetic
energy, realistic?

Analogous questions are studied by local simulations of the MRI in
accretion disks.  The answers may lead the way to formulate a
turbulence model to be used in global simulations.  The simplest model
would provide a parametrization of the angular momentum transport by
an $\alpha$ viscosity \citep{Shakura_Sunyaev__1973__AA__alpha_visco},
i.e., a turbulent viscosity proportional to the local sound speed and
the pressure scale height.  However, despite a large number of local
simulations, no unique formulation of an $\alpha$ model for accretion
disks has been found up to now.  Lacking similar comprehensive local
simulations, a turbulence model for the MRI in supernovae is even less
conceivable.  Our simulations intent to provide only a first step
towards these highly desired turbulence models.

The paper is organized as follows: after a discussion of the main
properties of the MRI in disks and stars (\secref{Sec:MRIds}), we
outline our numerical method in \secref{Sec:Num}, discuss our results
in \secref{Sec:Res}, and summarize our main results and give
conclusions in \secref{Sec:Sum}.

\section{MRI in discs and stars}
\label{Sec:MRIds}
%
\subsection{Physical model}
\label{sSec:Phys}
We work in the limit of ideal magnetohydrodynamics (MHD), solving the
the equations of ideal MHD in the presence of an external
gravitational potential $\varphi$,
\begin{equation}
  \label{Gl:Phys--MHD-rho}
  \partial_{t} \rho 
  + \nabla_j \left[ \rho v^j \right]
  = 0
  ,
\end{equation}
\begin{equation}
  \label{Gl:Phys--MHD-mom}
  \partial_{t} p^i 
  + \nabla_j 
  \left[ 
    p^i v^j + P_{\star} \delta^{ij} - b^i b^j
  \right]
  = \rho \nabla_i \varphi,
\end{equation}
\begin{equation}
  \label{Gl:Phys--MHD-erg}
  \partial_{t} e_{\star}
  + \nabla_j 
  \left[ 
    \left( e_{\star} + P_{\star} \right) v^j
    - b^i v_i b^j
  \right]
  = \rho v^j \nabla_j \varphi
  ,  
\end{equation}
\begin{equation}
  \label{Gl:Phys--MHD-ind}
  \partial_{t} \vec b 
  = 
  - c \ \vec \nabla \times \vec E,
\end{equation}
\begin{equation}
  \label{Gl:Phys--MHD-div}
  \nabla_j b^j = 0.
\end{equation}
Here, $\rho$, $\vec p$, $\vec v$, and $e_{\star}$ denote the mass
density, momentum density, velocity, total-energy density, of the gas,
respectively; $\vec b$ is the magnetic field.  The total-energy
density and the total pressure, $P_{\star}$, are composed of fluid and
magnetic contributions: $e_{\star} = \varepsilon + \frac{1}{2} \rho
\vec v^2 + \frac{1}{2} \vec b ^ 2$ and $P_{\star} = P + \frac{1}{2}
\vec b ^ 2$ with the internal-energy density $\varepsilon$ and the gas
pressure $P = P( \rho, \varepsilon, \dots)$.  The electric field,
$\vec E$, is given by $\vec E = - \frac{\vec v}{c} \times \vec b $.
Here, $c=2.998 \times 10^{10}\,\mathrm{cm}\ \mathrm{s}^{-1}$ is the
speed of light in vacuum and Einstein's summation convention applies.

We use the hybrid equation of state (EOS) due to
\cite{Keil_Janka_Mueller__1996__ApJL__NS-Convection} as a rough model
for neutron-star matter.  Following this EOS, the total gas pressure,
$P$, consists of a barotropic part, $P_\mathrm{b}$, and a thermal
part, $P_\mathrm{th}$.  The two parts are given by
\begin{eqnarray}
  \label{Gl:Phys--EOS-1}
  P_{\mathrm{b}}  & = & \kappa \rho^{\Gamma_{\mathrm{b}}}, \\
  P_{\mathrm{th}} & = & ( \Gamma_{\mathrm{th}} - 1 ) \varepsilon_{\mathrm{th}}.
\end{eqnarray}
Here, $\Gamma_\mathrm{b}$ and $\kappa$ refer to the barotropic
adiabatic index, and the polytropic constant of the EOS, respectively;
$\varepsilon_{\mathrm{th}} = \varepsilon - P_{\mathrm{b}} / (
\Gamma_{\mathrm{b}} - 1 )$ is the thermal part of the internal energy,
and $\Gamma_{\mathrm{th}}$ the corresponding adiabatic index.  Please
  note that we consider only sub-nuclear densities, $\rho <
  \rho_\mathrm{nuc} = 2 \times 10^{14}\, \mathrm{g\,cm^{-3}}$ here
  since the maximum density reached in our models is a few times
  $10^{13}\, \mathrm{g\,cm^{-3}}$.

We define a pseudo-entropy $S$ for this equation of state by
\begin{equation}
  \label{Gl:Phys--entropy}
  S = \frac{P_\mathrm{th}}{P_\mathrm{b}}.
\end{equation}
In the Schwarzschild criterion for convective stability (see below),
this quantity appears instead of the entropy of, e.g., an ideal gas.

A few quantities used frequently in the remainder of this paper are:
\begin{enumerate}
  \item the Alfv\'en velocity
    \begin{equation}
      \label{Gl:Phys--Alfven}
      c_{\mathrm{A}} = \frac{\left| \vec b \right|}{\sqrt{\rho}},
    \end{equation}
  \item the (local) magnetic energy density
    \begin{equation}
      \label{Gl:Phys--emag}
      \tens{e}_{\mathrm{mag}} = \frac{\vec b^2}{2},
    \end{equation}
    and the corresponding volumetric mean value
    \begin{equation}
      \label{Gl:Phys--emag_V}
      e_{\mathrm{mag}} = \frac{1}{\mathcal{V}} 
                        \int \mathrm{d} \mathcal{V}\, \tens{e}_{\mathrm{mag}},
    \end{equation}
  \item the (local) Maxwell stress tensor
    \begin{equation}
      \label{Gl:Phys--Maxwell}
      \tens{M}_{ij} = b_i b_j \,,
    \end{equation}
    and the corresponding volumetric mean value
    \begin{equation}
      \label{Gl:Phys--Maxwell_V}
      M_{ij} = \frac{1}{\mathcal{V}} 
               \int \mathrm{d} \mathcal{V}\, \tens{M}_{ij} \,.
    \end{equation}
We will use most frequently the component $M_{\varpi\phi}$ which
governs the transport of angular momentum in radial direction, and we
will sometimes refer to this component as \textit{the Maxwell stress}
for short.
\end{enumerate}

\subsection{General properties of the MRI}
\label{sSek:MRIds--gen}
The stability criteria for the MRI was first discovered by
\cite{Velikhov__1959__SovPhys__MRI,Chandrasekhar__1960__PNAS__MRI} and
further discussed by \cite{Balbus_Hawley__1991__ApJ__MRI} in a series
of papers.  These authors analyze wave-like (WKB) perturbations of the
form $\exp [i(\vec{k}\cdot\vec{r} + \omega t)]$ in a background
equilibrium of the MHD equations. For convenience cylindrical
coordinates $\left( \varpi,\phi, z \right)$ are used in the following
equations.  From the dispersion relation, they derive the criteria for
exponential growth and, if applicable, the growth rates of WKB
modes. Because the main astrophysical context of this series of papers
is accretion discs, some assumptions are made which considerably
simplify the analysis: i) weak magnetic fields where $|{\vec v}_{\rm
  A}| \ll \min (c_{\rm s}, |v_{\varphi}|)$, ii) incompressible gas
(Boussinesq approximation), and iii) angular velocity constant on
cylinders, $\Omega(\varpi)$. The discussion is mostly restricted to
thin discs (i.e., to equatorial regions and to a Keplerian rotation
law) and $5/3$-polytropes. Under these assumptions the stability
criterion for a differentially rotating magnetized fluid is
\citep{Balbus_Hawley__1991__ApJ__MRI}
\begin{equation}
  \label{Gl:MRIds--gen--BH-crit}
  \mathcal{R}_{\varpi} \equiv \varpi \partial_{\varpi} \Omega^2 > 0.
\end{equation}
If the criterion is not fulfilled only modes with (dimensionless)
wavenumber
\begin{equation}
  \hat{k}< \sqrt{-\hat{\mathcal{R}}_\varpi} = \hat{k}_{\rm crit}
\end{equation}
are unstable, where $\hat{k} \equiv \vec{k}\cdot{\vec v}_{\rm A} /
\Omega$, $\hat{k}_{\rm crit} \equiv \vec{k}_{\rm crit} \cdot{\vec
  v}_{\rm A} / \Omega$ and $\hat{\mathcal{R}}_\varpi \equiv
      {\mathcal{R}}_\varpi / \Omega^2$.  The (dimensional) growth rate
      of the fastest growing mode is
\citep{Balbus_Hawley__1992__ApJ__MRI_Oort}
\begin{equation}
  \hat{\omega}_{\rm FGM} \equiv \omega_{\rm FGM} / \Omega
                        =      -\hat{\mathcal{R}}_\varpi / 4
\end{equation}
which is independent of the magnetic field and corresponds to
(dimensionless) wave numbers close to $\hat{k}_{\rm crit}$.

However in the context of core collapse supernovae some of these
assumptions do not apply: entropy and composition gradients are
important, more general rotation laws $\Omega(\varpi,z)$ have to be
considered, and the analysis can no longer be restricted to equatorial
regions. In this general case the dispersion relation of WKB modes is
\citep[cf.][]{Balbus__1995__ApJ__stratified_MRI,
  Urpin_1996_MNRAS_MHD-stability},
\begin{eqnarray}
  \left (\hat{\omega}^2 - \hat{k}^2 \right )^2 
  &-&
  \left(\hat{\omega}^2 - \hat{k}^2 \right) \left ( \hat{\omega}_{\rm G}^2 
    + \hat{\omega}_{\rm R}^2 + 4 \cos^2{\theta_k} \right )
  \\
&-& 4 \, \hat{k}^2 \cos^2{\theta_k} = 0
\label{Gl:MRIds--gen--Bal-crit}
\end{eqnarray}
where $\hat{\omega} = \omega / \Omega$ is the dimensionless growth
rate of the instability, and $\theta_k$ is the angle between $\vec{k}$
and the $z$-axis.  The (dimensionless) frequencies related to buoyancy
terms and differential rotation are
\begin{equation}
 \hat{\omega}_{\rm G}^2 = \frac{1}{\Omega^2} 
                          \left[\mathcal{G}\cdot \mathcal{B}  - 
                                \frac{(\vec{k}\cdot\mathcal{B})
                                      (\vec{k}\cdot \mathcal{G})}{k^2}
                          \right]
\end{equation}
and
\begin{equation}
 \hat{\omega}_{\rm R}^2 = \frac{1}{\Omega^2} 
                          \left[\mathcal{R}\cdot\vec{e}_{\varpi} - 
                                \frac{(\vec{k}\cdot\vec{e}_{\varpi})
                                      (\vec{k}\cdot \mathcal{R})}{k^2}
                           \right] \,,
\end{equation}
respectively, where ${\vec e}_{\varpi}$ is the unit vector in
$\varpi$- direction,
\begin{eqnarray}
\mathcal{G} &\equiv & \frac{\vec \nabla P}{ \rho} \\
\mathcal{R} &\equiv & \varpi \vec \nabla \Omega^2 \\
\mathcal{B} &\equiv & \frac{\vec \nabla \rho}{ \rho} -  
                      \frac{\vec \nabla P}{ \Gamma_1 P} 
= -\frac{1}{\Gamma_1} \left . \frac{\partial \ln{P}}{\partial s} 
                      \right |_\rho \vec \nabla s
\end{eqnarray}
are the gravitational, rotational, and buoyancy terms, respectively,
and $\Gamma_1\equiv\partial \ln{P} / \partial \ln{\rho}|_{s}$. It is
convenient for the mode analysis and if $\cos \theta_k \ne 0$ to
define the quantity
\begin{eqnarray}
\mathcal{C} &\equiv& \frac{\hat{\omega}_{\rm G}^2 + 
                     \hat{\omega}_{\rm R}^2}{\cos^2{\theta_k}} 
\nonumber \\ &=& 
( \mathcal{G}_z \mathcal{B}_z \tan^2{\theta_k}
  - 2 \mathcal{B}_{\varpi} \mathcal{G}_z \tan{\theta_k}
  + \mathcal{G}_\varpi \mathcal{B}_\varpi 
  + \mathcal{R}_\varpi ) / \Omega^2,
\label{Gl:MRIds--C-param}
\end{eqnarray}
where the curl of \eqref{Gl:Phys--MHD-mom}, i.e. the vorticity
equation, has been used to simplify the expression of $\mathcal{C}$.
Note that this quantity depends on the direction of the perturbation
$\theta_k$, but not on $k^2$ itself. If $\cos \theta_k = 0$, which
corresponds to velocity perturbations parallel to the rotation axis,
the value of $\mathcal{C}$ diverges, but all length scales and growth
rates of the discussion below are finite, and can be computed by
taking the limit $\cos \theta_k \to 0$.

In the absence of a magnetic field, i.e. $\hat{k}^2=0$, the stability
condition is simply $\mathcal{C}+4 > 0$, which is equivalent to the
Solberg-H{\o}iland stability criteria for a non-magnetized rotating
fluid \citep{Tassoul_Book__Theo_rotating_stars}.

Because we want to study instabilities of magnetized fluids, we
consider hereafter only the case $\hat{k}\ne 0$. Then the stability
condition is $\mathcal{C}> 0$, which corresponds to that of
\cite{Balbus__1995__ApJ__stratified_MRI},
\begin{eqnarray}
 \mathcal{G}\cdot \mathcal{B} +  
 \mathcal{R}\cdot {\vec e}_{\varpi} &\ge& 0 \\
 ( \mathcal{G} \times {\vec e}_{\varpi} ) 
   \cdot (\mathcal{B} \times \mathcal{R} ) &\ge& 0 \,.
\end{eqnarray}
Modes with wave numbers smaller than the (dimensionless) critical
wavenumber
\begin{equation}  
  \hat{k}_{\rm crit} = \cos{\theta_k} \sqrt{ -\mathcal{C}}
\end{equation}
are unstable and grow. The critical wavenumber depends on the angle
$\theta_k$ in a complicated way involving, in general, the rotation
profile, $\mathcal{R}$, the thermal structure, $\mathcal{B}$, and the
stratification, $\mathcal{G}$.

\begin{figure}[htbp]
  \includegraphics[width=0.49\textwidth]{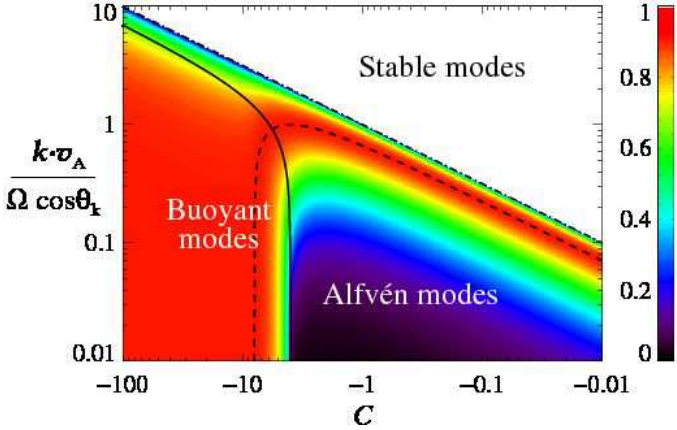}\\
  \caption{Imaginary part of the growth rate normalized to the
    imaginary part of the maximum growth rate, $\Im(\hat{\omega}) /
    \Im(\hat{\omega_{FGM}})$ as a function of $\mathcal{C}$ and
    $\hat{k}/\cos\theta_k \equiv \vec{k} \cdot \vec{v}_{\rm A} /
    (\Omega \cos\theta_k)$. The dashed line shows the value of
    $\hat{k}$ corresponding to the fastest growing mode,
    $\Im(\hat{\omega}) / \Im(\hat{\omega_{FGM}}) = 1$, the solid line
    gives the boundary between the two branches of unstable modes
    (Alfv\'en and Buoyant modes), and the dash-dotted line corresponds
    to the stability limit ($\hat{k} = \hat{k}_{\rm crit}$). For $-4
    <\mathcal{C}<0$ only Alfv\'en modes appear, with a narrow spectrum
    of fast growing modes close to $\hat{k}_{\rm crit}$ (dash-dotted
    line). For $\mathcal{C}<-4$ buoyant modes appear and become
    dominant for $\mathcal{C}<-8$. In the latter case the spectrum of
    fast growing modes is much wider covering the entire region from
    $\hat{k}_{\rm crit}$ to $0$.  }
  \label{Fig:MRI--Modes}
\end{figure}

Two branches of unstable modes arise from the dispersion relation with
$\hat{k}\ne 0$ \citep{Urpin_1996_MNRAS_MHD-stability}: the branch of
Alfv\'en modes appearing for $\mathcal{C}< 0$, and the branch of
buoyant modes which only appear for $\mathcal{C} + 4< 0$
(\figref{Fig:MRI--Modes}).

For a given $\theta_k$ the fastest growing mode is obtained from the
condition $\partial_{\hat{k}} \hat{\omega}=0$. For $-8 < \mathcal{C} <
0$ it has a (dimensionless) wavenumber
\begin{equation}
  \hat{k}_{\rm FGM} = \cos{\theta_k}
                     \frac{\sqrt{-\mathcal{C} (\mathcal{C} + 8)}}{4} \,,
    \label{Gl:MRIds--kMRI-Alfv}
\end{equation}
and a (dimensionless) growth rate
\begin{equation}
  \hat{\omega}_{\rm FGM} = \cos{\theta_k} 
                          \frac{\sqrt{-\mathcal{C}^2}}{4} \,.
  \label{Gl:MRIds--oMRI-Alfv}
\end{equation}
If $\mathcal{C} \le -8$, the fastest growing mode corresponds to
$\hat{k}_{\rm FGM} = 0$, i.e. it is dominated by buoyant modes with a
(dimensionless) growth rate
\begin{equation}
  \hat{\omega}_{\rm FGM} = 
  \cos{\theta_k} \sqrt{\mathcal{C} + 4 } .
\end{equation}

Thus, there exist two different instability regimes depending on the
value of $\mathcal{C}$.  For $-4 < \mathcal{C} < 0$ only Alfv\'en
modes are possible. This {\it magneto-shear regime} was discussed by
\cite{Balbus_Hawley__1991__ApJ__MRI}. A mixed regime is found for $-8
< \mathcal{C} < -4$, where both Alfv\'en and buoyant modes compete.
For $\mathcal{C} < -8$ the buoyant modes completely dominate the
growth of the instability, and this regime is thus called {\it
  magneto-convective} regime. It is similar to the convective regime,
but the critical wavenumber is determined by the strength of the
magnetic field.

Note that for a given fluid element the behavior of the unstable modes
depends on the angle $\theta_k$. Thus, different regimes can hold in
different directions. To find the absolute fastest growing mode of a
fluid element, i.e. not considering a fixed angle $\theta_k$, one has
to determine the zeros of $\partial \hat{\omega} / \partial \theta_k$,
which involves the solution of a quartic equation. This fact makes a
more detailed study of the instability difficult.

\begin{figure}[htbp]
  \includegraphics[width=0.45\textwidth]{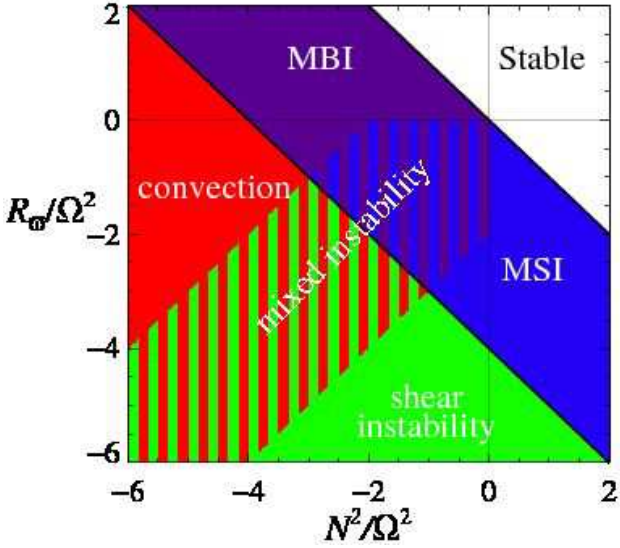}\\
  \caption{Stability regions in the plane
    $\mathcal{R}_{\varpi}/\Omega^2$ vs $N^2/\Omega^2$. The solid thick
    line separating the stable region from the magneto-rotational
    instabilities (MSI and MBI) corresponds to $\mathcal{C}=0$, and
    the solid thick line separating the magneto-rotational
    instabilities from the hydrodynamic instabilities (convection and
    shear instability) corresponds to the $\mathcal{C}=-4$. The mixed
    zone is arbitrarily defined by $|\mathcal{R}_{\varpi}/\Omega^2 -
    N^2 / \Omega^2|<2$.}
  \label{Fig:MRI--Regimen}
\end{figure}

To simplify the analysis, we restrict ourselves in the following
discussion to regions near the equator, where it is reasonable to
assume only a radial dependence of the hydrodynamic quantities and a
vertical magnetic field.  Therefore,
\begin{equation}
  \mathcal{C}_{\rm 90}= (N^2 + \mathcal{R}_{\varpi})/ \Omega^2 \,,
\end{equation}
where $N^2 \equiv \mathcal{B}\cdot \mathcal{G}$ is the square of the
Brunt-V\"ais\"al\"a or buoyancy frequency. Because $\mathcal{C}_{\rm
  90}$ does not depend on $\theta_k$, all modes of the considered
fluid element belong to the same branch of modes, i.e. they are either
buoyant modes or Alfv\'en modes.  All modes with wavelengths shorter
than
\begin{equation}
\lambda_{\rm crit} \equiv \frac{2 \pi}{k_{\rm crit}} 
                  =      \frac{2\pi \, |\vec{v}_{\rm A}|}{
                               \sqrt{-(N^2 + \mathcal{R}_{\varpi})}},
\label{Gl:MRI--lamda_crit}
\end{equation}
are stabilized by magnetic tension.  It is easy to show that the modes
grow faster when $\vec{k}$ is parallel to the $z$-axis ($\theta_k=0$),
i.e. velocity and magnetic field perturbations grow in direction
perpendicular to the rotation axis. The stability criterion, $N^2 +
\mathcal{R}_{\varpi}>0$, can easily be interpreted according to the
relative size of the buoyancy term, $N^2$, and the shear term,
$\mathcal{R}_{\varpi}$. Several different regimes result
(\figref{Fig:MRI--Regimen}):
\begin{itemize}
\item
\emph{Magneto-shear instability (MSI):} $\mathcal{R}_{\varpi}\ll N^2$ and
$-4 \Omega^2 < N^2 + \mathcal{R}_{\varpi}< 0$.
All modes longer than $\lambda_{\rm crit}$ are unstable albeit with a
vanishing growth rate as their wavelength $\lambda$ approaches
infinity. The growth rate peaks for
\begin{equation}
\lambda_\mathrm{\rm MRI} \equiv 2 \pi / k_{\rm FGM}
                         \sim   \sqrt{2} \, \lambda_\mathrm{\rm crit}, 
\label{Gl:MRIds--gen--la_MRI}
\end{equation}  
where the limit $|\mathcal{C}| \ll 8$ is used to obtain the second
expression.  It is important to note that $\lambda_\mathrm{MRI}$,
which scales with the background field strength $b_0$, becomes small
for weak initial fields.  Hence, in the limit of a pure shear
instability, only relatively strong initial fields are accessible by
numerical simulations due to the restrictive constraint on the grid
size imposed by the requirement to resolve $\lambda_\mathrm{MRI}$ by
at least several grid zones.
\item
\emph{Magneto-buoyant instability (MBI)}: 
\footnote{ The reader should not confuse this instability with the
  magnetic buoyancy or Parker instability
  \citep{Parker__1966__ApJ__Interstellar-gas-field}, related to the
  magnetic field strength gradients.}
$N^2 \ll \mathcal{R}_{\varpi}$ and $-4 \Omega^2 < N^2 +
\mathcal{R}_{\varpi}< 0$.  
This regime resembles the magneto-shear regime, but the instability is
not driven by the shear, but rather by the unstable stratification.
\item
\emph{Magneto-convective instability:} $N^2 \ll \mathcal{R}_{\varpi}$
and $N^2 + \mathcal{R}_{\varpi}< -4 \Omega^2$.
This regime corresponds to magnetized convective flow. The main
difference is the stabilization of short modes ($\lambda <
\lambda_{\rm crit}$) due to the magnetic tension. The more important
the negative entropy gradient becomes with respect to the angular
velocity gradient, the faster is the growth of infinitely long modes
compared to the growth rate at $\lambda_\mathrm{MRI}$.
\item
\emph{Hydrodynamic shear instability:} $\mathcal{R}_{\varpi}\ll N^2$
and $N^2 + \mathcal{R}_{\varpi}< -4 \Omega^2$.
This case is not of interest in core collapse since for the
differential rotation of PNS we always find $\mathcal{R}_{\varpi} >
1.5 \Omega^2$.
\end{itemize}
Core collapse occurs in general in a \emph{mixed regime}, where
Alfv\'en modes and buoyant modes compete. Therefore, none of the above
mentioned pure regimes holds for the MHD instabilities appearing
during core collapse.

\section{Method}
\label{Sec:Num}
%
\subsection{Code}
\label{sSec:Code}
We use a newly developed three-dimensional Eulerian MHD code
(Obergaulinger et al., in preparation) to solve the MHD equations,
\eqsref{Gl:Phys--MHD-rho}--(\ref{Gl:Phys--MHD-div}).  The code is
based on a flux-conservative finite-volume formulation of the MHD
equations and the constrained-transport scheme to maintain a
divergence-free magnetic field \citep{Evans_Hawley__1998__ApJ__CTM}.
Using high-resolution shock capturing methods
\citep[e.g.,][]{LeVeque_Book_1992__Conservation_Laws}, it employs
various optional high-order reconstruction algorithms including a
total-variation diminishing piecewise-linear (TVD-PL) reconstruction
of second-order accuracy, a fifth-order monotonicity-preserving (MP5)
scheme \citep{Suresh_Huynh__1997__JCP__MP-schemes}, and a fourth-order
weighted essentially non-oscillatory (WENO4) scheme
\citep{Levy_etal__2002__SIAM_JSciC__WENO4}, and approximate Riemann
solvers based on the multi-stage (MUSTA) method
\citep{Toro_Titarev__2006__JCP__MUSTA}.  The simulations reported here
are performed with the MP5 scheme and a MUSTA solver based on the HLL
Riemann solver \citep{Harten_JCP_1983__HR_schemes}.

The simulations are performed using cylindrical coordinates, and
include both three-dimensional and two-dimensional (i.e.,
axisymmetric) models.  The computational grid covers a region of a few
(typically one or two) kilometers aside resolved by at least 26 and at
most 800 zones per dimension, corresponding to a resolution between 40
and 0.625\,m.

\begin{figure}[t]
  \sidecaption
  \includegraphics[width=9.0cm]{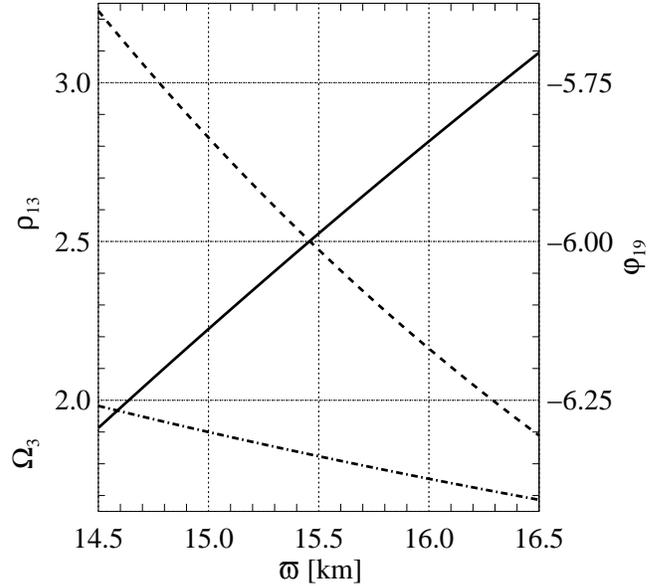}
  \caption{Hydrostatic structure of the initial models.  
    The diagram shows the gravitational potential $\varphi_{19} =
    \varphi / (10^{19}~\mathrm{erg}~\mathrm{cm}^{-3})$ (solid lines,
    right $y$ axis), the density $\rho_{13} = \rho / (10^{13} ~
    \mathrm{g}~\mathrm{cm}^{-3})$ (dashed line, left $y$ axis), and
    the angular velocity $\Omega_3 = \Omega / ( 10^3 ~
    \mathrm{s}^{-1})$ (dash-dotted line, left $y$ axis).  The entropy
    profile of this specific model is assumed to be flat. }
  \label{Fig:Init--Profiles}
\end{figure}

\subsection{Boundary conditions}
\label{sSec:BC}
In local simulations, the choice of boundary conditions is a crucial
issue, with possibly subtle effects on the flow geometry.  The
standard technique for local simulations of the MRI in accretion disks
is the \emph{shearing-sheet} method due to
\cite{Balbus_Hawley__1992__ApJ__MRI_3}.  This approach consists of two
important ingredients: (i) a transformation into a frame of reference
co-rotating at the mean angular velocity of the shearing box,
$\Omega_0$, and the linearizion of the rotation profile around
$\Omega_0$; (ii) the use of shearing-sheet boundaries in the radial
direction, and (in most cases) periodic boundary conditions in the
perpendicular directions.

Periodic boundary conditions are often used in simulations of a small,
representative sub-volume of a larger system.  These boundary
conditions are based on the idea that the entire system is covered by
a homogeneous (e.g., cubic) lattice of identical sub-volumes.
Consequently, the, e.g., left boundary of the simulated sub-volume is
identified with the right boundary of an identical sub-volume
translated by one lattice width.

A shearing box represents only a small part of the entire system.  The
influence of larger scales is considered by suitable boundary
conditions, the most natural choice being periodic ones.  These
boundary conditions, however, do not allow one to impose global
gradients throughout the shearing box, e.g., for differential rotation
($\partial_{\varpi} \Omega \neq 0$). This shortcoming is eliminated by
the linearization of the rotation profile and the transformation into
the co-rotating frame since, in this case, the deviation from the
background profile, $\delta \Omega$, is the dynamical variable rather
than $\Omega$ itself.  Thus, it is possible to use periodic boundary
conditions in the radial direction accounting for differential
rotation by an offset $\delta \phi (t) = t \left( \Omega_{out} -
\Omega_{inn}\right)$, as described by
\citep{Balbus_Hawley__1992__ApJ__MRI_3}, where $\Omega_{out,inn}$ are
the angular velocities at the outer and inner radial surface of the
shearing box, respectively.

In contrast to accretion disks, thermodynamic variables in stars may
have global gradients both in the direction perpendicular and parallel
to the gradient of $\Omega$.  Thus, standard shearing-sheet boundaries
cannot be used.  Instead, we follow
\cite{Klahr_Bodenheimer__2003__ApJ__Global-baroclinic-inst-disc} and
employ \emph{shearing-disc} boundary conditions.  We abandon the
transformation into the co-rotating frame and assume radial
periodicity of the deviation, $\delta q \equiv q - q_0$, of a variable
$q$ from a given background distribution $q_0$, instead of periodicity
of $q$ itself.  We define the background distribution $q_0$ by its
distribution at the initial time $t=0$, i.e. $q_0 \equiv q (\varpi;
t=0 )$.  This recipe is applied to density, momentum, and entropy.  As
\cite{Klahr_Bodenheimer__2003__ApJ__Global-baroclinic-inst-disc}, we
observe the development of resonant radial oscillations which are
suppressed, however, by damping the radial velocity in the first $n$
(we use $n=2$) computational zones at both radial boundaries.  We
point out that shearing-disc simulations allow for large-scale
modifications of global gradients.  In particular, angular momentum
transport may modify the global rotation profile, and change the
angular momentum and rotational energy of the matter in the
computational volume.  This process can eliminate the differential
rotation causing the instability, and thus, terminate the growth of
the MRI.

As we will show later, the evolution of our models depends crucially
on whether we do or do not apply this damping term.  However, we note
here that the artificial oscillations prevented by the damping do not
have a strong influence on the evolution of the MRI. We use, if at
all, a damping of $v_\varpi$ by $1.25\%$ in the innermost and
outermost two zones of the grid, which is a considerably weaker
damping than in the simulations of
\cite{Klahr_Bodenheimer__2003__ApJ__Global-baroclinic-inst-disc}.
Despite its relative weakness, the damping term is able to suppress
weak radial motions across the boundary.  Thus, it introduces a
preferred length scale (the radial size of the box) into the otherwise
shear-periodic simulation.  Comparing simulations with and without
damping (we will refer to these boundary conditions by \textit{d} and
\textit{p}, respectively), we can study the influence of a preferred
scale on the MRI.

The box size of standard shearing-sheet simulations does not define a
preferred length scale, i.e., these simulations are scale free and
entirely local.  In shearing-disc simulations, in contrast, the scale
height of the thermodynamic variables introduces a physical length
scale into the simulation.  If this preferred length scale is smaller
than the entire size of the star or disk, the simulations can be
characterized as being semi-global.

The semi-global approach falls in between a purely local and a global
one sharing merits and drawbacks with both methods.  Similar to local
simulations, semi-global simulations allow one to resolve a small part
of the entire system better.  Because they rely on a fixed lab frame
and do not eliminate the mean rotation, the basic time scales are the
same as in a global simulation of the same resolution.  In a Keplerian
disk dominated by rotation this might add a major difficulty to the
numerical treatment of the problem.  On the other hand, with pressure
dominating over kinetic energy, the time step of our simulations is
governed by the sound speed rather than the rotational velocity.  As
there is no way of eliminating the sound speed, we do not feel a need
to use a shearing-sheet transformation.

We expect the MRI in core collapse to grow and reach saturation within
several tens of milliseconds.  The time step $\delta t \lesssim \delta
x / c_\mathrm{S}$, on the other hand, is much smaller because of the
high value of the sound speed in a post-collapse core ($c_\mathrm{S}
\sim 10^{10}\,\mathrm{cm \, s}^{-1}$), where $\delta x$ is the width
of the computational zone.  Thus, we have to perform a large number
(typically several millions) of time steps, which implies a limit on
the grid resolution we can afford in the simulations, although the
resolution is still much better than that of a global simulation.

\subsection{Initial conditions}
\label{sSec:Init}

We use equilibrium initial models based on post-bounce cores from
\cite{Obergaulinger_Aloy_Mueller__2006__AA__MR_collapse}.  Several
tens of milliseconds after the core bounce, the shock wave has reached
distances of a few hundred kilometers, the post-shock region
exhibiting a series of damped oscillations as the proto-neutron star
relaxes into a \emph{nearly} hydrostatic configuration.  We extract
the radial profile of the gravitational potential along the equator of
their model A1B3G3\footnote{This model experiences a core collapse
  that is halted by the stiffening of the equation of state above
  nuclear matter density.}, and construct from that the density
stratification within our shearing box solving the equation of
hydrostatic equilibrium
\begin{equation}
  \label{Gl:Init-HS}
  0 =   \rho \partial_{\varpi} \varphi 
      - \partial_{\varpi} P 
      + \varpi \rho \Omega ^2 
\end{equation}
for a given rotation profile
\begin{equation}
  \label{Gl:Init-Omega}
  \Omega ( \varpi ) = \Omega_0 \varpi ^ { \alpha_\Omega } \,,
\end{equation}
where $\Omega_0$ and $\alpha_\Omega$ are constants.  The pressure is
determined using the hybrid equation of state in the form
\begin{equation}
  \label{Gl:Init-EOS}
  P = ( S + 1 ) \rho ^ {\gamma}  \,,
\end{equation}
assuming an entropy profile of the form
\begin{equation}
  \label{Gl:Init-Entropy}
  S ( \varpi ) = S_0 + S_1 ( \varpi - \varpi_0 )
\end{equation}
with constants $S_0$ and $S_1$.  \eqref{Gl:Init-HS} is solved in a
radial domain of size, $\Delta \varpi$, which is either one or two
kilometers large, centered at $\varpi_0 = 15.5\,\mathrm{km}$.  The
structure of an initial model, characterized by the set of parameters
$\Omega_0 = 1900 \,\mathrm{s}^{-1}$, $\alpha_\Omega = - 1.25$, $S_0 =
0$, and $S_1 = 0$, is shown in \figref{Fig:Init--Profiles}.  This
model has a radial density scale height of $H_{\rho} =
\frac{P}{\partial_{\varpi} P} \approx 3.8\,$km, i.e., our
computational grid covers a significant fraction of a density scale
height.  The rotation rate of $\sim 2\,000\,\mathrm{s}^{-1}$
corresponds to that of a rapidly rotating \pns~with a rotational
period of $\sim 3\,$ms.

Assuming that the background gravitational potential is a function of
$\varpi$ only, we construct cylindrically symmetric initial models.
This approximation is justified by the small size of the simulation
box in $z$-direction ($1\,$km) compared to its radial position
($15\,$km).

We added three different types of initial magnetic fields to the
initial hydrostatic model:
\begin{description}
\item[Model \emph{UZ}:] a uniform B-field in $z$-direction, $\vec b =
  \left(0,0,b^z_0 \right)^T$.
\item[Model \emph{VZ}:] a B-field in $z$ direction with vanishing net
  flux, $\vec b = \left( 0, 0, b^z_0 \sin \left( 2 \pi (\varpi -
  \varpi_0 ) / \Delta \varpi\right) \right)^T$.
\end{description}
In all models, the initial field is weak, both in compaison with the
thermal and the rotational energy of the models.  The weakness of the
associated Lorentz force justifies the use of \emph{hydrostatic}
instead of \emph{hydromagnetic} equilibria as initial conditions.

From \eqref{Gl:MRIds--gen--la_MRI} and the values of typical model
parameters we obtain the following estimate for the MRI wavelength
\begin{equation}
  \label{Gl:Init--Lambda_MRI}
  \lambda_\mathrm{\mathrm{FGM}}
  \sim
  6.9 \,\mathrm{km} 
  \left( \frac{b}{10^{15}\,\mathrm{G}}  \right)
  \left( \frac{\rho}{2.5\,10^{13}\,\mathrm{g}\,\mathrm{cm}^{-3}} 
  \right)^{-\halb}
  \left( \frac{\Omega}{1900\,\mathrm{s}^{-1}} \right)^{-1}
  \,.
\end{equation}
To properly simulate the evolution of the MRI, $\lambda_\mathrm{MRI}$
should be resolved by at least a few grid zones.  Using a grid
resolution of $10$\,m or $20$\,m, we thus can follow the growth of the
MRI for magnetic fields exceeding several $10^{12}\,\mathrm{G}$
\citep{Obergaulinger_Aloy_Mueller__2006__AA__MR_collapse,
  Obergaulinger_et_al__2006__AA__MR_collapse_TOV}.  To trigger the
instability, we impose a small random radial velocity perturbation
with an amplitude of a few times $10^{-4..-2}$ of the rotational
velocity.

\section{Results}
\label{Sec:Res}
%
\subsection{General considerations}
\label{sSek:Res--Gen}
In axisymmetry, the growth of the MRI requires a non-vanishing
poloidal initial field.  Axisymmetry restricts the dynamics of the
MRI, suppressing a class of instabilities that affect the evolution of
MRI-unstable modes (see below).  Consequently, the predictive power of
axisymmetric simulations for the evolution of the MRI is limited, and
we cannot rely on them in determining the saturation amplitude of the
instability in supernova cores.  The growth of the instability does,
however, not differ strongly from full 3D models.  Thus, we can use 2D
models to determine growth rates, while detailed conclusions can only
be drawn from 3D models.

In axisymmetry, the flow is dominated by \emph{channel modes}, a pattern
of predominantly radial
\footnote{In general, the channels are oriented parallel to the
  gradient of $\Omega$, wherever it points to.}
flows of alternating direction stacked in $z$ direction
\citep{Balbus_Hawley__1991__ApJ__MRI}.  As the MRI grows, the channels
start to merge and their characteristic length scales increase, but
they survive as coherent flow structures throughout the entire
evolution and, particularly, do not dissolve into turbulence.

The analysis of \cite{Goodman_Xu__1994__ApJ__MRI-parasitic} shows that
channel modes are an exact nonlinear solution of the axisymmetric MHD
equations, which explains their stability observed in many numerical
simulations.  They are, on the other hand, unstable to genuinely 3D
\emph{parasitic} instabilities of, e.g., Kelvin-Helmholtz type.
Consequently, in 3D, the channel modes appearing during the early
growth phase of the MRI, do not persist until saturation.  Instead,
the channels decay due to the growing parasitic instabilities, and
turbulence develops.

This basic picture emerged from many simulations of the MRI in
accretion disks.  As we will discuss in the following, our simulations
confirm this result for the MRI in supernova cores.

\subsection{Axisymmetric models with no entropy gradient}
\label{sSek:Res--2d}
%
\subsubsection{Uniform initial magnetic fields}
\label{sSek:Res--2d-uf}
\begin{figure}[htbp]
  \centering
  \includegraphics*[width=7cm]{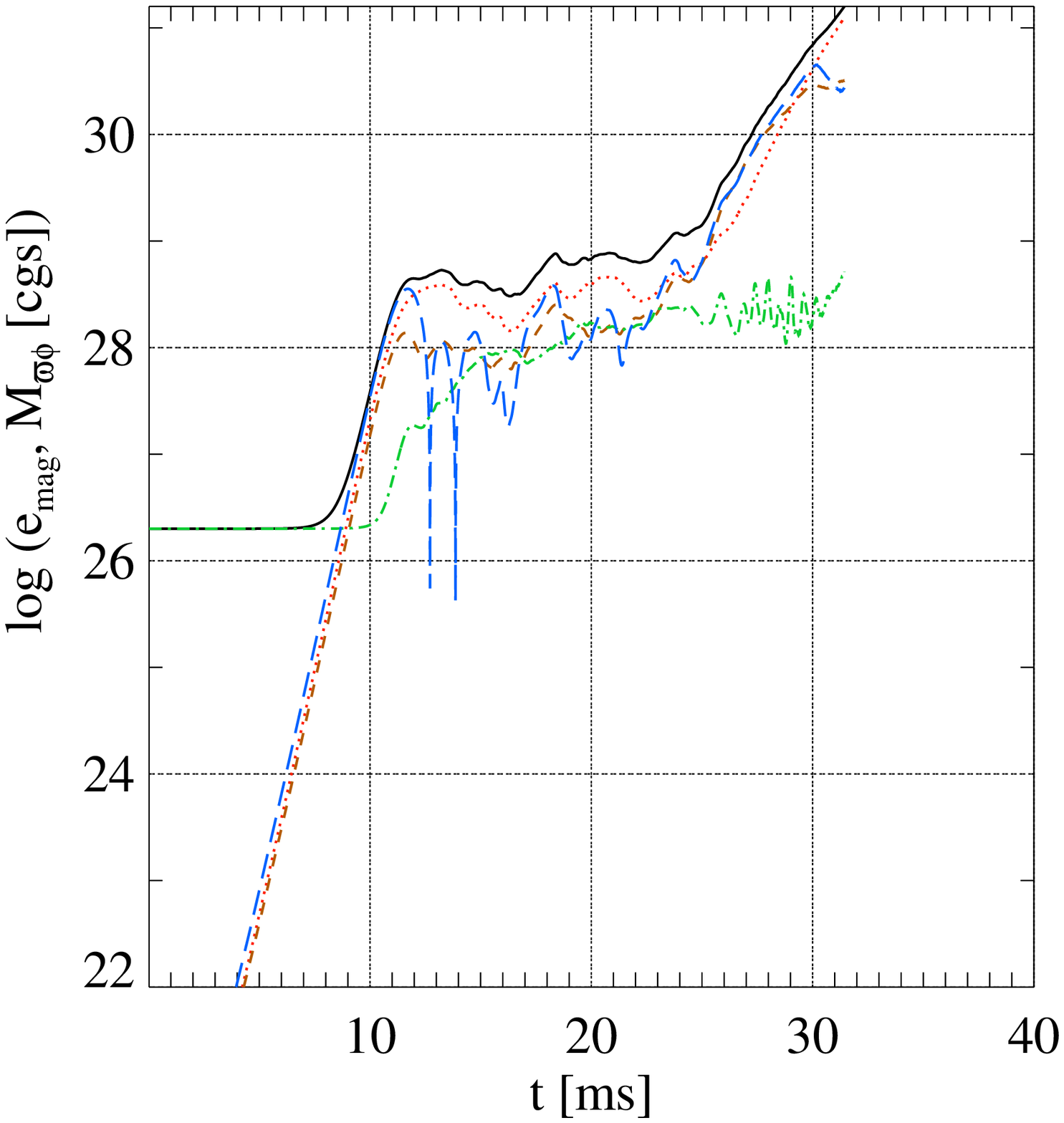}
  \caption{Evolution of the mean magnetic energy density
    $e^\mathrm{mag}$ (solid black line), the mean energy densities
    corresponding to the $\varpi$ (dotted red), $\phi$ (dashed brown),
    and $z$ (dash-dotted green) component of the magnetic field, and
    the absolute value of the mean Maxwell stress component
    $M_{\varpi,\phi}$ (dashed blue line) for an axisymmetric model
    with an initially uniform magnetic field $b_0^z = 2 \times
    10^{13}\,$G in z-direction, and a rotation law given by $\Omega_0
    = 1900\,$s$^{-1}$ and $\alpha_\Omega = -1.25$. The model was
    computed in a box of $L_\varpi \times L_z = 1 \mathrm{km} \times
    1\mathrm{km}$ with a grid resolution of $5\mathrm{m}$.  }
  \label{Fig:Res-2d--UZ2-4-tevo}
\end{figure}

Our models having no entropy gradient show the same dynamics as that
observed in previous simulations of the MRI in accretion discs
\citep[see, e.g.,][]{Balbus_Hawley__1998__RMP__MRI}.  We discuss first
the models with a uniform initial field $b_0^z$ in z-direction (model
series UZ2) focusing on models with a rotational law given by
$\Omega_0 = 1900\,$s$^{-1}$ and $\alpha_\Omega = -1.25$ (see
Eq.\ref{Gl:Init-Omega}).  The evolution of these models is
characterized by an exponential growth of the magnetic field, see
e.g., \figref{Fig:Res-2d--UZ2-4-tevo} for a model with $b_0^z = 2
\times 10^{13}\,$G.  The fastest growing MRI mode is well resolved in
this model, and its growth rate $\sigma_{\mathrm{MRI}} =
1.08\,$ms$^{-1}$ is found to be close to the theoretical prediction
$\sigma_{\mathrm{MRI}} \equiv \Im( \omega_{FGM} ) \approx \left|
\alpha_\Omega \Omega_0 / 2 \right| = 1.14\,$ms$^{-1}$ (see
\eqref{Gl:MRIds--oMRI-Alfv}).  The magnetic field reaches a maximum
value of about $10^{15}\,$G at $t \approx 15\,$ms, and the mean
Maxwell stress component $M_{\varpi \phi}$ (see
\eqref{Gl:Phys--Maxwell_V}) becomes large enough to alter the rotation
profile considerably within a few tens of milliseconds.  Consequently,
the angular momentum of the gas drops drastically at $t \approx
25\,$ms.

\begin{figure}[htbp]
  \centering
  \includegraphics[width=7cm]{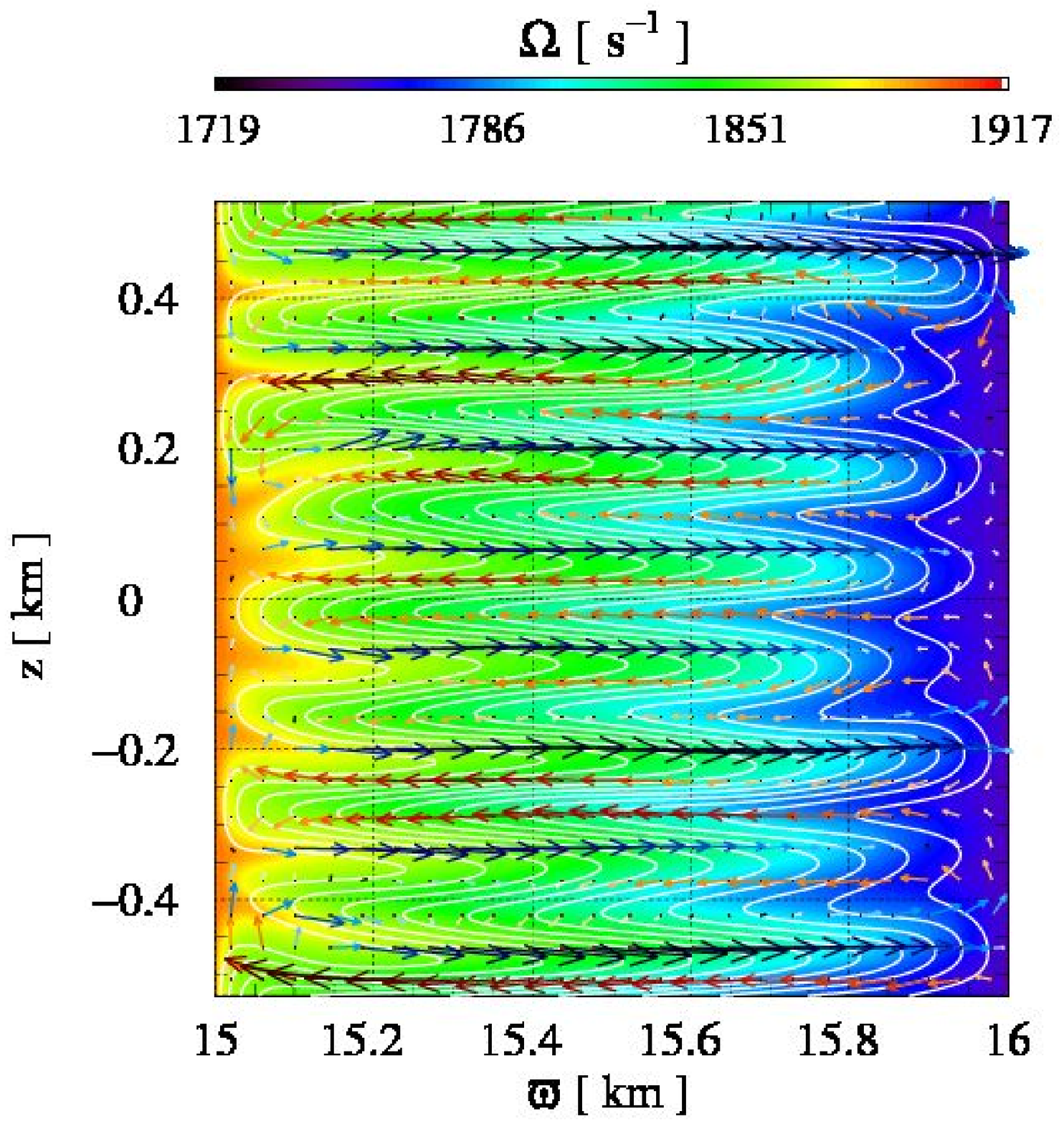}
  \includegraphics[width=7cm]{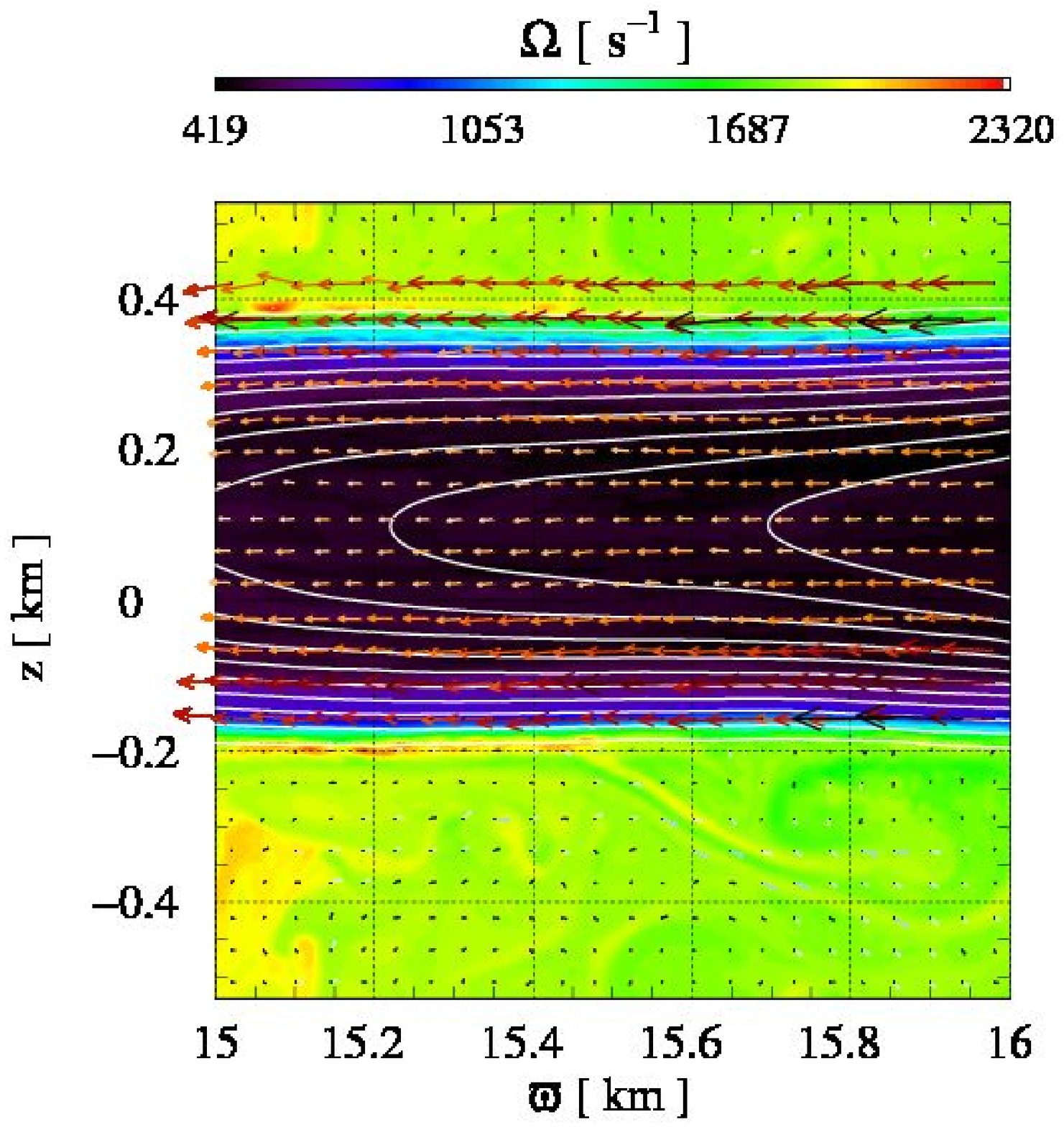}
  \caption{The channel modes present in two snapshots taken from the
    model for which \figref{Fig:Res-2d--UZ2-4-tevo} shows the time
    evolution.  The snapshots are taken at $t = 10.6\,$ms (upper
    panel) and $t = 30.7\,$ms (lower panel), respectively.  The panels
    show the color coded angular velocity $\Omega$, the magnetic field
    lines (white), and the flow field.  The colors of the velocity
    vectors indicate the magnitude and the direction of the flow: up-
    and down-flows are represented by blue and red vectors,
    respectively, their color intensity corresponding to the absolute
    value of the (poloiodal) velocity (the darker the larger). The
    maximum velocities are $2.7\times 10^7 \,$cm\,s$^{-1}$ (upper
    panel), and $8.8\times 10^8\,$cm\,s$^{-1}$ (lower panel),
    respectively.
  }
  \label{Fig:Res:channel-modes}
\end{figure}

The growth of the MRI proceeds via channel flows, whose vertical
extent and number depends on the initial magnetic field.  Two typical
channel flows are shown in \figref{Fig:Res:channel-modes}.  During the
early phase of the instability ($t = 10.6\,$ms; upper panel) eight
distinct channels are present each one consisting of a pair of up- and
down-flows in radial direction. The magnetic field is organized into
eight elongated radial sheets, and this pattern is also imprinted onto
the distribution of $\Omega$, as the magnetic field enforces
co-rotation along field lines.

A flow topology dominated by channel modes implies a phase of
exponential growth of the magnetic field, which ends when the channel
modes are disrupted and a less organized, more turbulent state ensues
(in \figref{Fig:Res-2d--UZ2-4-tevo} this happens at $t\approx
11\,$ms).  In most axisymmetric models, the turbulent state is only of
transient nature, because after some time coherent channel flows form
again leading to a secondary phase of exponential growth (see
\figref{Fig:Res-2d--UZ2-4-tevo} at $t \approx 23\,$ms).

Very late in the evolution ($t = 30.7\,$ms; lower panel of
\figref{Fig:Res:channel-modes}) we find only one large-scale channel
flow which extends across the entire domain in radial direction.  The
magnetic field is now predominantly radial and is concentrated near
the channel boundary.  This coherent flow pattern is the result of a
strong transport of mean angular momentum by Maxwell stresses.  The
stresses enforce co-rotation along field lines, and consequently turn
the rotation profile, initially constant on cylinders $\varpi =
\mathrm{const.}$, by 90 degrees, so that $\Omega$ becomes a function
of $z$ only.  We can distinguish two regions of slow and fast rotation
inside and outside $z \in [-0.15; 0.25 ]$, respectively.  Inside the
slowly rotating channel matter is accreting towards the center with
$v_\varpi \sim 4\times 10^{8}\,$cm\,$^{-1}$, while the rapidly rotating
gas outside the channel has much slower, random velocities.

\begin{figure}[htbp]
  \centering
  \includegraphics*[width=7cm]{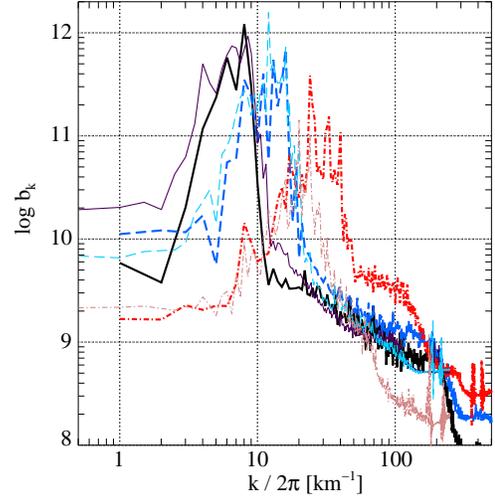}
  \caption{Radially averaged Fourier spectra of the radial component
    of the magnetic field, $b_\varpi (z)$, for different uniform-field
    models at $t \approx 7.5\,$ms.  Models with $b_0 = 4\times
    10^{12}\,$G, $10^{12}\,$G, and $2\times 10^{13}\,$G are shown by
    the red dash-dotted, the blue dashed, and the black solid line,
    respectively. Thick and thin lines refer to a computational domain
    of $1\,$km$^2$ and $2\,$km$^2$, respectively. For all models a
    grid of $400^2$ zones is used.
  }
  \label{Fig:UZ2-4-spectra}
\end{figure}

To investigate the dependence of the channel geometry on the initial
magnetic field and the grid resolution we compute Fourier spectra of
the radial component of the magnetic field, $b_\varpi$, for models
with $\Omega_0 = 1900\,$s$^{-1}$ and $\alpha_\Omega = -1.25$, and an
initial field strength of $4$, $10$, and $20\times 10^{12}\,$G,
respectively. The simulations are performed in a box of either $1
\times 1\,$km$^2$ or $2 \times 2\,$km$^2$ using a $400^2$ grid
(\figref{Fig:UZ2-4-spectra}).  At each radius we Fourier-transform
$b_\varpi(z)$, and the resulting spectra $b_{\varpi} (k_z)$ (where
$k_z$ is the vertical wave number) are then averaged over radius.  We
applied this procedure to the models during the growth phase of the
instability at $t \approx 7.5 \, \mathrm{ms}$.  The first set of
models ($1\,$km$^2$ domain, 2.5\,m spatial resolution) exhibits growth
rates close to the theoretical values, while this only partially holds
for the models of the second set of models ($4\,$km$^2$ domain, 5\,m
spatial resolution). Due to insufficient spatial resolution the MRI in
the model with the weakest initial field ($b_0^z = 4\times
10^{12}\,$G) grows slower than theoretically predicted. However, for
the two more strongly magnetized models of this set ($b_0^z = 10$, and
$2\times 10^{13}\,$G) the fastest growing modes are well resolved, and
the MRI growth rates agree with the theoretical ones.

For each model the spectrum shows a distinct maximum corresponding to
a dominant vertical length scale given by the width of one channel
mode.  The position of this maximum is a function of the initial
magnetic field only, $k_\mathrm{max} \propto b_0^{-1}$, and thus does
neither depend on the size of the computational domain nor on the
resolution.  A dependence on the last quantities is only observed, if
the fastest growing mode is under-resolved.  In this case, we recover
the low-$k$ wing of the spectral peak, but find a truncated spectral
distribution at higher wave numbers/smaller length scales.

\begin{figure}[htbp]
  \centering
  \includegraphics[width=7cm]{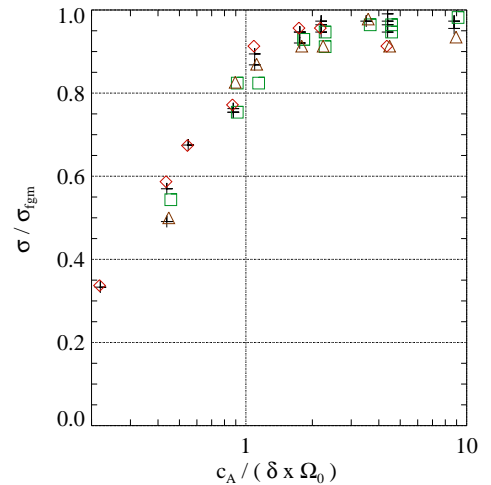}
  \caption{Growth rate $\sigma$ of the MRI for axisymmetric models
    with uniform initial field as a function of the initial Alfv\'en
    speed normalized to the rotational velocity and the grid
    resolution, $c_\mathrm{A} / (\delta x\, \Omega_0)$.  The colored
    symbols distinguish different initial rotational laws, where
    $\left( \Omega_0, \alpha_\Omega \right)$ are equal to
    $(1900\,\mathrm{s}^{-1}, -1.25)$ [black plus signs], 
    $(1900\,\mathrm{s}^{-1}, -1)   $ [red diamonds], 
    $( 950\,\mathrm{s}^{-1}, -1.25)$ [green squares], 
    and
    $( 950\,\mathrm{s}^{-1}, -1)   $ [brown triangles], 
    respectively.
 }
  \label{Fig:UZ2-n--growthrates}
\end{figure}

MRI theory predicts that the growth rate is independent of the initial
field strength.  Neglecting magneto-convective modes, we can expect to
observe this behavior in numerical simulations only if the grid is
sufficiently fine to resolve the fastest growing modes close to
$\lambda_\mathrm{MRI}$. Otherwise, if the grid is too coarse the
growth rate should be much smaller.  Our simulations reproduce this
behavior.  We show a comparison of the maximum growth rates from
linear analysis ($\sigma_\mathrm{FGM} = \Im(\omega_\mathrm{FGM} )
\approx |\frac{1}{2} \alpha_\omega \Omega_0|$) and the numerical ones
for models with different initial rotation laws ($\Omega_0$ ranging
from 950\,s$^{-1}$ to 1900\,s$^{-1}$, and $\alpha_\Omega$ from -1 to
-1.25) in \figref{Fig:UZ2-n--growthrates}.  If $\lambda_\mathrm{MRI}$
is under resolved for a given initial field $b_0$, the growth rate
increases with $b_0$, but once the MRI wavelength is well resolved,
the growth rate becomes constant as theoretically predicted.
\figref{Fig:UZ2-n--growthrates} implies the following criterion for a
sufficient resolution of the MRI: $\Delta \varpi \gtrsim 2 \Omega_0 /
c_{\mathrm{A}}$.  The growth rate of the instability does not depend
on the size of the computational domain.  For models with strong
initial fields the computed MRI growth rate is smaller than
$\sigma_{\mathrm{FGM}}$, because the MRI wavelength, i.e., the
wavelength of the fastest growing mode, exceeds the box size. Thus, we
can only properly simulate the slower growth of shorter modes.

\subsubsection{Channel disruption and MRI termination}
\label{subsub-channel}
As long as the dynamics of the model is dominated by channel modes,
the MRI grows exponentially.  We observe a termination of its initial
exponential growth -- henceforth called \emph{MRI termination} -- as
soon as the coherent channels are disrupted. Further MRI growth occurs
after an eventual reformation of the channel flows.  To understand
these processes better, we study MRI termination in a large number of
axisymmetric models with different initial magnetic fields, boxes of
different size and grid resolution, and different boundary conditions.

\begin{figure}[htbp]
  \centering
  \includegraphics[width=7cm]{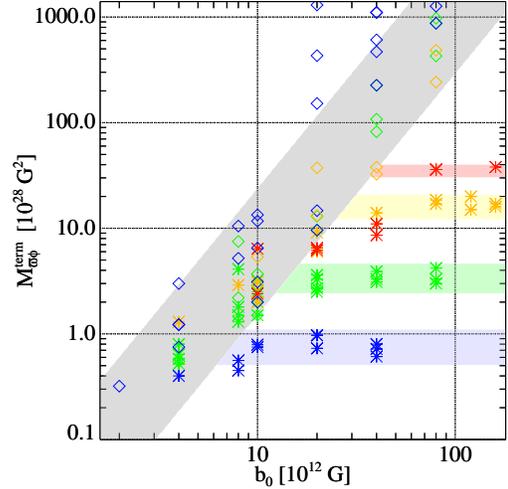}
  \caption{Volume-averaged Maxwell stress component
    $M^{\mathrm{term}}_{\varpi \phi}$ at MRI termination as a function
    of the initial magnetic field strength, $b_0$, for axisymmetric
    models with uniform initial magnetic field in $z$-direction , a
    rotational profile $\Omega= 1900\,$s$^{-1}\, \varpi^{-1.25}$, and
    a vanishing entropy gradient for a set of axisymmetric models.
    Blue, green, orange, and red symbols correspond to models computed
    in a square box having an edge size of 0.5, 1, 2, and $4\,$km,
    respectively.  Models computed with and without velocity damping
    at the radial boundaries are denoted by asterisks and diamonds,
    respectively.  The latter models show a box size independent
    scaling $M^{\mathrm{term}}_{\varpi \phi} \sim (b_0^z)^{16/7}$
    (gray band), while in models with damping $M^{term}_{\varpi \phi}$
    saturates at high field strengths the saturation value depending
    on the box size (colored horizontal bands).
  }
  \label{Fig:UZ2-R4S0--termination}
\end{figure}

\figref{Fig:UZ2-R4S0--termination} shows the value of the mean Maxwell
stress component $M^{\mathrm{term}}_{\varpi,\phi}$ at MRI termination
as a function of the initial magnetic field strength, $b_0$ for models
with a rotational law $\Omega= 1900\,$s$^{-1} \varpi^{-1.25}$ and a
vanishing entropy gradient.  We can distinguish two classes of models
according to the boundary conditions applied in the simulations (see
section\,\ref{sSec:BC}), the qualitative difference between the models
with and without velocity damping near the boundaries being quite
remarkable given the weak damping we apply. In models with damping
$M^{\mathrm{term}}_{\varpi,\phi}$ grows with increasing initial field
strength until it levels off at a grid size dependent value (colored
bands in \figref{Fig:UZ2-R4S0--termination}).  For the same models,
when simulated without damping, we find that
$M^{\mathrm{term}}_{\varpi,\phi} \propto b_0^{16/7}$ (gray band in
\figref{Fig:UZ2-R4S0--termination}) independent of the grid size.  The
``outliers'' in the upper part of the figure correspond to models
computed with a higher grid resolution than most other models. We will
discuss this fact below.

To determine whether the radial or the vertical size of the
computational grid is responsible for the leveling off of the Maxwell
stress in the runs with radial damping we simulate two models with a
grid of $0.5\, \mathrm{km} \times 2\,\mathrm{km}$ (for short called
\emph{high} models in the following), and $2\,\mathrm{km} \times
0.5\,\mathrm{km}$ (\emph{long} models), respectively.  Our results
show that the determining factor for the growth is primarily the
radial rather than the vertical box size, as both models follow the
behavior of $M_{\varpi,\phi}$ as a function of $b_0$ for the
respective radial grid sizes.  The two classes of models also exhibit
remarkably different post-growth dynamics.  In the high models, a few
channel modes reappear from the turbulent state, and a secondary phase
of exponential growth of $M_{\varpi,\phi}$ sets in.  Eventually, two
of the newly formed channel modes merge.  By this process, which
occurs repeatedly, the number of channels decreases, and the final
state of the flow is dominated by one short but wide channel mode.  In
the long models, on the other hand, no secondary exponential growth is
observed, and the Maxwell stress remains approximately constant,
albeit oscillating considerably due to the temporary presence of
coherent flow patterns.

To interpret these results, one has to analyze the mechanism
responsible for the disruption of the channel modes.  We discuss this
mechanism for an undamped model with $\Omega_0 = 1900\,$s$^{-1}$,
$\alpha_\Omega = -1.25$, and an initial magnetic field strength $b_0 =
4\times 10^{13}\,$G using a box of $0.5\,\mathrm{km} \times 0.5\,$km and a
resolution of $100 \times 100$ zones.  During the growth of the
instability a few large channels are present, which are disrupted at
MRI termination (at $\approx 15.9\,$ms).

\begin{figure*}[htbp]
  \centering
  \hspace{-0.4cm}
  \includegraphics[width=6cm]{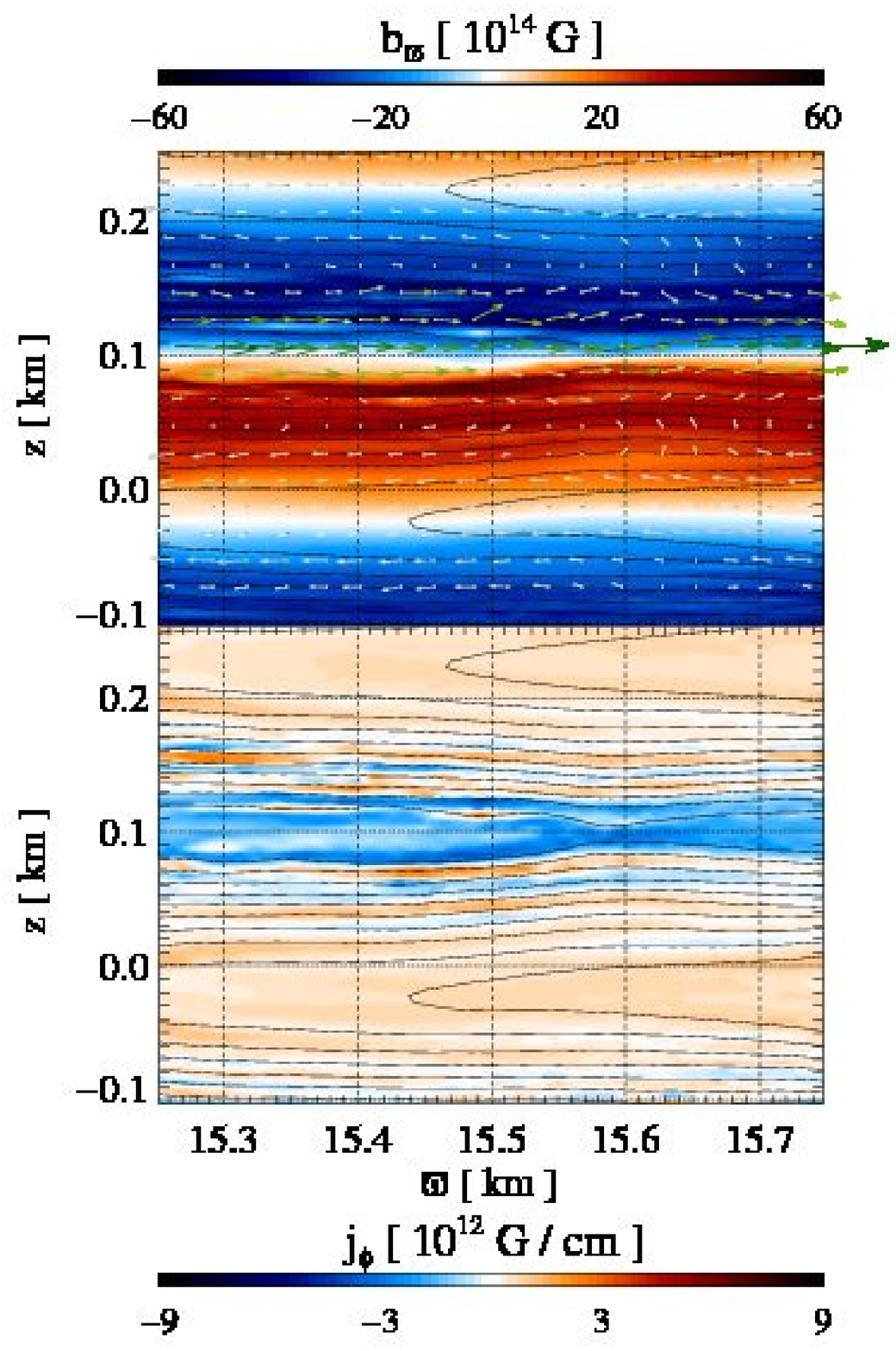}
  \includegraphics[width=6cm]{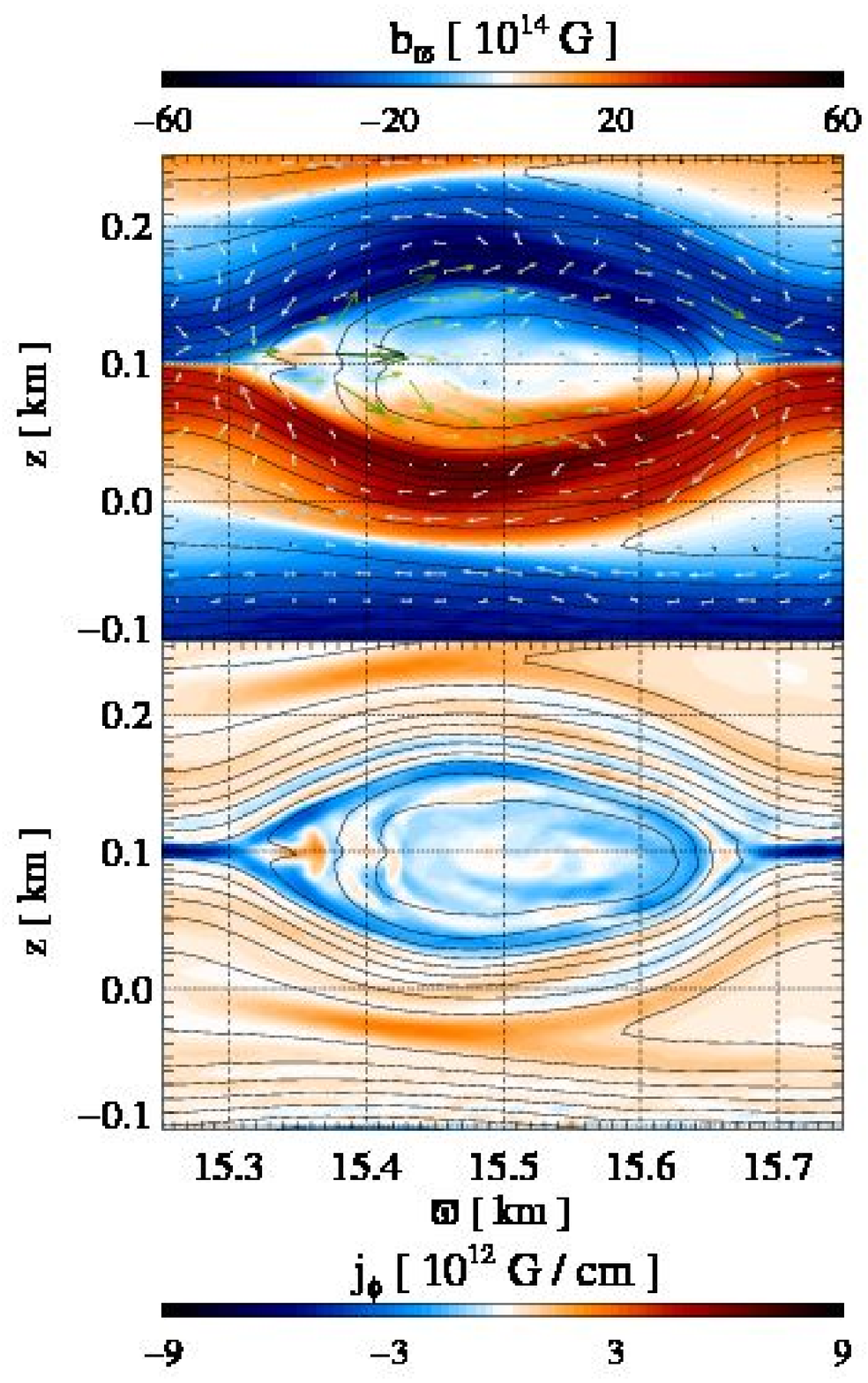}
  \includegraphics[width=6cm]{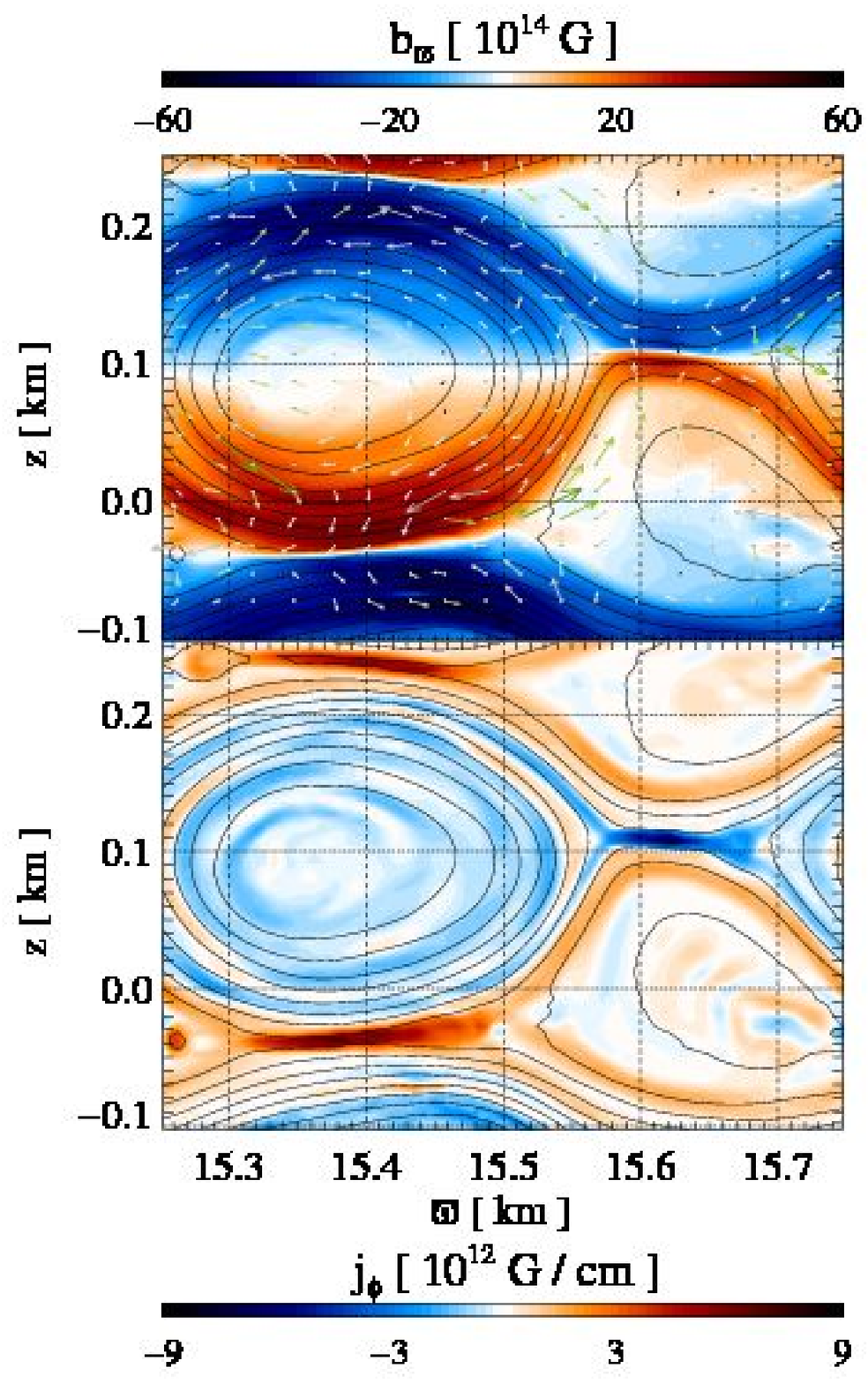}
  \caption{The disruption of a channel mode in an axisymmetric
    uniform-field model.  We show a section of a model with an initial
    field $b_0 = 4\times 10^{13}\,$G computed on a grid of $0.5 \times
    0.5\,$km$^2$.  The left, middle, and right panels display the
    color-coded radial component of the magnetic field $b_\varpi$
    (top) and the current density $j_\phi = \left(\vec \nabla \times
    \vec b\right)_\phi$ (bottom) before ($t = 15.850\,$ms), during ($t
    = 15.983\,$ms), and after ($t = 16.078\,$ms) the violent
    disruption of the channel flow, respectively. Additionally,
    magnetic field lines (black lines), and the velocity field
    (arrows; top only) are shown.  The arrows are color-coded
    according to the magnitude and direction of the flow. Inflows and
    outflows are shown by gray and green vectors, respectively. The
    longest vector corresponds to a velocity of $|v| = 7\times
    10^8\,$cm/s.
  }
  \label{Fig:UZ2-R4S0--channel-disr}
\end{figure*}

\figref{Fig:UZ2-R4S0--channel-disr} illustrates the disruption of one
of the channel flows in some detail. At $t = 15.850\,$ms, the channel
flow is still intact (left panel), and one recognizes two broad
streams of in-flowing and out-flowing gas both permeated by a strong
radial magnetic field of opposite polarity.  A broad current sheet
separates the two flow regions.  Owing to small-scale fluctuations in
the flow, the field lines are not perfectly (anti-)parallel, and the
current sheet is slightly deformed.  These deformations act as seed
perturbations for resistive instabilities of the \emph{tearing-mode}
type.  Although we evolve the equations of \emph{ideal} MHD neglecting
resistivity, the presence of \emph{numerical} resistivity enables the
growth of these instabilities, leading to a reconnection of
anti-parallel field lines.  As a consequence, the elongated current
sheet dissolves into a configuration of \emph{X} and \emph{O} points
(located at $\varpi \sim 15.25\,$km and $\sim 15.5\,$km, respectively;
see middle panel of \figref{Fig:UZ2-R4S0--channel-disr}).  When field
lines reconnect near the X point, the fluid is accelerated away from
the reconnection point towards the O point. This causes the intense
gas flow in positive radial direction at $(\varpi,z) \sim (15.35,
0.1)\,$km.  The change of the topology of the magnetic field and of
the flow continues shortly afterward (at $t = 16.078\,$ms; right panel
of \figref{Fig:UZ2-R4S0--channel-disr}).  The O point has grown in
size, and the fluid is in vortical motion.  As the vortex grows, field
lines in the vortex are advected towards field lines of opposite
polarity belonging to an adjacent channel flow (centered at $z \approx
-0.15\,$km), where reconnection occurs. Note the formation of a second
X point at $(\varpi,z ) \approx (15.38, -0.03)\,$km
(\figref{Fig:UZ2-R4S0--channel-disr}, right panel).

\begin{figure}[htbp]
  \centering
  \includegraphics[width=7cm]{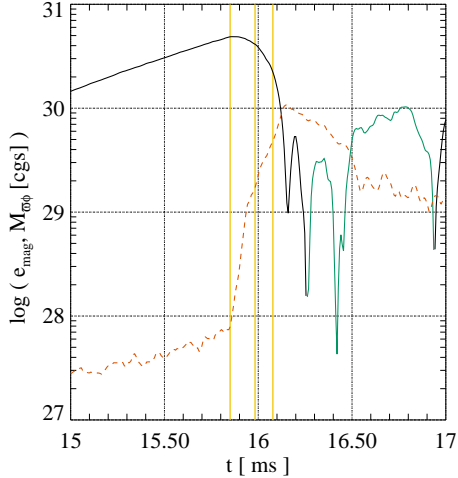}
  \caption{Temporal evolution of the absolute value of the mean
    Maxwell stress component $M_{\varpi \phi}$ (solid line; the line
    is colored black where $M_{\varpi \phi} < 0$, and green
    otherwise), and of the magnetic energy density of the z-component
    of the magnetic field, $e_{\mathrm{mag}}^z$ (dashed red line) of
    the model shown in \figref{Fig:UZ2-R4S0--channel-disr}.  The
    vertical yellow lines mark the times of the snapshots shown in
    \figref{Fig:UZ2-R4S0--channel-disr}.
  }
  \label{Fig:UZ2-R4S0--channel-disr-tevo}
\end{figure}

To demonstrate the growth of the tearing-mode instability and to
support its importance for MRI termination, we compare the evolution
of the mean Maxwell stress component $M_{\varpi \phi}$ and of the
magnetic energy density of the z-component of the magnetic field,
$e_{\mathrm{mag}}^z$ of the model (see
\figref{Fig:UZ2-R4S0--channel-disr-tevo}).  Before the tearing mode
grows $M_{\varpi \phi}$ and $e_{\mathrm{mag}}^z$ increase, their
growth rates being similar to that of the MRI.  At $t = 15.850\,$ms,
the growth rate of $e_{\mathrm{mag}}^z$ becomes larger than that of
the MRI by one order of magnitude within less than 0.2\,ms, whereas
$M_{\varpi \phi}$ approaches a maximum.  Once the tearing mode is
fully operative ($t = 15.983\,$ms), the growth of $e_{\mathrm{mag}}^z$
becomes slower but still continues due to the appearance of more
tearing modes (see, e.g., the right panel of
\figref{Fig:UZ2-R4S0--channel-disr-tevo} at $(\varpi,z ) \approx
(15.38, -0.03)\,$km).  Subsequently, $e_{\mathrm{mag}}^z$ begins to
decrease as the tearing modes saturate.

According to the previous discussion the dynamics of the channel flows
is dominated by the interplay between their growth due to the MRI and
their destruction by resistive instabilities.  Channel flows are
unstable against tearing-mode-type instabilities at any point in their
evolution.  To study these instabilities in more detail, we have
performed a set of simulations (see App.\,\ref{Sec:App-Resi}) using
simplified models of channel flows.  We recapitulate our results,
summarized in \eqref{Gl:App-Resi-sigma-result}, here:
\begin{equation}
  \label{Gl:Res-resi-sigma}
  \sigma_{\mathrm{r}} \propto
  \left( c_\mathrm{A} \right)^{7/4}
  \left( c_\mathrm{S} \right)^{-3/4}
  \left( a \right)^{-2}
  \left( \delta x \right)^{1} ,
\label{GL:Resi-sigma}
\end{equation}
where $\sigma_\mathrm{r}$, $c_\mathrm{A}$, $c_\mathrm{S}$, $a$, and
$\delta x$ are the growth rate of the instability, the Alfv\'en
velocity corresponding to the channel magnetic field, the sound speed,
the width of the channel, and the grid resolution, respectively.
because the instability is not based on a physical resistivity, but is
of purely numerical origin, no physical transport coefficient appears
in \eqref{Gl:Res-resi-sigma}. However, our results can be interpreted
in terms of an effective resistivity $c_\mathrm{A} \delta x$, as
detailed in App.\,\ref{Sec:App-Resi}.  In our models the width of the
channel flow, $a$, is set by the MRI wavelength corresponding to the
initial vertical magnetic field:
\begin{equation}
  \lambda_\mathrm{MRI} \propto b_0^z / \sqrt{\mathcal{R}_{\varpi}}
                      \propto b_0^z / ( \sqrt{\alpha_\Omega} \Omega_0)
\end{equation}
(see \eqref{Gl:MRI--lamda_crit}, \eqref{Gl:MRIds--gen--la_MRI}, and
\eqref{Gl:Init-Omega}).  The width remains constant during the growth,
as only mergers of adjacent channels occurring as a result of resistive
instabilities can change the field topology. 

Our basic proposition for MRI termination is that channel flows are
disrupted once the growth rate of the resistive instability exceeds
the MRI growth rate:
\begin{equation}
  \label{Gl:Res--channel-term-cond}
  \sigma_{\mathrm{r}} > \sigma_{\mathrm{MRI}} \;
  \Rightarrow \mathrm{MRI~termination}\, .
\end{equation}

\begin{figure*}
  \centering
  \includegraphics[width=14cm]{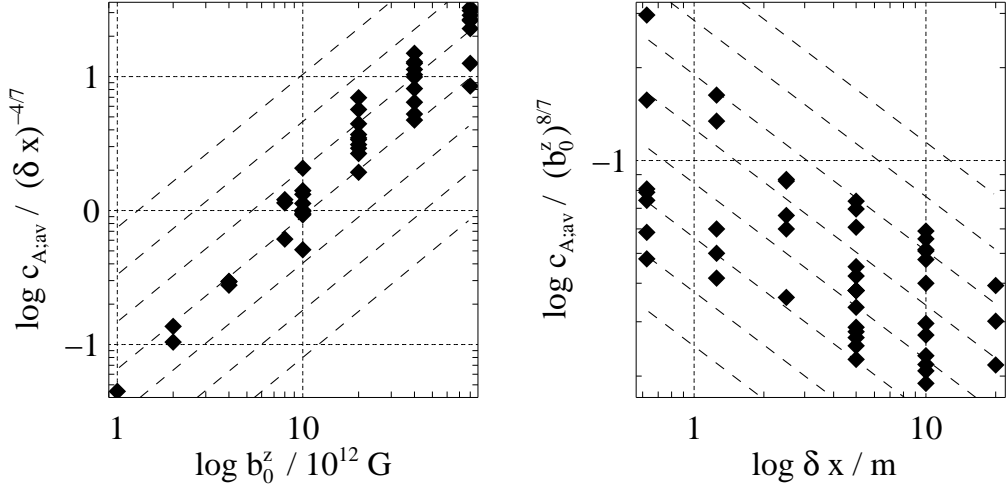}
  \caption{Average Alfv\'en velocity -- of models with uniform initial
    magnetic field and without velocity damping near the radial
    boundaries -- corresponding to the radial magnetic field at MRI
    termination normalized to $(\delta x)^{-4/7}$ (left) and
    $(b_0^z)^{8/7}$ (right) as a function of the initial magnetic
    field strength, $b_0^z$ (left), and the grid resolution $\delta x$
    (right).  The dashed lines represent the power laws expected
    from our analysis of resistive instabilities (see
    \eqref{Gl:Res--channel-term-cA-Omega}).  Note that we only
    consider well-resolved models ($\sigma_{\mathrm{MRI}} \ge
    0.95\,$ms$^{-1}$) here.
  }
  \label{Fig:Res--MRI-term-scaling}
\end{figure*}

Using in addition the functional dependence of $\sigma_{\mathrm{r}}$
(\eqref{GL:Resi-sigma}), we can establish scaling laws for MRI
termination for a given hydrodynamic background model. As the channel
width scales with the MRI wavelength, $a \propto
\lambda_\mathrm{MRI}$, and as the MRI growth rate is given by
$\sigma_{\mathrm{MRI}} \equiv \Im( \omega_{FGM} ) \propto
\alpha_\Omega \Omega_0$ (see \eqref{Gl:MRIds--oMRI-Alfv}), we find for
the Alfv\'en speed at MRI termination
  \begin{equation}
    \label{Gl:Res--channel-term-cA-Omega}
    c_{\mathrm{A}}^{\mathrm{term}}
    \propto
    \left( c_{\mathrm{S}} \right)^{3/7}
    \left( b_0^z \right)^{8/7}
    \left( \Omega_0 \right) ^ {-4/7}
    \left( \delta x \right) ^ { -4/7}
    ,
  \end{equation}
and for the corresponding Maxwell stress
  \begin{equation}
    \label{Gl:Res--channel-term-Mxy-Omega}
    M_{\varpi\phi}^{\mathrm{term}}
    \propto
    \left( c_{\mathrm{S}} \right)^{6/7}
    \left( b_0^z \right)^{16/7}
    \left( \Omega_0 \right) ^ {-8/7}
    \left( \delta x \right) ^ { -8/7}
    .
  \end{equation}
The latter equation implies that $M_{\varpi\phi}^{\mathrm{term}}$
decreases with faster rotation.  Two effects play a role in
explaining this behavior. Firstly, slower rotation leads to slower MRI
growth ($\sigma_{MRI} \propto \Omega_0$), and hence weaker magnetic
fields are required for the tearing modes to overcome the MRI growth.
Secondly, slower rotation implies wider channel flows ($a \propto
(\Omega_0)^{-1}$), i.e., resistive instabilities grow slower as
$\sigma_r \propto a^{-2} \propto (\Omega_0)^2 $ (see
\eqref{GL:Resi-sigma}).

The qualitative features of these scaling relations are:
\begin{enumerate}
\item Stronger initial vertical fields and correspondingly wider
  channels tend to suppress resistive instabilities.  Consequently,
  MRI termination requires more strongly magnetized channel flows.
\item Finer grid resolution implies less numerical viscosity, and
  hence larger values for $c_{\mathrm{A}}^{\mathrm{term}}$ and 
  $M_{\varpi\phi}^{\mathrm{term}}$.
\item The scaling of $M_{\varpi\phi}^{\mathrm{term}}$ with the sound
  speed implies a proportionality between the Maxwell stress and the
  background pressure: $P \propto c_{\mathrm{S}}^2$, and thus
  $M_{\varpi\phi}^{\mathrm{term}} \propto P^{3/7}$.  This scaling is
  reminiscent of the $\alpha$-law in accretion discs according to
  which the (MRI-generated) viscosity is proportional to the gas
  pressure.
\end{enumerate}

\begin{figure}
  \centering
  \includegraphics[width=6.5cm]{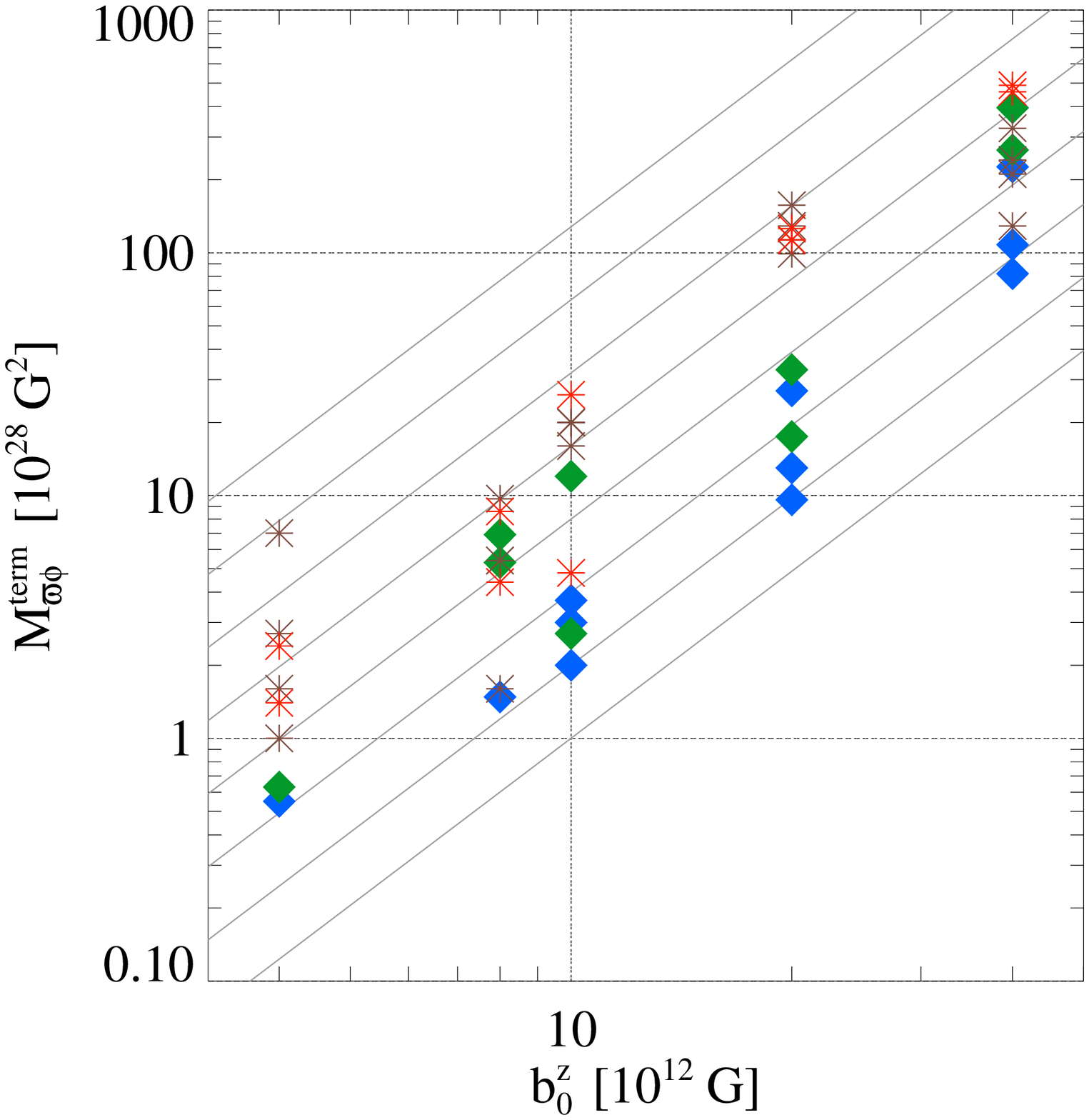}
  \includegraphics[width=6.5cm]{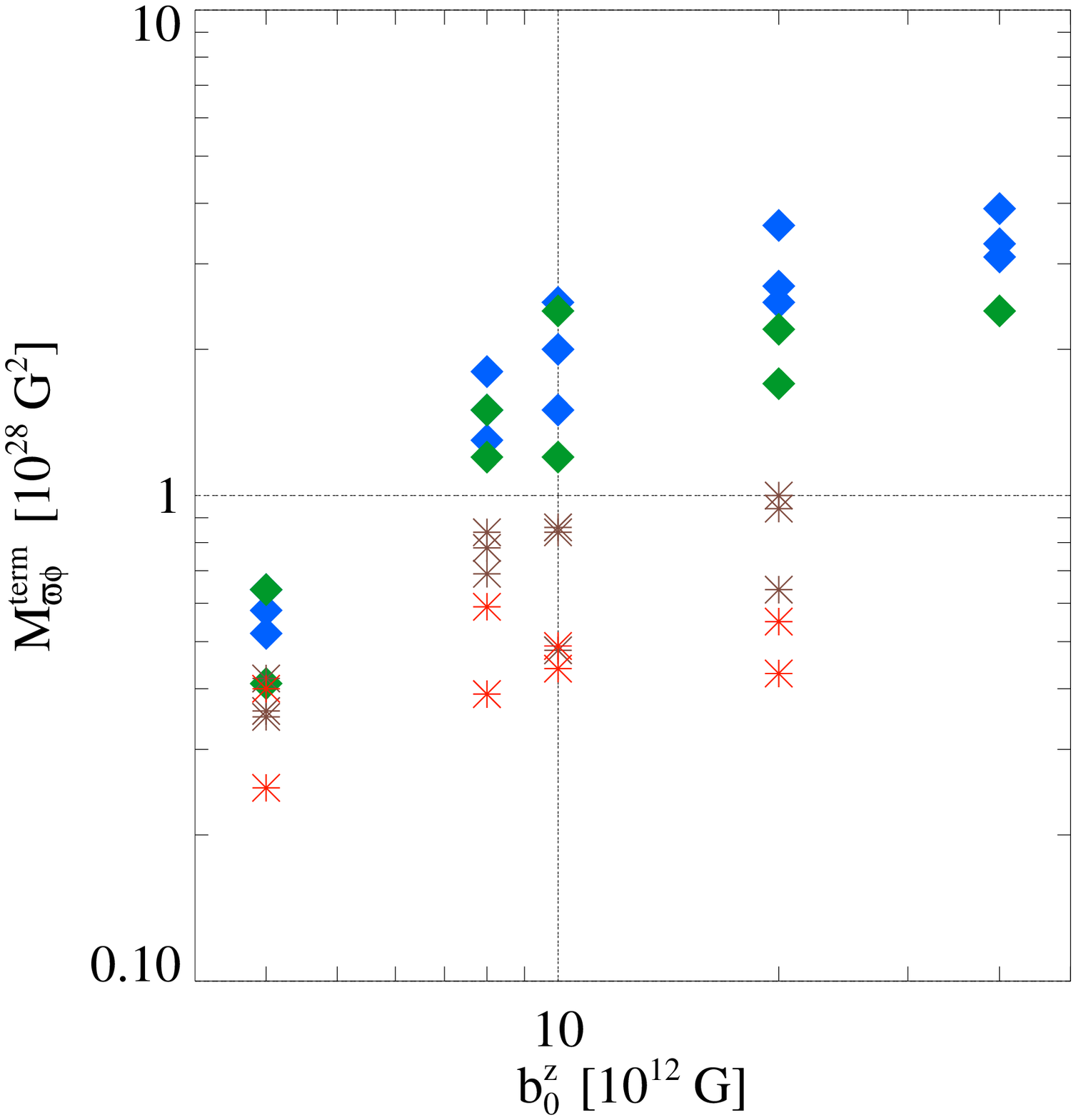}
  \caption{ 
    Maxwell stress, $M_{\varpi\phi}^{\mathrm{term}}$ at MRI
    termination as a function of the initial magnetic field strength,
    $b_0^z$ for models with different initial rotational profiles.
    The upper and lower panels show models with non-damping and
    damping boundary condition, respectively.  The colored symbols
    distinguish different initial rotational laws where $\left(
    \Omega_0, \alpha_\Omega \right)$ are equal to
    $(1900\,\mathrm{s}^{-1}, -1.25)$ [blue diamonds], 
    $(1900\,\mathrm{s}^{-1}, -1)   $ [green diamonds], 
    $( 950\,\mathrm{s}^{-1}, -1.25)$ [brown asterisks], 
    and
    $( 950\,\mathrm{s}^{-1}, -1)   $ [red asterisks], 
    respectively.  Note that only models with a box size $L_{\varpi}
    \times L_z = 1 \times 1 ~ \mathrm{km}^2$ and a resolution of 50,
    100, or 200 zones (per dimension) are considered here.  The light
    gray lines in the upper panel illustrate power laws $\propto
    \left( b_0^z \right)^{16/7}$.
  }
  \label{Fig:U2-Rdp-termination}
\end{figure}

\figref{Fig:Res--MRI-term-scaling} shows that the average Alfv\'en
velocity corresponding to the radial\footnote{The radial field is
  typical for all three components.  Thus the Alfv\'en velocity
  corresponding to the total magnetic field shows the same
  dependence.}  magnetic field at MRI termination is well described by
the scaling law given in
\eqref{Gl:Res--channel-term-cA-Omega}. Similarly,
\figref{Fig:UZ2-R4S0--termination} and \figref{Fig:U2-Rdp-termination}
(upper panel) confirm that the data for
$M_{\varpi\phi}^{\mathrm{term}}$ obey the corresponding scaling law
(\eqref{Gl:Res--channel-term-Mxy-Omega}), too.  The upper panel of the
latter figure shows $M_{\varpi\phi}^{\mathrm{term}}$ as a function of
the initial magnetic field strength for models with different initial
rotational laws.  Obviously, the proportionality
$M_{\varpi\phi}^{\mathrm{term}} \propto \left( b_0^z \right)^{16/7}$
(light gray lines) provides a good approximation to the behavior of
the models.  Due to the small number of models the results should be
taken with care, but a strong anti-correlation of
$M_{\varpi\phi}^{\mathrm{term}}$ with $\Omega_0$ is suggested. The
data also do not support any dependence of
$M_{\varpi\phi}^{\mathrm{term}}$ on $\alpha_\Omega$.  Finally, in
\figref{Fig:UZ2-R4S0--termination} we noticed earlier some outliers at
large values of $M_{\varpi\phi}^{\mathrm{term}}$ which correspond to
models computed on a fine grid.  However, considering that MRI
termination depends on grid resolution, all models lie within a narrow
band which corroborates our scaling laws and provides more evidence of
the importance of resistive instabilities in understanding the MRI.
Consequently, physical (instead of numerical) transport coefficients
should be used in MRI simulations, which may give rise to different
scaling laws considering the growth rate of tearing modes.

\paragraph{Models with velocity damping}
In models with uniform initial magnetic fields where the radial
velocity is damped near the inner and outer radial boundary, i.e., in
models located in the horizontal bands in
\figref{Fig:UZ2-R4S0--termination}, MRI termination happens earlier
than predicted by our scaling laws, and the Maxwell stresses saturate
for strong initial magnetic fields, the saturation value of
$M_{\varpi\phi}^{\mathrm{term}}$ being smaller for slower rotation
(see lower panel of \figref{Fig:U2-Rdp-termination}).  This is due to
the reconnection instability occurring close to the radial boundaries
well before the theoretically predicted time of MRI termination.  This
premature reconnection is caused by the field geometry: due to the
suppressed motion across the inner and outer radial boundary field
lines must bent there in $z$-direction.  Consequently, field lines of
opposite polarity approach each other much earlier than in models
without velocity damping, and efficient reconnection ensues.  In this
case, the onset of reconnection is determined by the field geometry
rather than by the initial field strength.  With reconnection
occurring in the bent flux sheets near the radial boundaries instead
between parallel flow sheets in the bulk volume as for non-damping
boundaries, the width of a flux sheet is less important in determining
the resistive growth rates. Thus, a slower MRI growth and smaller
Maxwell stresses are found for slower rotating models when velocity
damping is imposed.  Apart from the dependence of
$M_{\varpi\phi}^{\mathrm{term}}$ on $\Omega_0$, we also find a
dependence on $\alpha_\Omega$.  Both dependences together give rise to
a monotone relation between the strong-field limit of
$M_{\varpi\phi}^{\mathrm{term}}$ and the MRI growth rate,
$\sigma_{\mathrm{MRI}}$, which qualitatively agrees with the above
reasoning.

Once the initial channel flows are disrupted and the field geometry is
changed by reconnection, the mean magnetic and kinetic energies, and
the absolute value of the Maxwell stresses begin to fluctuate strongly
around roughly constant values (see the phase between 11\,ms and
23\,ms in \figref{Fig:Res-2d--UZ2-4-tevo}).  Subsequently, a second
phase of exponential MRI growth is possible, exhibiting a similar
dynamics but involving less channel flows than the previous growth
phase.  The reduced number of channels is probably due to the strong
increase in the vertical magnetic field during the growth of the
tearing modes.  Similarly to their predecessors, the newly formed
channels are also unstable against resistive instabilities, but due to
their larger width their disruption requires much higher Alfv\'en
velocities, i.e., the MRI can lead to much higher Maxwell stresses in
the second generation of channels.  In principle, this process of
formation and merger of channels can continue until only one single
channel flow remains covering the entire box.  We note that in later
growth phases, the radial velocity and the magnetic field strength are
typically so large that damping at the radial boundaries, if applied,
does not lead to early saturation.

\begin{figure}[htbp]
  \centering
  \includegraphics[width=7cm]{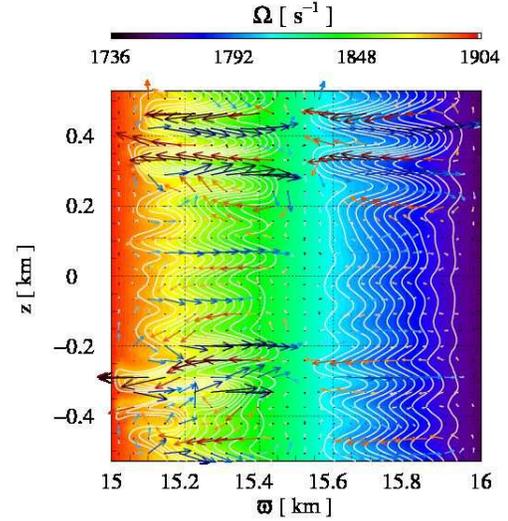}
  \caption{ An early state ($t= 12.1\, \mathrm{ms}$) in the evolution
    of a model with $b_0^z = 2\times 10^{13}\, \mathrm{G}$ computed in
    a box of $L_{\varpi} \times L_z = 1\,$km$^2$ covered by $200
    \times 200$ grid zones.  Shown are the same variables as in
    \figref{Fig:Res:channel-modes}.  The maximum velocity is
    $1.1\times 10^7 \mathrm{cm~s^{-1}}$.
  }
  \label{Fig:Res-2d--Z2-4-channels}
\end{figure}

\subsubsection{Models with non-uniform initial magnetic fields}
Models having a non-uniform initial magnetic field exhibit a different
evolution \citep[see also][]{Balbus_Hawley__1998__RMP__MRI}.  To study
this evolution we simulated a set of models varying the initial
magnetic field configuration and the boundary condition (applying
velocity damping or not; see previous subsection).  All models rotate
initially according to the law given in \eqref{Gl:Init-Omega} with
$\Omega_0 = 1900\, \mathrm{s}^{-1}$ and $\alpha_\Omega = -1.25$.

We considered three types of non-uniform initial magnetic fields all
of which have only a $z$-component. The first one
\begin{equation}
  \label{Gl:Init--zero-flux-field}
  b^z_{\mathrm{ZNF}} = 
  b^z_0 \sin \left(
    \frac{2\pi \varpi}{\lambda^b_\varpi}
  \right)
  \times \frac{\varpi}{\varpi_0}
\end{equation}
varies sinusoidally with radius and scales in addition as $\propto
\varpi$ to guarantee that the net magnetic flux through the surfaces
of the computational box at $z = z_{0,1}$ vanishes.  This is the
standard zero flux field used in most MRI simulations.  The second
type of a non-uniform initial field considered by us is given by 
\begin{equation}
  \label{Gl:Init-b-abs}
   b^z_{\mathrm{Abs}} = \left| b^z_{\mathrm{ZNF}} \right| \, . 
\end{equation}
Finally, the third type also has a vanishing net magnetic flux, as
$b^z_{\mathrm{ZNF}}$, but a step-like dependence on $\varpi$, i.e.,
\begin{equation}
  b^z_{\mathrm{step}} = b^z_0 \Theta 
                       \left( \varpi - \varpi_{\mathrm{c}} 
                       \right)\, \frac{\varpi}{\varpi_0} \, ,
\label{Gl:Init-b-step} 
\end{equation}
where $\Theta$ and $\varpi_{\mathrm{c}}$ denote the Heaviside step
function and the radial coordinate of the center of the box,
respectively.

For the first type of models the MRI starts growing via a multitude of
channel modes giving rise to less amplification of the magnetic field
than in models with a uniform initial field.  Separate channels
develop in the two radial regions of negative and positive $b^z$ (see
\figref{Fig:Res-2d--Z2-4-channels}), whereas the channels span the
full radial extent if the initial field is uniform.  The channel flows
do not merge to form a few large-scale channels, but are destroyed by
turbulence.  After reaching a transient maximum, the magnetic energy
and the Maxwell stress level off at values much less than for
uniform-field models.  The magnetic field becomes strongest right
after MRI termination ($\sim 10^{15}\, \mathrm{G}$). After $60\,$ms
the maximum field strengths are about $2\times 10^{14}\, \mathrm{G}$,
and decrease to $10^{14}\, \mathrm{G}$ until the end of the
simulation.  Fields of this strength can change the $\Omega$ profile
significantly only on time scales of many tens of milliseconds, i.e.,
at the end of the simulation the rotation profile is basically
unchanged.

\begin{figure}
  \centering
  \includegraphics[width=7cm]{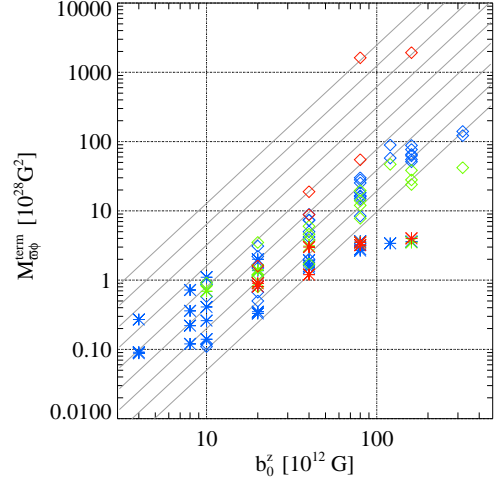}
  \caption{Maxwell stress $M_{\varpi\phi}^{\mathrm{term}}$ at MRI
    termination as a function of the initial magnetic field strength
    $b_0^z$ for models with non-uniform initial fields.  Models with
    and without velocity damping are shown by asterisks and diamonds,
    respectively. The blue, green and red symbols denote models where
    the $z$-component of the initial magnetic field is given by
    $b_{\mathrm{ZNF}}^z$  (see \eqref{Gl:Init--zero-flux-field}),
    $b_{\mathrm{step}}^z$ (see \eqref{Gl:Init-b-step}), and
    $b_{\mathrm{Abs}}^z$  (see \eqref{Gl:Init-b-abs}), respectively.  
    The light gray lines illustrate power laws $\propto
    \left( b_0^z \right)^{16/7}$.
  }
  \label{Fig:VZ2-Mterm}
\end{figure}

\figref{Fig:VZ2-Mterm} shows a comparison of the maximum Maxwell
stress at MRI termination for the models with a non-uniform initial
magnetic field. When velocity damping is applied the value of
$M_{\varpi\phi}^{\mathrm{term}}$ does not depend on the initial field
geometry, and is determined by reconnection of anti-parallel field
lines occurring close to the boundaries.  If no velocity damping is
applied, the evolution is similar to that of uniform field models, but
depends on the field geometry. This finding can be understood in the
light of our previous discussion of re-connective instabilities, and by
the fact that $M_{\varpi\phi}^{\mathrm{term}}$ is determined by
reconnection in the bulk volume, and not by reconnection near the
boundaries.

Models without velocity damping and the initial field,
$b^z_{\mathrm{Abs}}$ develop large Maxwell stresses which increase
with the initial field strength.  The evolution of these models and
the geometry of their channel flows are similar to those of models
with uniform initial fields, i.e., the growth rates of tearing-modes
are similar for both classes of models. For models in which the net
magnetic flux vanishes initially we find that
$M_{\varpi\phi}^{\mathrm{term}}$ is roughly constant for sufficiently
strong initial fields, the stress being slightly larger for sinusoidal
($b^z_{\mathrm{ZNF}}$) than for step-like initial fields
($b^z_{\mathrm{step}}$). The models develop a more complex field
morphology with more intense current sheets and more potential sites
for reconnection than uniform field models.  Thus, the growth rates of
the resistive instabilities are comparable to those of the MRI for
much weaker fields than in the uniform-field models. MRI termination
also occurs at smaller values of $M_{\varpi\phi}^{\mathrm{term}}$.

\subsection{Axisymmetric models with entropy gradients}
\label{sSek:Res--2ds}
We also simulated some axisymmetric models imposing an additional
entropy gradient.  In this case, the instability criterion is more
complicated and various instability regimes exist (see
\secref{Sec:MRIds}): (magnetic) shear instability, convection, and
magneto-buoyant instability.  Otherwise unstable modes may be
stabilized by a stable thermal stratification or by fast (not
necessarily differential) rotation.  Tables\,\ref{Tab:U2-entr-MSI}
(models with a positive entropy gradient) and
\ref{Tab:U2-entr-convection} (models with a negative entropy gradient)
provide a list of the simulated models. Note that all models discussed
in this subsection have a uniform initial magnetic field.

\subsubsection{Positive entropy gradients}
We first discuss differentially rotating models having a stabilizing
entropy gradient comparing models with and without magnetic field.
The non-magnetic models are stable due to the positive entropy
gradient, i.e. initial perturbations do not grow.

The models with a magnetic field belong to the MSI regime
(cf.~\figref{Fig:MRI--Regimen}), their MRI growth rates being reduced
compared to models with no entropy gradient.  We simulated models with
different entropy gradients ($\partial_{\varpi} S = 0.02, 0.04, 0.08 ~
\mathrm{km}^{-1}$) and different adiabatic index of the equation of
state ($\Gamma_{\mathrm{b}} = 1.31, 5/3$).  Generally, we find a good
agreement between the analytic predictions and the numerical results.
For $\partial_{\varpi} S < 0.08$ the models are unstable belonging to
the MSI regime, whereas an entropy gradient of $\partial_{\varpi} S =
0.08$ suffices to stabilize the model.  The growth rates agree well
with the analytic ones, and the numerical models show the typical
dependence of the growth rate of the MRI on the initial magnetic field
strength. However, there exists one interesting difference: $\sigma$
increases from small values for weak initial fields for which
$\lambda_{\mathrm{MRI}}$ is under resolved and converges to the
correct growth rate for strong fields for which
$\lambda_{\mathrm{MRI}}$ exceeds the grid resolution significantly.
Unlike for models without entropy gradient, the growth rate becomes
largest for magnetic fields for which the MRI wavelength is similar to
the grid resolution, and at these field strengths the numerical growth
rates can exceed the theoretical ones.

\begin{figure}
  \centering
  \includegraphics[width=7cm]{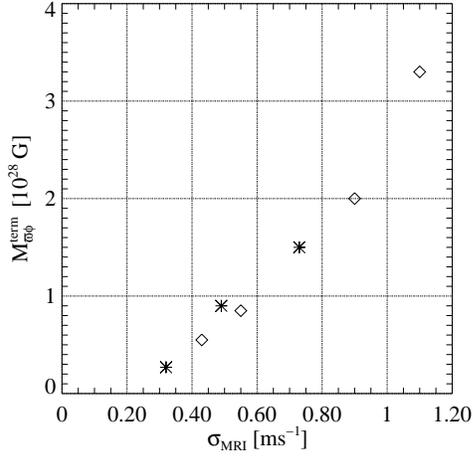}
  \caption{ Maxwell stress at MRI termination,
    $M_{\varpi\phi}^{\mathrm{term}}$, as a function of the MRI growth
    rate, $\sigma_{\mathrm{MRI}}$, for models with zero (diamonds) and
    positive (plus signs) entropy gradients.  }
  \label{Fig:U2d-Sigma-Mxy}
\end{figure}

Dynamically the models behave similarly as models without an entropy
gradient.  Channel flows develop during the growth phase of the MRI,
their width being set by the MRI wavelength.  MRI termination occurs
due to the growth of tearing-mode-like resistive instabilities.  When
velocity damping is applied, the maximum Maxwell stress at MRI
termination, $M_{\varpi\phi}^{\mathrm{term}}$, is determined by the
box size.  Comparing models with a positive entropy gradient and with
no entropy gradient we find a common linear relation between
$M_{\varpi\phi}^{\mathrm{term}}$ and $\sigma_{\mathrm{MRI}}$,
indicating a common reason for MRI termination (see
\figref{Fig:U2d-Sigma-Mxy}).

According to \eqref{Gl:MRIds--kMRI-Alfv}, the channels are wider in
models with larger entropy gradients.  As wider channels are less
prone to resistive instabilities, they can support stronger fields
before being disrupted (see discussion above), i.e.
$M_{\varpi\phi}^{\mathrm{term}}$ is larger in models with larger
entropy gradients.  The MRI growth rate, on the other hand, is smaller
for larger entropy gradients (see \eqref{Gl:MRIds--oMRI-Alfv})
implying that $M_{\varpi\phi}^{\mathrm{term}}$ decreases with
increasing entropy gradient.  Both effects taken together suggest a
weak anti-correlation of $M_{\varpi\phi}^{\mathrm{term}}$ with the
size of the (positive) entropy gradient. An anti-correlation is also
suggested by our numerical results, although more models are needed to
confirm it. It is unclear, for example, whether the growth rates of
resistive instabilities derived in \secref{subsub-channel} also hold
for stably stratified media, and whether the boundary conditions have
an influence in models with large entropy gradients.  Small
perturbations of the quasi-periodic (because of global gradients)
radial entropy distribution may leave their imprint on MRI termination
by enforcing a preferred length scale, thus clouding effects due to a
variation of $\partial_{\varpi}S$.

We have also simulated a few of the models using an ideal-gas equation
of state, $P = ( \Gamma - 1 ) \varepsilon$, instead of the hybrid EOS
finding, however, no effect on the evolution of the models.

\subsubsection{Negative entropy gradients}
Convection can develop in models having a negative entropy gradient,
but it can be suppressed by rapid rigid or differential rotation.  If
a magnetic field is added to a convectively unstable system which
cannot be stabilized by rotation, the system is located in the
``convection'' regime of \figref{Fig:MRI--Regimen}.  In a situation
where rotation suffices to suppress the convective instability, the
addition of a weak magnetic field puts the system into the ``MBI''
regime, and a magneto-buoyant instability similar to the standard
(i.e., $\partial_{\varpi} S = 0$) MRI develops. When
$\partial_{\varpi} S < 0$ the equivalent of the standard MRI
corresponds to the ``MSI'' regime in \figref{Fig:MRI--Regimen}.

Before discussing our results (see \tabref{Tab:U2-entr-convection} for
a list of the simulated models), we need to comment on the boundary
conditions.  Allowing for radial transport of energy shearing-disc
boundaries can, in principle, lead to a transport of entropy across
the pseudo-periodic boundaries, thus modifying the initial entropy
profile.  By comparing shearing-disc models and models with reflecting
boundary conditions, we verified that none of these boundary
conditions suppresses the growth of the MRI, and that both give
similar results.

\begin{figure}[htbp]
  \centering
  \includegraphics[width=7cm]{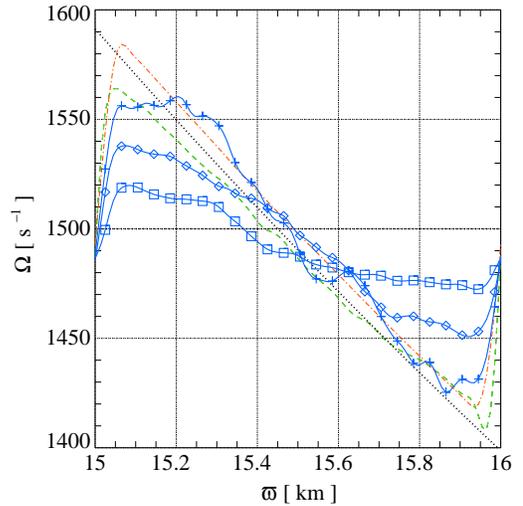}
  \caption{ $\Omega$, averaged over $z$, as a function of $\varpi$ for
    non-magnetic convective models at $t = 51.9\, \mathrm{ms}$ (green
    dashed), with an initial magnetic field $b_0^z = 10^{12}\,
    \mathrm{G}$ at $t = 51.9\, \mathrm{ms}$ (red dash-dotted), and an
    initial field $b_0^z = 10^{13}\, \mathrm{G}$ (blue solid lines),
    respectively.  For the latter model the different symbols indicate
    different epochs: $t = 11.5\, \mathrm{ms}$ (plus signs), $t =
    20.8\, \mathrm{ms}$ (diamonds), and $t = 41.7\, \mathrm{ms}$
    (squares).  The dotted black lines show the initial rotation law
    ($\Omega_0 = 1500\, \mathrm{s}^{-1}$, and a power-law profile
    $\Omega \propto \varpi^{-2}$.  }
  \label{Fig:UZ2-S-3-O15--rotprof}
\end{figure}

Let us first consider a non-magnetic model which rotates rigidly
with an angular velocity $\Omega_0 = 1000\, \mathrm{s}^{-1}$ and has
an entropy profile given by $S_0 = 0.2$ and $\partial_{\varpi} S =
-0.075\, \mathrm{km}^{-1}$ (see \eqref{Gl:Init-Entropy}).  With $N^2 /
\Omega_0^2 \approx -14$ and $\mathcal{R}_{\varpi} / \Omega_0^2 = 0$,
the model belongs to the convective regime.  Buoyant modes are
unstable and grow at a theoretical rate $\sigma_{\mathrm{th}} = 3.4\,
\mathrm{ms}^{-1}$.  Simulated on a grid with $2.5\,$m spatial
resolution, the model is unstable with a numerical growth rate $\sigma
= 2.7\, \mathrm{ms}^{-1}$, and convection sets in quickly.  The flow
is dominated by a few (one or two) fairly circular convective rolls.
Due to the transport of entropy and angular momentum by the
overturning fluid, the model develops complementary entropy and
rotation profiles characterized by ``cold'' (i.e., low-entropy),
rapidly rotating matter in down-flows and ``hot'' (i.e.,
high-entropy), slowly rotating matter in up-flows.  The redistribution
of angular momentum and entropy leads to an average (with respect to
the $z$-coordinate) rotational profile of the form $\Omega \propto
\varpi ^ {-2}$, i.e., constant specific angular momentum (see
\figref{Fig:UZ2-S-3-O15--rotprof}), and a flat entropy profile.  

For a faster rotation rate of $\Omega_0 = 1500\, \mathrm{s}^{-1}$,
corresponding to $N^2/\Omega_0^2 \approx -5.7$, i.e., still in the
convective regime, the evolution is similar except for a reduced
growth rate ($\sigma \approx 1.6\, \mathrm{ms}^{-1}$) due to
rotational stabilization.  The model develops differential rotation
with the same $\varpi$-dependence as in the case of slower rotation.
If we increase the rotation rate to $\Omega_0 = 1900\,
\mathrm{s}^{-1}$ ($N^2 / \Omega_0^2 \approx -3.2$) buoyant modes are
stabilized by rotation.

The above results also hold if the initial model is rotating
differentially. In particular, convection (i.e., the negative initial
entropy gradient) gives also rise to a rotation law of the form
$\Omega \propto \varpi ^ {-2}$, and a flat entropy profile.

\begin{figure}
  \centering
  \includegraphics[width=7cm]{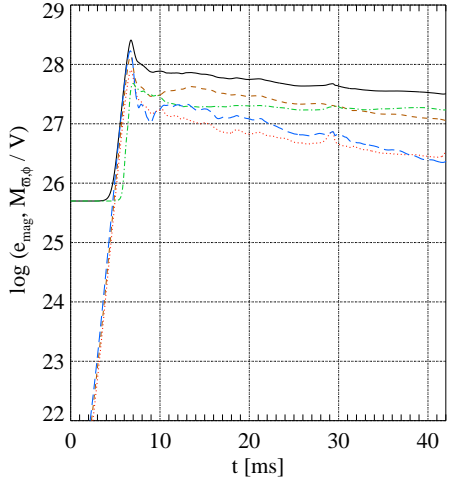}
  \caption{
    Evolution of the mean magnetic energy density $e^\mathrm{mag}$
    (solid black line), the mean energy densities corresponding to the
    $\varpi$ (dotted red), $\phi$ (dashed brown), and $z$ (dash-dotted
    green) component of the magnetic field, and the absolute value of
    the mean Maxwell stress component $M_{\varpi,\phi}$ (dashed blue
    line) for the model whose rotational profiles at different times
    are shown in \figref{Fig:UZ2-S-3-O15--rotprof}, i.e., a rigidly
    rotating model ($\Omega_0 = 1500~\mathrm{s}^{-1}$) with an initial
    magnetic field of $b^z_0 = 10^{13}~\mathrm{G}$.}
  \label{Fig:UZ2-R0-15Sm3-B10-tevo}
\end{figure}

\begin{figure*}
  \centering
  \includegraphics[width=7cm]{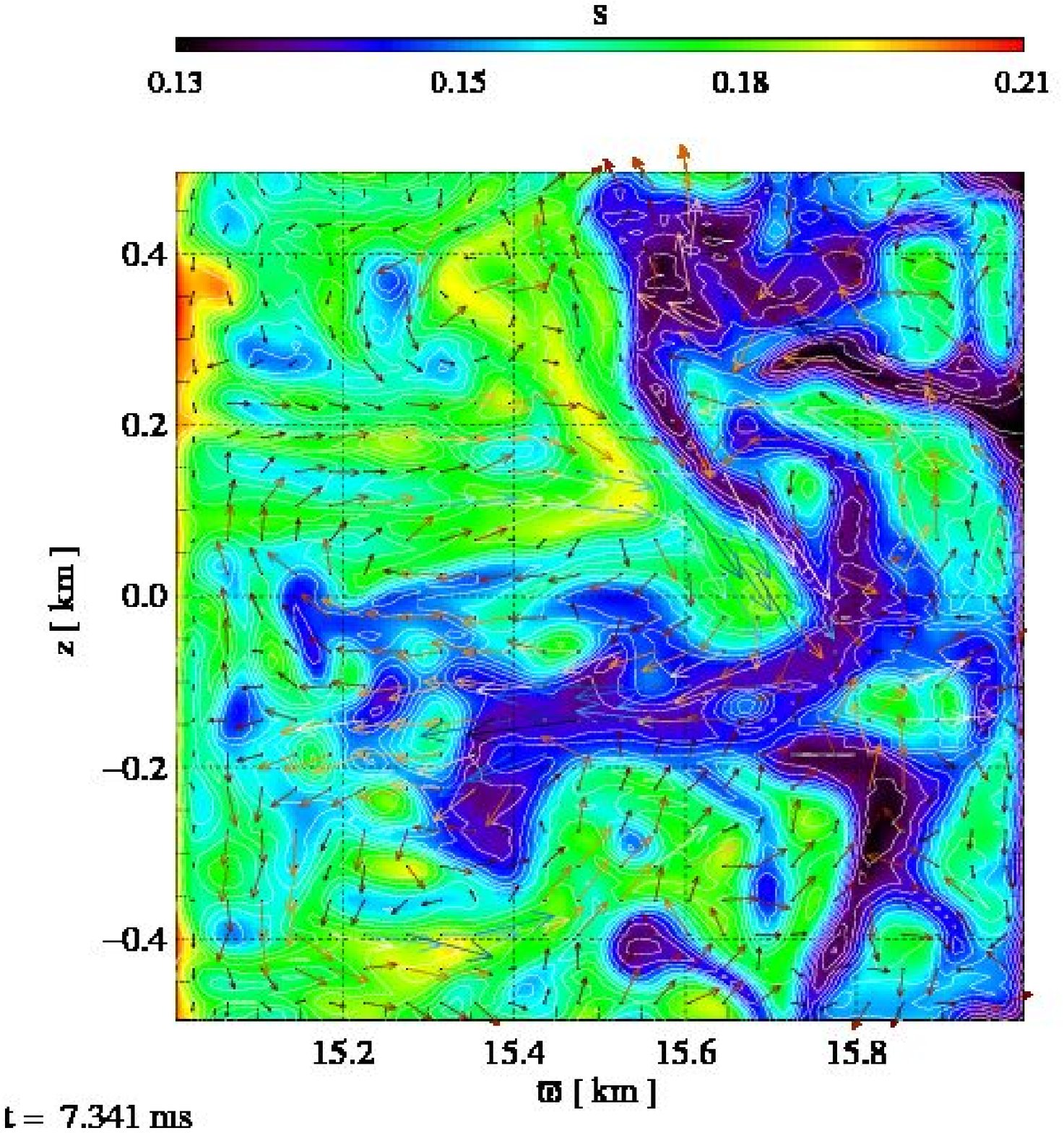}
  \includegraphics[width=7cm]{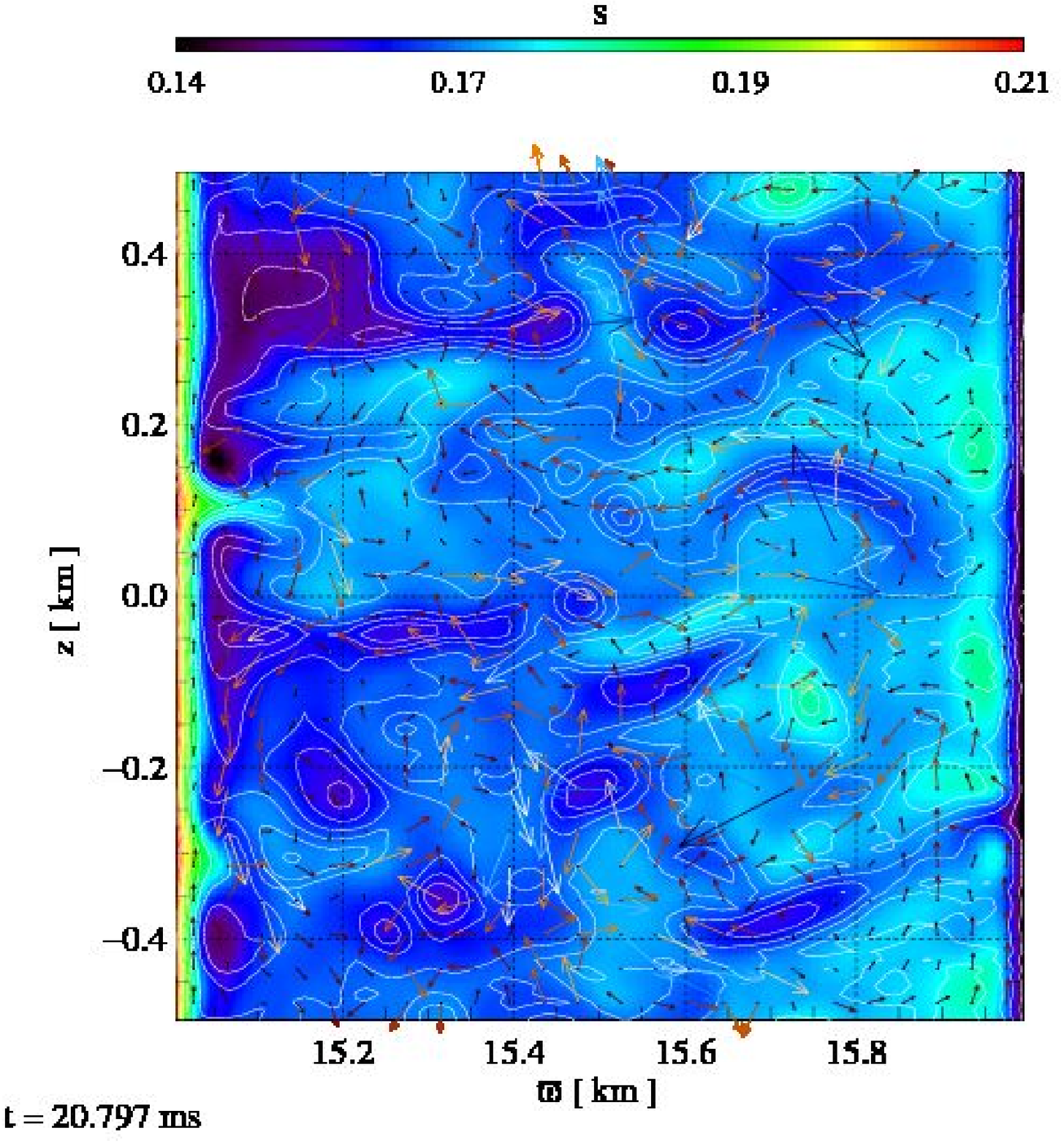}
  \caption{
    The entropy distribution (color coded), the poloidal velocity
    field (arrows), and the magnetic field lines of the model
    displayed in \figref{Fig:UZ2-R0-15Sm3-B10-tevo} at $t =
    7.3~\mathrm{ms}$ (left panel) and $t = 20.8~\mathrm{ms}$ (right
    panel), respectively, i.e., at the begin of the saturation phase
    and during saturation.}
  \label{Fig:UZ2-S-3-O15--2dp}
\end{figure*}

Next, we add a magnetic field of $b_0^z = 10^{13} \mathrm{G}$ to a
convective model, i.e., to a model with a negative entropy gradient
rotating too slowly (in our case, for rigid rotation, with $\Omega_0
\la 1500 \mathrm{s}^{-1}$) for convection to be stabilised by
rotation.  The temporal evolution of the magnetic energy and the
Maxwell stress of this model is shown in
\figref{Fig:UZ2-R0-15Sm3-B10-tevo}, while
\figref{Fig:UZ2-S-3-O15--2dp} gives the spatial distribution of the
entropy of the model at two different times.  Initial perturbations
are amplified rapidly, but saturation sets in within 6 milliseconds
the growth rate being slightly higher than in the non-magnetic models.
With $N^2/\Omega_0^2 \approx - 5.7$, the model is dominated by buoyant
modes, but there still exists some influence of the Alfv\'en modes.
In particular, although infinitely long modes grow rapidly, the
fastest growing modes are Alfv\'en modes of finite wavenumber which
depends on $b_0^z$.  For sufficiently strong initial fields (or
sufficiently fine resolution), these modes are numerically resolved,
and have a growth rate exceeding that of the corresponding
non-magnetic model.  The magnetic field strength increases
exponentially as the instability develops, and at the onset of
saturation large convective rolls develop (left panel of
\figref{Fig:UZ2-S-3-O15--2dp}).  In the saturation phase (right panel
of \figref{Fig:UZ2-S-3-O15--2dp}) the flow geometry differs
considerably from that of the corresponding non-magnetic model.  It
consists of down-drafts of cold material and up-flows of hot gas
forming small-scale structures rather than large circular convective
rolls.  Like in the non-magnetic model, cold and hot regions
correspond to regions of low and high angular velocity, respectively.
Differential rotation with constant specific angular momentum, $\Omega
\propto \varpi ^{-2}$, develops due to hydrodynamic transport of
angular momentum in convective overturns.  The magnetic energy related
to the radial component of the magnetic field and the Maxwell stress
component $M_{\varpi\phi}$ remain high during saturation, i.e.
angular momentum transport converts the $\varpi^{-2}$ rotation law
prevailing at early epochs into nearly rigid rotation
(\figref{Fig:UZ2-S-3-O15--rotprof}).

If for the same initially rigidly rotating model the initial magnetic
field is too weak to resolve $\lambda_{\mathrm{MRI}}$ (we simulated
two models with $b_0 = 10^{4}$ and $10^{12}$, respectively) convection
develops.  The growth rates are similar to those of non-magnetic
models.  The weakest initial field, $b_0^z = 10^4\, \mathrm{G}$, has
no impact at all, and the evolution of the model with an initial field
of $10^{12}\, \mathrm{G}$ differs only slightly from that of the
non-magnetic model.  The magnetic field increases exponentially as the
instability grows. Large persistent convective rolls form, and
differential rotation develops.  After an initial exponential growth
the mean magnetic energy remains large, but the contribution of the
radial component of the magnetic field and the mean Maxwell stress
component $M_{\varpi\phi}$ decrease almost by four and two orders of
magnitude, respectively.  Consequently, no significant angular
momentum transport occurs due to magnetic stresses, and similar to the
non-magnetized model a $\Omega \propto \varpi ^ {-2}$ rotation law
develops.  \figref{Fig:UZ2-S-3-O15--rotprof}).

We now consider the \emph{MBI} regime (see \figref{Fig:MRI--Regimen})
and discuss models rotating initially rigidly with $\Omega_0 = 1900\,
\mathrm{s}^{-1}$ and having an initial entropy gradient
$\partial_{\varpi} S = - 0.1\, \mathrm{km}^{-1}$ (see
\tabref{Tab:U2-entr-convection}), which implies $N^2 / \Omega_0^2
\approx -3.6$.  Without magnetic field, the instability of the buoyant
modes is suppressed by the fast rotation.  However, if a weak magnetic
field is added, an instability of the MBI type develops, i.e., the
Alfv\'en modes become unstable.  The numerical growth rates show a
similar dependence on the magnetic field strength as in case of the
standard (i.e., $\partial_{\varpi} S = 0$) MRI, because
$\lambda_{\mathrm{MRI}}$ is resolved.  The instability grows rapidly
($\sigma \sim 1.4\, \mathrm{ms}^{-1}$, similar to the theoretical
value $\sigma_{\mathrm{th}} \approx 1.7\, \mathrm{ms}^{-1}$).  During
the growth phase channel modes appear, which lead to a transport of
both angular momentum and entropy.  After an exponential initial
growth and some decrease after MRI termination the mean magnetic
energies contained in the total magnetic field and all three field
components remain large (corresponding to field strengths of $\sim
10^{14}\, \mathrm{G}$), but the mean Maxwell stress component
$M_{\varpi\phi}$ drops to zero within ten milliseconds oscillating
afterward with decreasing amplitude between positive and negative
values.  Hence, large-scale angular momentum transport is limited.  At
the end of the simulation, the model shows considerable variations in
$\Omega$, but there is no clear indication of a mean differential
rotation of the form $\Omega = \Omega ( \varpi )$.  The entropy
profile in the saturated state is almost flat.

Finally, we summarize a few common features of the models having a
negative entropy gradient (see \tabref{Tab:U2-entr-convection}).  All
models develop instabilities in accordance with the flow regime to be
expected from their model parameters. The growth rates of all models
are, within the uncertainties, similar to the theoretical predictions.
As for the dynamics, we have to distinguish models in the convective
regime from those in the mixed and MBI regimes.  The former class of
models shows convective mushrooms and large-scale overturns with only
little influence of the magnetic field, whereas the last class is
dominated by channel flows.  Consequently, angular momentum transport
by hydrodynamic flow leads to a rotational profile $\Omega =
\varpi^{-2}$ for models in the convective regime, while models in the
mixed-type regime tends towards rigid rotation the angular momentum
transport being dominated by magnetic fields.  Termination of the
instability growth occurs for models both in the MBI and mixed-type
regime analogously to that of models in the MSI regime without entropy
gradient, i.e., by reconnection in resistive instabilities altering
the topology of the channel flows.  Consequently, we find similar
dependencies on the initial field strength, the grid resolution, and
the type of boundary conditions.  The instability in convective
models, on the other hand, saturates when the initial entropy gradient
is removed by vigorous entropy transport due to overturning fluid
motions.

\subsection{Three-dimensional models}
\label{sSek:Res--3d}
The results of the axisymmetric simulations discussed in the previous
section demonstrate the possibility of MRI-driven field amplification
in core collapse supernovae, and provide some insight into the
evolution of MRI unstable layers in the core. However, to address the
MRI problem in full generality, one has to consider three-dimensional
models, because the assumption of axisymmetry implies severe
restrictions for the dynamics of the magnetic and kinetic fields.  The
most important limitations are that, in axisymmetry, a toroidal field
cannot be converted into a poloidal one, and that the disruption of
the channel flows requires non-axisymmetric parasitic instabilities
\citep{Goodman_Xu__1994__ApJ__MRI-parasitic}.

As 3D simulations are computationally much more expensive than 2D
ones, we could not perform a comprehensive study, but had to focus on
a few selected models. We simulated models with different field
geometries and varied the initial field strength, the entropy profile,
and the grid size (see \tabref{Tab:3d-models-1} and
\tabref{Tab:3d-models-2}).

\subsubsection{Uniform initial magnetic fields, no entropy gradient}
We first discuss models which have a uniform initial magnetic field
$b^z_0$ in $z$-direction, no entropy gradient, and rotate
differentially with $\Omega_0 = 1900\, \mathrm{s}^{-1}$ and
$\alpha_\Omega = -1.25$ (see \eqref{Gl:Init-Omega}).  If the MRI
wavelength is well resolved (e.g., for models with initial field
strengths of $2\times 10^{13}\, \mathrm{G}$ and $4\times 10^{13}\,
\mathrm{G}$ simulated at a grid resolution of $\delta \varpi = 20\,
\mathrm{m}$), the growth rate is high and independent of $b^z_0$ .
Under-resolved models (e.g., models with $b_0^z = 10^{13}\,
\mathrm{G}$ simulated at the same resolution) exhibit a slower growth
of the MRI.  From the growth rates of the MRI, we infer that in 3D the
same resolution criterion applies as in the case of axisymmetry.

\begin{figure*}[htbp]
  \centering
  \includegraphics[width=6.0cm]{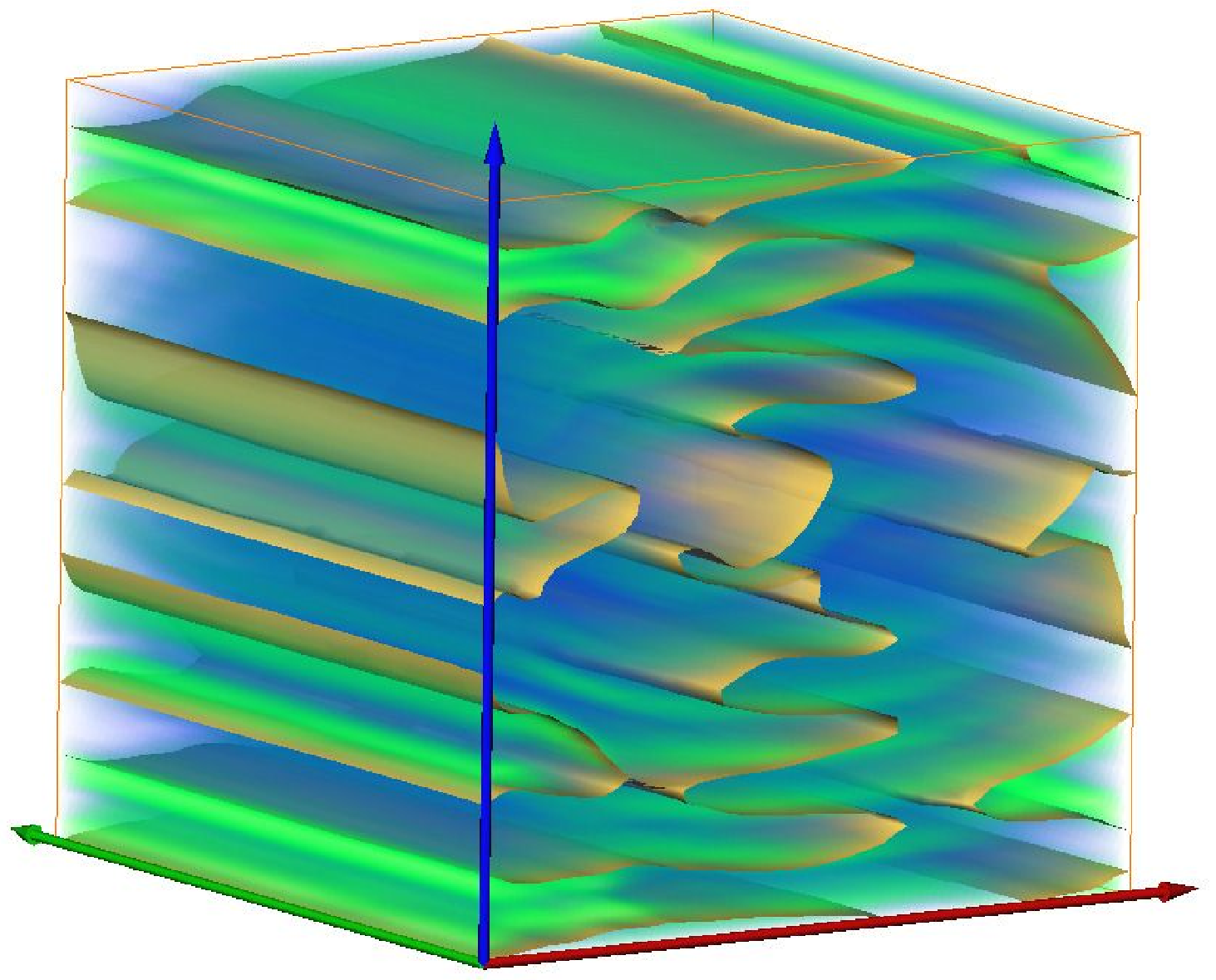}
  \includegraphics[width=6.0cm]{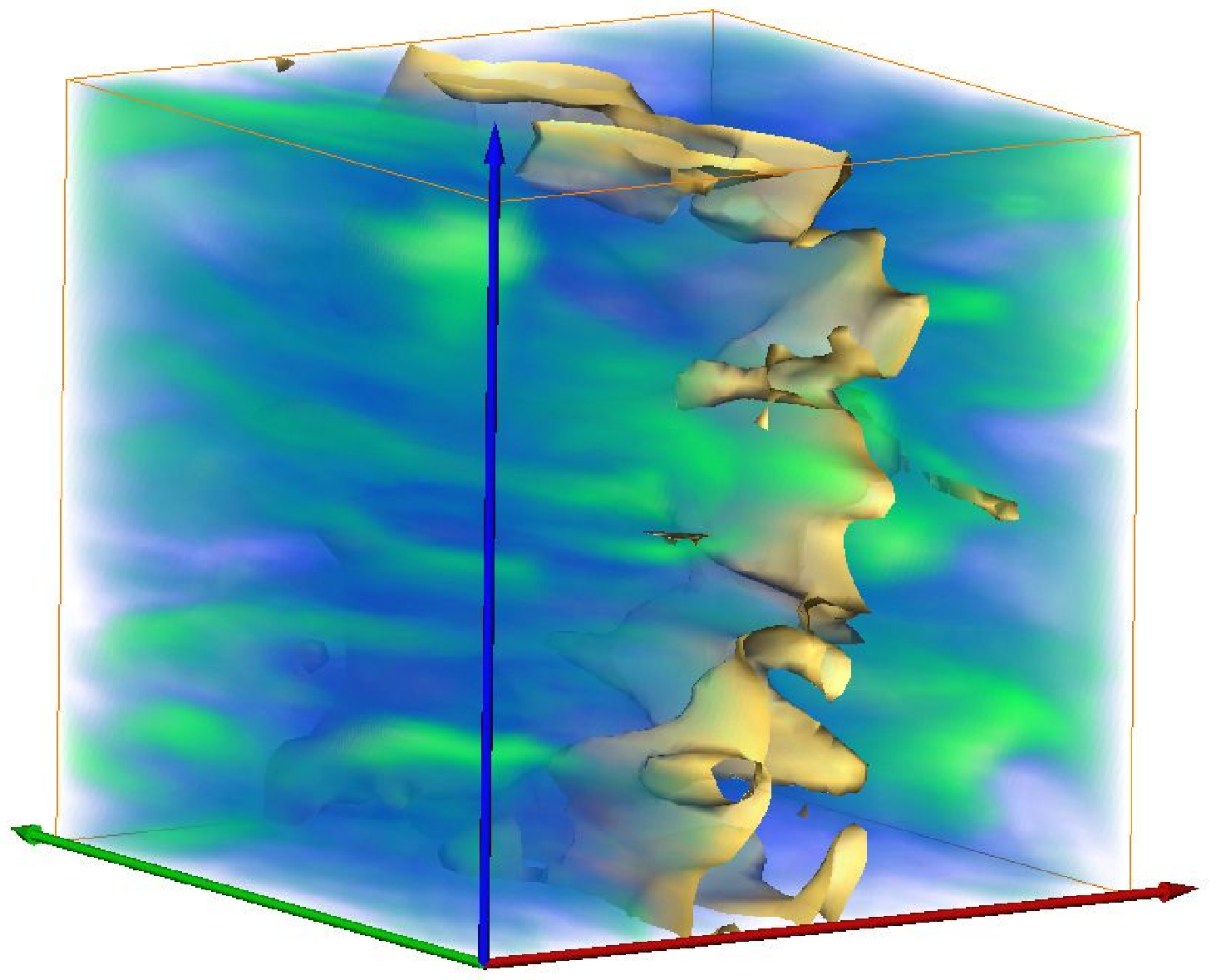}
  \includegraphics[width=6.0cm]{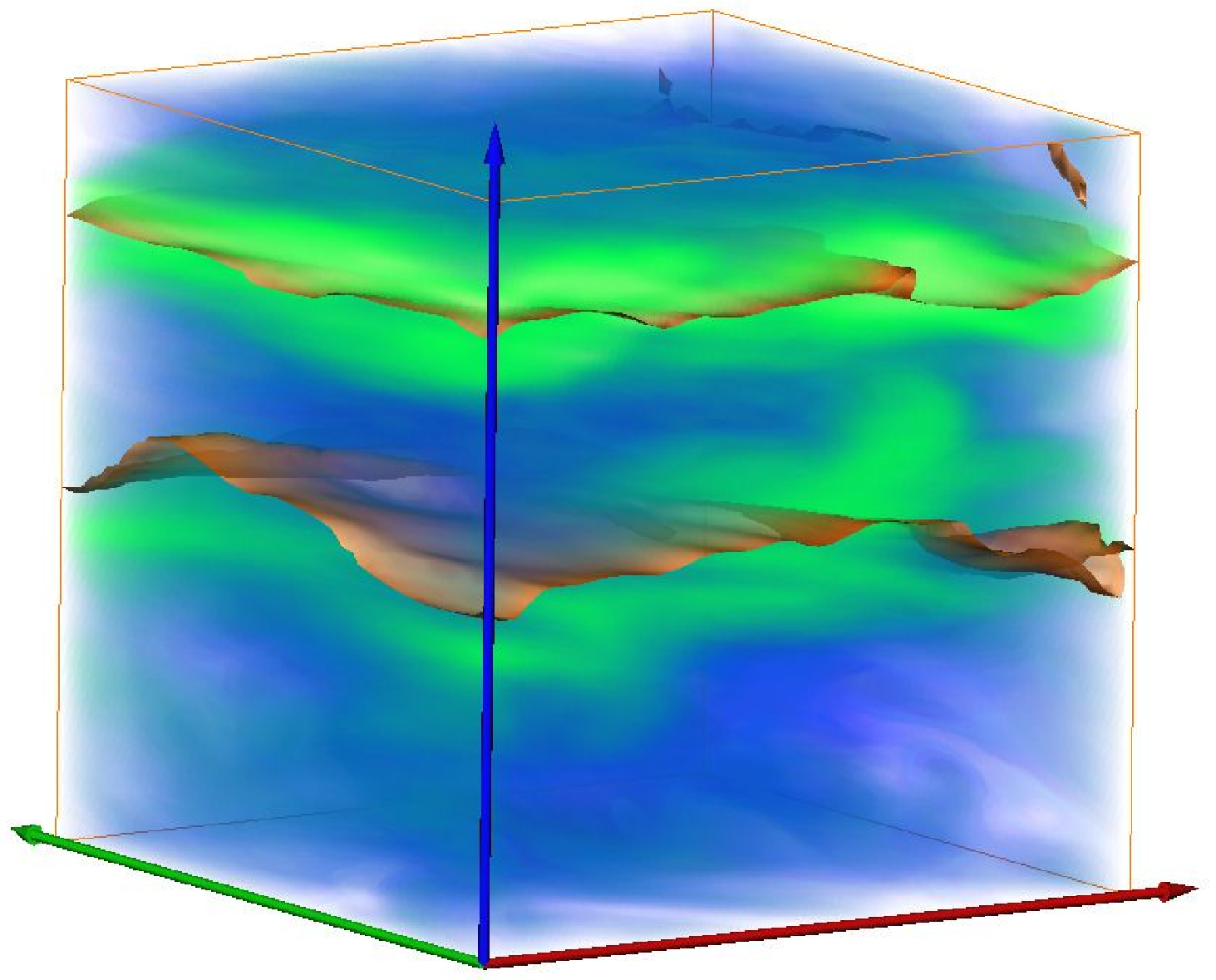}
  \caption{ Structure of a 3D model with $b_0^z = 2\times 10^{13}\,
    \mathrm{G}$ computed on a grid of $1\, \mathrm{km}^3$ at a
    resolution of $20\, \mathrm{m}$ at $t = 16.2\, \mathrm{ms}$
    (left), $t = 26.8\, \mathrm{ms}$ (middle), and $t = 42.5\,
    \mathrm{ms}$ (right), respectively.  Shown is the volume rendered
    magnetic field strength (blue to green), and a red-orange
    iso-$\Omega$ surface corresponding to $\Omega = 1820\,
    \mathrm{s}^{-1}$ (left and middle) and $\Omega = 1680\,
    \mathrm{s^{-1}}$ (right), respectively.  The red, green, and blue
    axes point into $\varpi$, $\phi$, and $z$ direction, respectively.
    Channel flows can be identified as green sheet-like structures.
  }
  \label{Fig:UZ3-R4S0-r150p1-R050-B13.2--z0x}
\end{figure*}

During early epochs the evolution is similar to that of the
corresponding axisymmetric models: a number of radially aligned
channels appear.  Strong differential rotation causes significant
wind-up of flow features leading to structures elongated in
$\phi$-direction, i.e., there exists only a modest variation of the
MHD variables with azimuthal angle at this stage. Sheet-like
structures dominate the field geometry.  The rotational profile begins
to show distortions due to the transport of angular momentum by
Maxwell stresses (see left panel of
\figref{Fig:UZ3-R4S0-r150p1-R050-B13.2--z0x}).  At later epochs the
flow in 3D is more complex than in axisymmetry. Although coherent
structures, i.e., flux sheets, are still present, their geometry is
more tangled and twisted, and less isotropic than earlier in the
evolution (see middle panel of
\figref{Fig:UZ3-R4S0-r150p1-R050-B13.2--z0x}).

\begin{figure*}[htbp]
  \sidecaption
  \includegraphics[width=6cm]{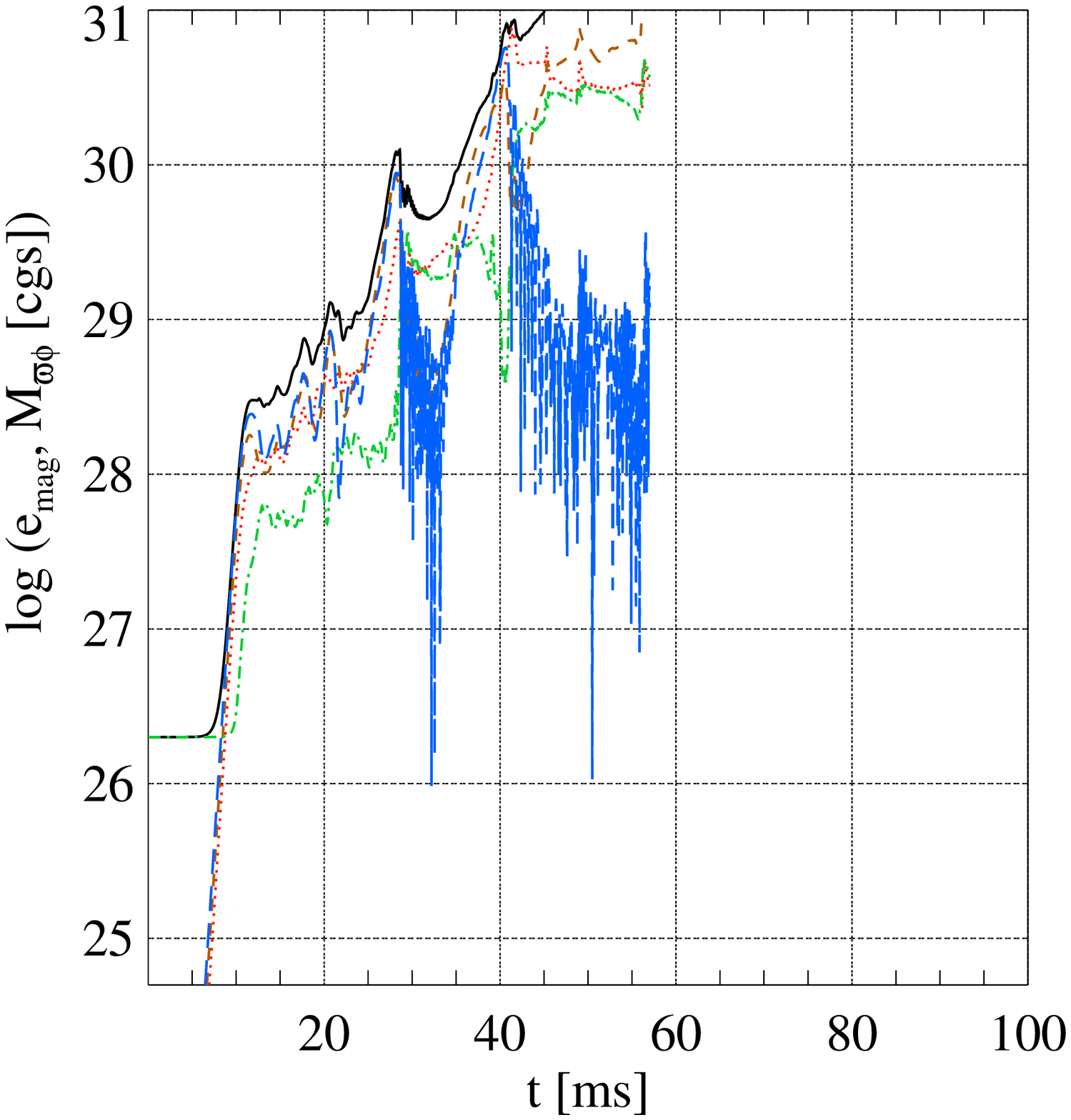}
  \includegraphics[width=6cm]{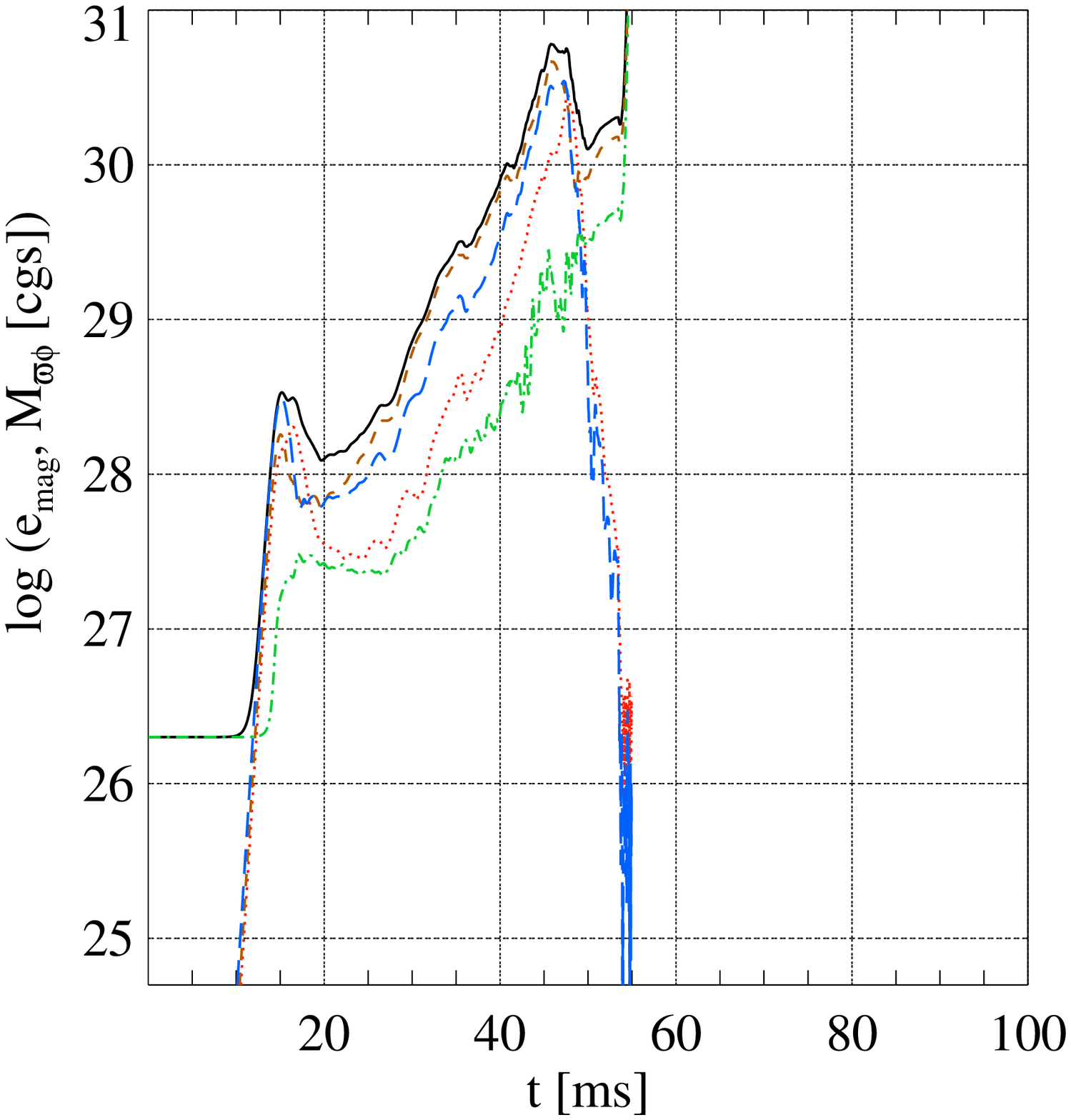}
  \includegraphics[width=6cm]{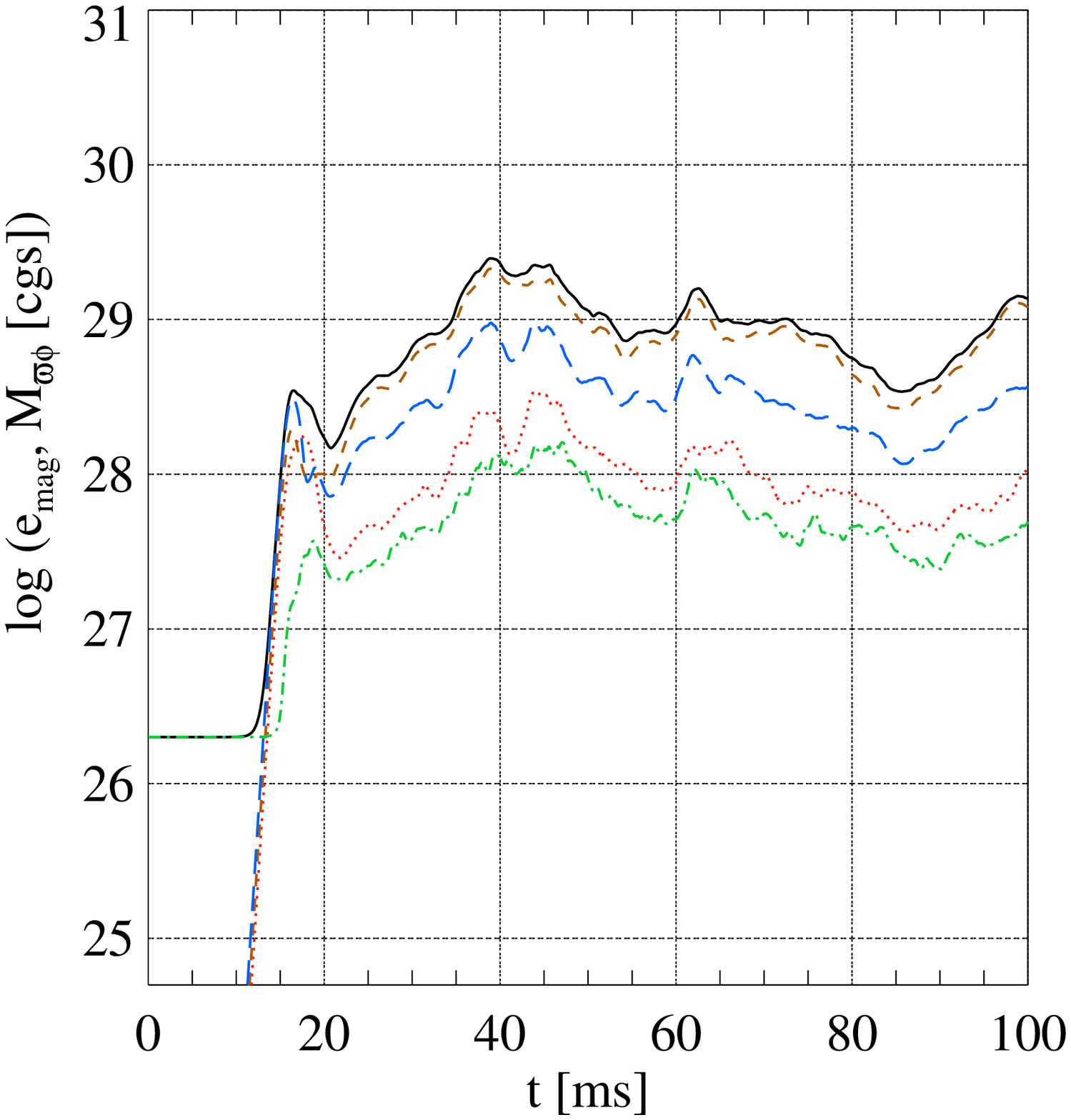}
  \caption{Evolution of the mean magnetic energy density
    $e^{\mathrm{mag}}$ (solid black line), the mean energy densities
    corresponding to the $\varpi$ (dotted red), $\phi$ (dashed brown),
    and $z$ (dash-dotted green) component of the magnetic field, and
    the absolute value of the mean Maxwell stress component
    $M_{\varpi,\phi}$ (dashed blue line) for models with an initially
    uniform magnetic field $b_0^z = 2\times 10^{13}\,$G in
    z-direction, and a rotation law given by $\Omega_0 =
    1900\,$s$^{-1}$ and $\alpha_\Omega = -1.25$. The panels show a 2D
    model computed in a box of $L_\varpi \times L_z = 1 \mathrm{km}
    \times 1\mathrm{km}$ (left), a 3D model computed in a box of
    $L_\varpi \times L_\phi \times L_z = 1\mathrm{km} \times
    1\mathrm{km} \times 1\mathrm{km}$ (middle), and a 3D model
    computed in a box of $L_\varpi \times L_\phi \times L_z =
    1\mathrm{km} \times 2\mathrm{km} \times 1\mathrm{km}$ (right),
    respectively. The grid resolution is $20 \mathrm{m}$ in all three
    cases.
    }
  \label{Fig:UZ3-4-p2--tevo}
\end{figure*}

An evolution from coherent channel flows to a more turbulent state is
characteristic for all three-dimensional models with a uniform initial
magnetic field.  However, as pointed out by
\cite{Sano_Inutsuka__2001__ApJL__MRI-recurrent-channels}, channel
flows can develop again from the turbulent state.  Consequently, the
magnetic field can continue growing, and the angular momentum
transport will be enhanced strongly.  In the most extreme cases, the
evolution is similar to that of a corresponding axisymmetric
model. This is exactly what we observe for some models at late times,
$t \gtrsim 30\, \mathrm{ms}$ (see right panel of
\figref{Fig:UZ3-R4S0-r150p1-R050-B13.2--z0x}), when a dominant channel
flow forms.  These model enter again a state of exponential growth,
and a large part of the angular momentum is extracted by Maxwell
stresses. The field strengths reach several $10^{15}\, \mathrm{G}$,
peaking at $10^{16}\, \mathrm{G}$, and the mean Maxwell stress
component $M_\mathrm{\varpi\phi}$ exceeds $10^{30}\, \mathrm{erg\,
  cm^{-3}}$ (see middle panel of \figref{Fig:UZ3-4-p2--tevo}), and
compare with the corresponding axisymmetric model in the left panel).
Despite a qualitative similarity between the evolution of the 3D and
axisymmetric models, we note that the secondary exponential growth is
slower in three dimensions.

\begin{figure*}[htbp]
  \centering
  \includegraphics[width=8cm]{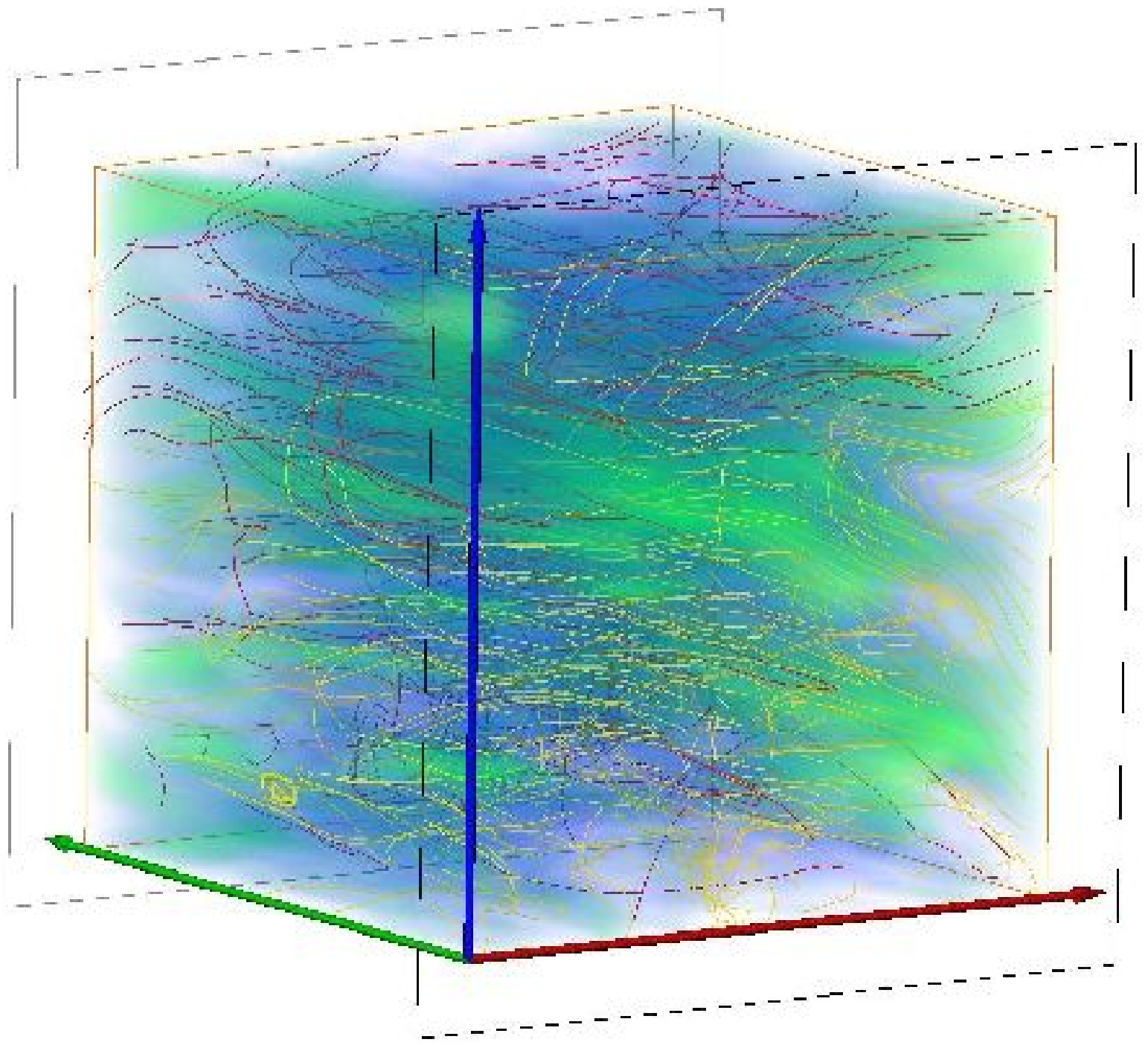}
  \includegraphics[width=8cm]{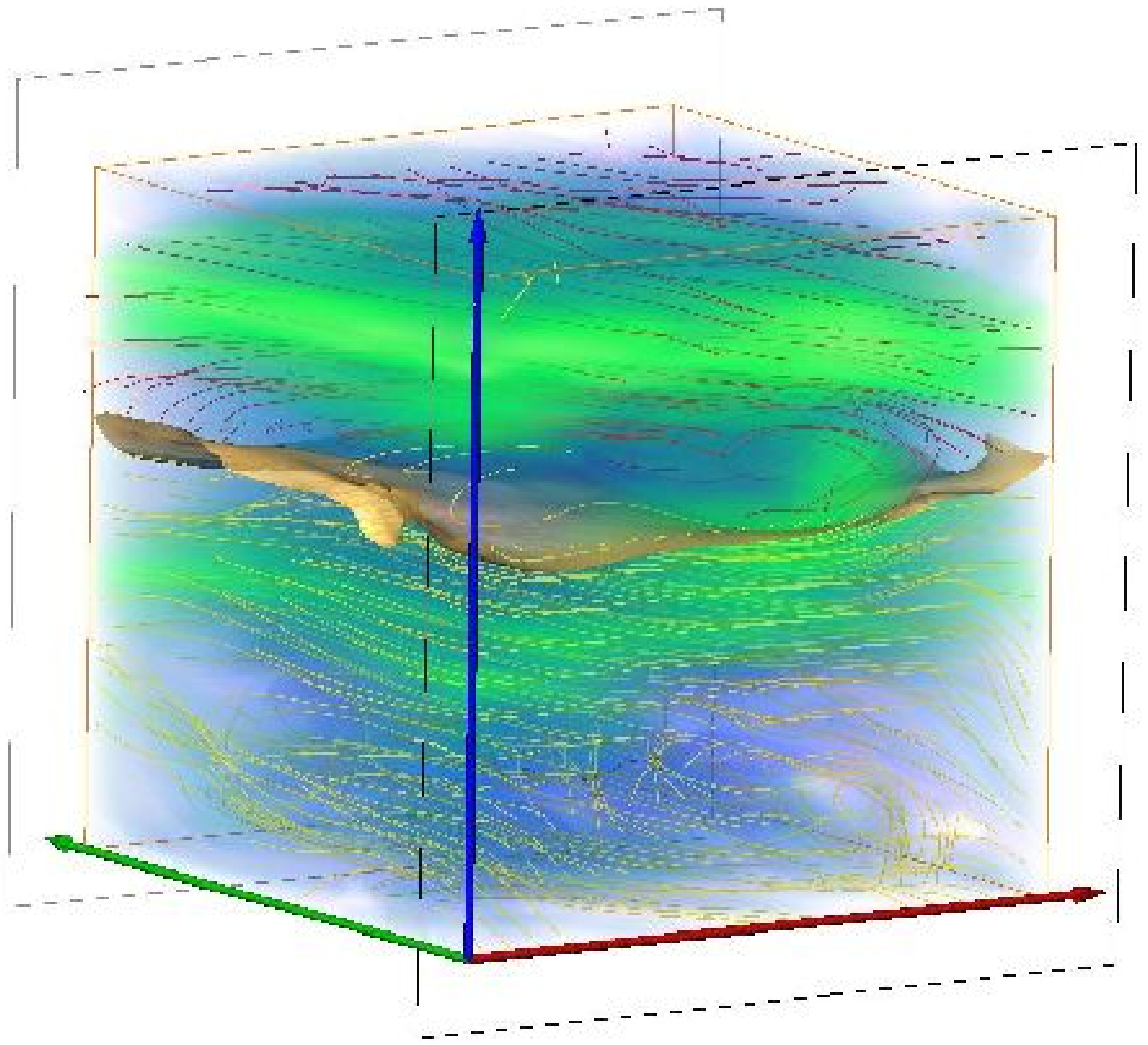}
  \caption{Same as middle and right panel of
    \figref{Fig:UZ3-R4S0-r150p1-R050-B13.2--z0x}, but showing besides
      the volume rendered magnetic field strength (blue to green) also
      the magnetic field lines, which are obtained by starting the
      integration of the magnetic field at two surfaces of constant
      $\phi$-coordinate (i.e., orthogonal to the green axis) at the
      left and right hand side of the domain.  The field lines
      originating from the left and right surface are plotted in red
      and yellow, respectively.  The right panel shows, in addition,
      the isosurface $b_\phi = 0$ (i.e. the magneto-pause).
  }
  \label{Fig:UZ3-R4S0-r150p1-R050-B13.2--bstruct}
\end{figure*}

The emergence of a large-scale structure of the magnetic field from a
turbulent state can be seen in
\figref{Fig:UZ3-R4S0-r150p1-R050-B13.2--bstruct} comparing the field
structure at $t = 26.8 \mathrm{ms}$ and $t=42.5 \mathrm{ms}$,
respectively.  At the earlier time (left panel), we find a small-scale
field dominated by slender flux tubes. Field lines of different
polarity (indicated by different colors) are lying close to each
other.  After the development of the channel flow (right panel), the
field is dominated by a large-scale pattern.  A smooth surface
permeating the box at nearly constant $z$-coordinate separates in two
large regions field lines of different polarity from each other.  In
each of the two regions, we find one broad flux sheet where most of
the magnetic energy is concentrated.  The separation layer is filled
by gas rotating nearly uniformly at a ow angular velocity ($\Omega
\sim 1500 \mathrm{s}^{-1}$).  The surrounding gas rotates uniformly as
well, but at a much higher velocity ($\Omega \sim 1800
\mathrm{s}^{-1}$).  The two flux sheets form a thin transition region
between both rotational states.  Thus, the dynamics is similar to that
of the corresponding axisymmetric model.

Because our boundary conditions allow for a loss of angular momentum,
and thus for the total disruption of the differential rotation profile
by transport through the radial boundaries, this stage represents the
end of the evolution, just as it did in axisymmetry: the instability
has used up its free-energy reservoir.  Hence, the later evolution
consists only of violent oscillations.

\begin{figure*}[htbp]
  \centering
  \includegraphics[width=8cm]{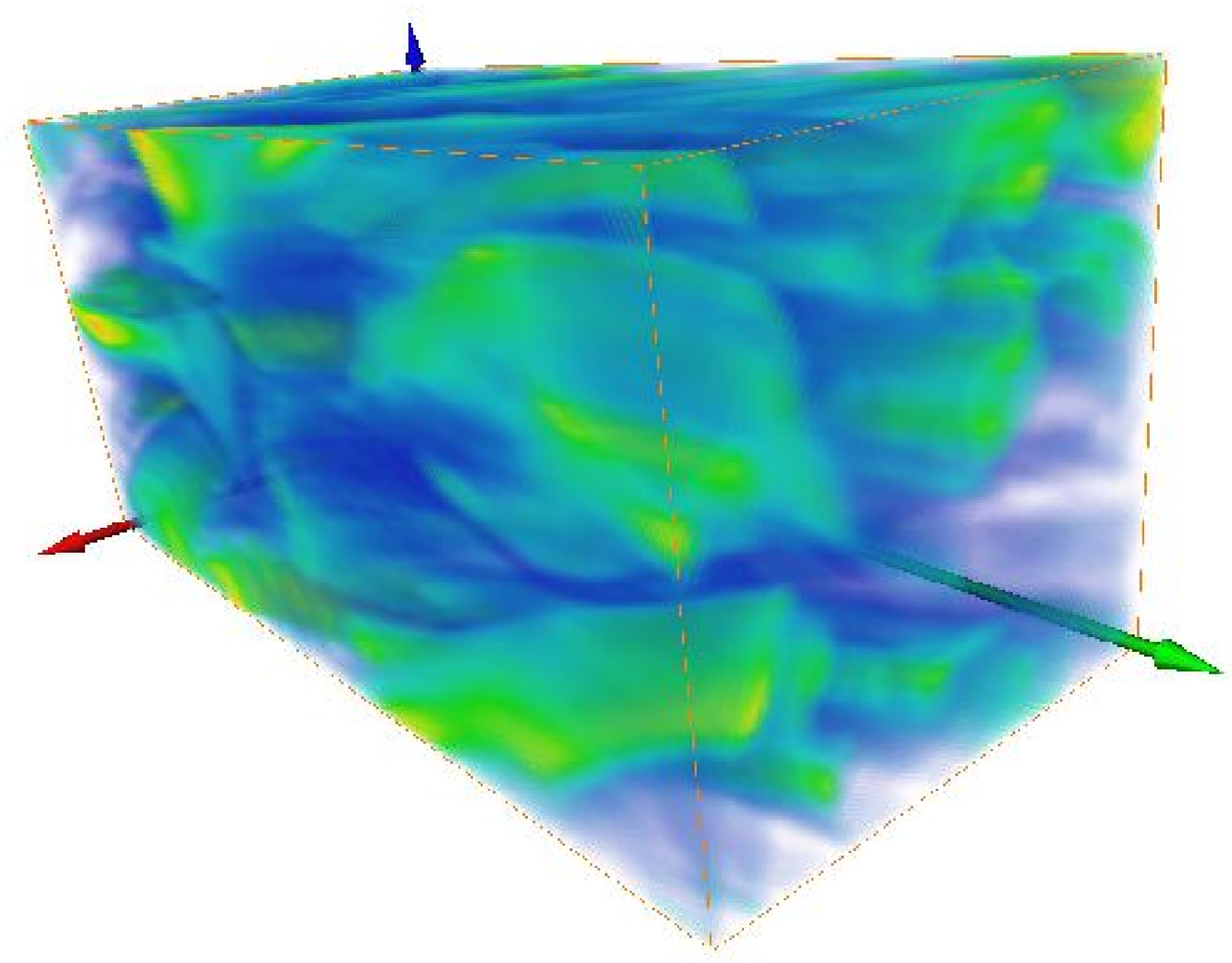}
  \includegraphics[width=8cm]{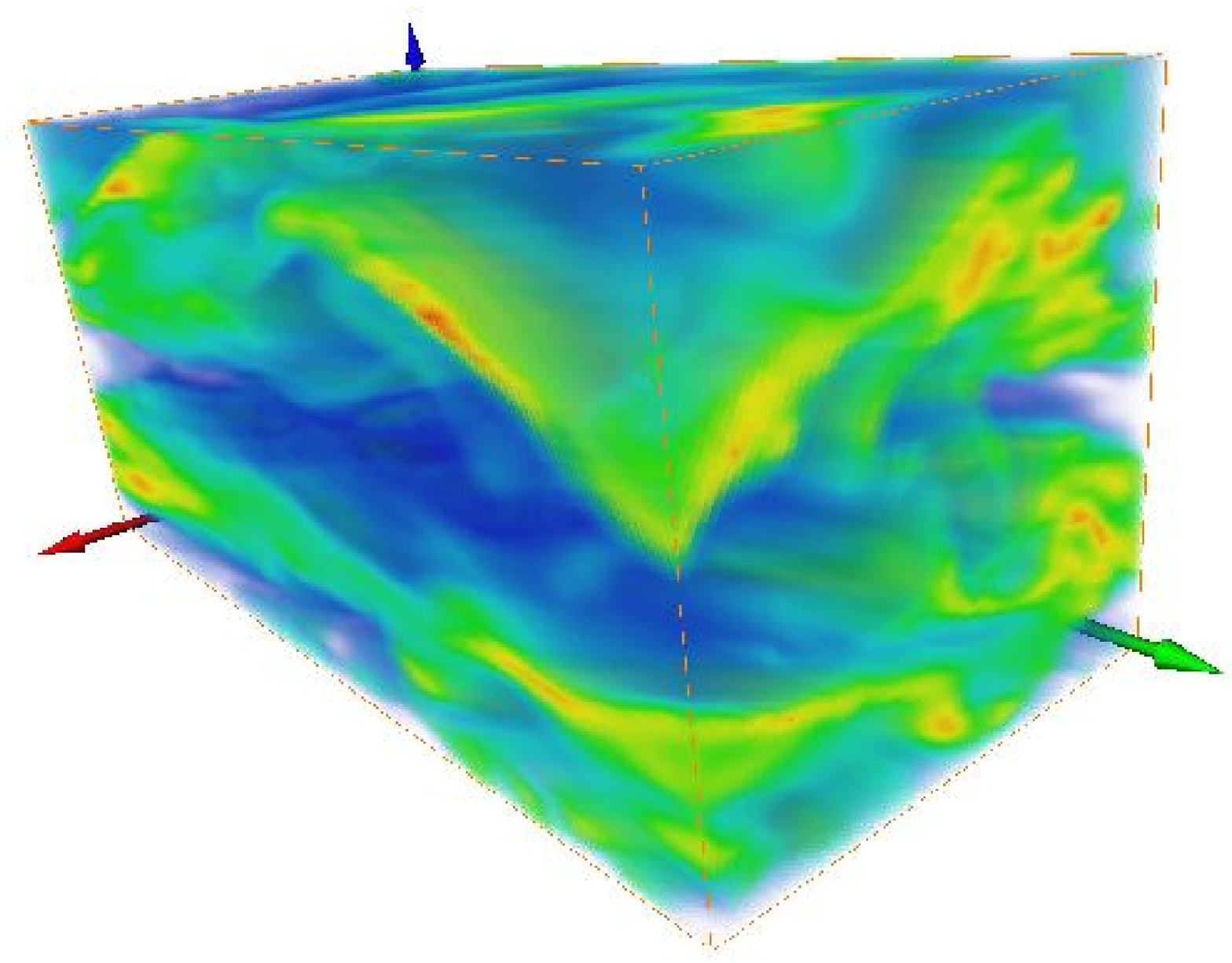}
  \caption{Volume rendered magnetic field strength of a model with
    $b_0^z = 4\times 10^{13}\, \mathrm{G}$ computed in a box of $1
    \times 2 \times 1\, \mathrm{km}^3$ with a resolution of $20
    \mathrm{m}$ at $t = 21.5 \mathrm{ms}$ (left) and $t= 37.2
    \mathrm{ms}$ (right), respectively.  The coordinate directions are
    indicated as in \figref{Fig:UZ3-R4S0-r150p1-R050-B13.2--z0x}.  }
  \label{Fig:UZ3-R4S0-r150p2-R050p2-B13.4--bstruct}
\end{figure*}

Only a subset of our models show a prominent re-appearance of single
channel flows, and most of them do not exhibit a secondary exponential
growth phase.  Instead, the mean magnetic energy and the Maxwell
stress remain roughly constant during saturation, albeit fluctuating
strongly (see the right panel of \figref{Fig:UZ3-4-p2--tevo}). Angular
momentum transport is less efficient for these models, and their
initial rotation profiles remain nearly unchanged.  A turbulent flow
and magnetic field persist during saturation, and coherent,
channel-like structures develop transiently.  The structure of the
magnetic field of a model with $b_0^z = 4\times 10^{13} \mathrm{G}$
computed on a grid of $50 \times 100 \times 50$ zones is displayed at
two different epochs in
\figref{Fig:UZ3-R4S0-r150p2-R050p2-B13.4--bstruct}.  At $t = 21.5\,
\mathrm{ms}$ (left panel) one recognizes a turbulent state, while
large-scale patterns (right panel; yellow structures) dominate the
flow at $t = 37.2 \mathrm{ms}$ when the magnetic field strength is
largest, and the Maxwell stress is strongest.  Unlike in the model
discussed above, the coherent flow is unstable and becomes turbulent
within a few milliseconds, and the absolute value of the Maxwell
stress $\left| M_{\varpi\phi} \right|$ decreases.

\paragraph{Channel modes and parasitic instabilities:}
The appearance and stability of single large-scale flows that lead to
a secondary exponential growth phase and eventually to the disruption
of the rotation profile depend on the geometry of the simulated domain
as well as on the ratio of the grid resolution and the fastest growing
mode.

Models, which are computed in a box of $1 \mathrm{km^{3}}$ with a
resolution of $20 \mathrm{m}$ and where velocity damping is applied,
develop secondary stable channels, if the initial magnetic field is
stronger than $2\times 10^{13} \mathrm{G}$.  The MRI growth rates
found for these models ($\sigma = \{0.76, 1.03, 1.10\}\,
\mathrm{ms^{-1}}$ for $b_0 = \{1, 2, 4 \} \times 10^{13} \mathrm{G}$,
respectively) indicate that the grid resolution is sufficiently fine
to resolve the fastest growing MRI mode for the two most strongly
magnetized models.  However, it is too coarse for the model with the
weakest initial field, because the theoretical growth rate for the
fastest growing MRI mode is $\sigma_{\mathrm{MRI}} = 1.14\,$ms$^{-1}$
for these models (see \secref{sSek:Res--2d-uf}).

\begin{figure}
  \centering
  \includegraphics[width=7cm]{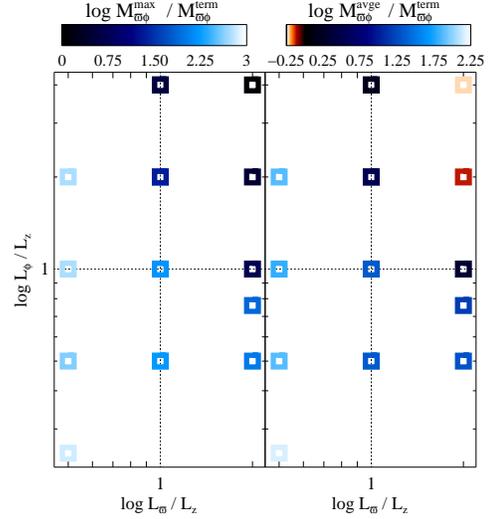}
  \caption{The left panel shows the ratio of the maximum Maxwell
    stress per unit volume, $M_{\varpi\phi}^{\mathrm{max}}$, and its
    value at MRI termination, $M_{\varpi\phi}^{\mathrm{term}}$, as a
    function of the toroidal and radial aspect ratios, $L_{\phi} /
    L_z$ and $L_{\varpi} / L_z$ for the models listed in
    \tabref{Tab:3d-models-1}.  The right panel shows the ratio of
    $M_{\varpi\phi}$ averaged over the saturation phase and the value
    at MRI termination.  Each model is represented by a symbol its
    color reflecting its maximum Maxwell stress. All models are
    computed imposing velocity damping at the radial grid boundaries.
}
  \label{Fig:U3-aspect-ratio}
\end{figure}

\begin{figure}
  \centering
  \includegraphics[width=7cm]{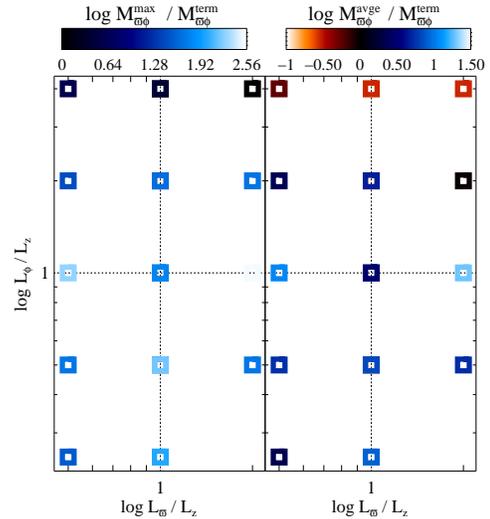}
  \caption{Same as \figref{Fig:U3-aspect-ratio}, but for models where
    no velocity damping is applied at the radial grid boundaries.
  }
  \label{Fig:U3-aspect-ratio-p}
\end{figure}

To investigate the stability properties of large-scale channel modes
as a function of the box geometry, we simulated models with an
initially uniform magnetic field using boxes of different size and
shape.  The models were rotating according to $\Omega_0 = 1900\,
\mathrm{s^{-1}}$ and $\alpha_\Omega = -1.25$, and their initial
magnetic field was $b_0^z = 4\times 10^{13} \mathrm{G}$ when applying
velocity damping boundaries, and $b_0^z = 2\times 10^{13} \mathrm{G}$
otherwise.  We varied both the ratio between the radial and vertical,
$L_\varpi / L_z$, and the radial and toroidal size, $L_\phi /
L_\varpi$, size of the box. The grid resolution was 20\,m (see
\tabref{Tab:3d-models-1}).  Plotting the stress ratios
$M_{\varpi\phi}^{\mathrm{max}} / M_{\varpi\phi}^{\mathrm{term}}$
(\figref{Fig:U3-aspect-ratio}; damping boundaries) and $\langle
M_{\varpi\phi} \rangle / M_{\varpi\phi}^{\mathrm{term}}$
(\figref{Fig:U3-aspect-ratio-p}; non-damping boundaries) as a function
of the aspect ratio of the computational box, provides some indication
of the range of $M_{\varpi\phi}$ values prevailing during the
post-growth phase. The ratios allow one to distinguish models with a
strong variability due to the dominant re-appearance of channel modes
from those models exhibiting a smooth evolution without dominant
large-scale coherent structures.

We find that models with a radial aspect ratio $L_{\varpi} / L_z = 1$
and a toroidal aspect ratio $L_{\phi} / L_z \ge 2$ are unstable
against parasitic instabilities, independent of the grid resolution in
toroidal direction.  Turbulence develops and leads to a flow structure
as shown in \figref{Fig:UZ3-R4S0-r150p2-R050p2-B13.4--bstruct}.
Models having the same radial aspect ratio, but a smaller toroidal one
are stable and evolve similarly as axisymmetric models, i.e.,
parasitic instabilities do not grow and a dominant large-scale channel
flow develops, which gives rise to a morphology of the type presented
in \figref{Fig:UZ3-R4S0-r150p2-R050p2-B13.4--bstruct}.  These findings
do not depend on how the growth of the MRI ends, i.e., whether
velocity damping is applied and reconnection between adjacent channels
occurs inside the box, or whether no damping is imposed and
reconnection occurs near the surface of the computational box.

These results can be understood from the analysis of parasitic
instabilities by \cite{Goodman_Xu__1994__ApJ__MRI-parasitic}, who
argued that three-dimensional flows are unstable against parasitic
instabilities, but these instabilities can be suppressed by the
geometry of the computational box.  According to their analysis, the
growth rate of the parasitic instabilities is highest for modes with
half the wave number of the unstable MRI modes they are feeding off.
Hence, if a channel flow forms at late times with a wavelength equal
to the box size in $z$-direction, $L_z$, unstable parasitic modes must
have a toroidal wavelength $\sim 2 L_z$ to grow rapidly. Thus results
in the criterion for the channel flow instability we have found in our
simulations.

In accordance with simulations presented recently by
\cite{Bodo_etal__2008__AA__MRI-aspect-ratrio}, we find that models
with a radial aspect ratio $L_{\varpi}/L_z \le 1$ experience a second
exponential growth phase as described in \secref{sSek:Res--2d} (note
the large ratios of $M_{\varpi\phi}^{\mathrm{max}}$ and
$M_{\varpi\phi}^{\mathrm{term}}$ for the corresponding models in
\figref{Fig:U3-aspect-ratio}), whereas a larger radial aspect ratio
appears to favor a less violent post-growth phase where coherent
channel modes can appear but are disrupted after a short time.
\cite{Bodo_etal__2008__AA__MRI-aspect-ratrio} obtained this result for
simulations performed with a toroidal aspect ratio $L_{\phi} / L_z =
4$.

We confirm a similar dependence of the dynamics on the radial aspect
ratio also in axisymmetry and for three-dimensional boxes with smaller
$L_{\phi} / L_z$. In this case, parasitic instabilities are unable to
disrupt the channel modes. Consequently, the MRI experiences a second
exponential growth phase dominated by just one (two in a few cases)
large channel mode of width $a$ which is determined by the size of the
computational box in $z$-direction. The maximum Maxwell stress that
can be reached is limited by the onset of resistive instabilities.
The dependence on the channel width (\eqref{Gl:App-Resi-sigma-result})
explains why the maximum Maxwell stress varies with $L_z$: larger boxes
allow for wider channels for which the resistive instabilities grow
slower, thus requiring higher Alfv\'en velocities for a growth rate
comparable to the one of the MRI.  Hence, the MRI reaches stronger
fields for larger (in $z$ direction) boxes.

Despite the differences in the MRI termination process, the behavior
of models with and without velocity damping is quite similar, because
the velocity damping does not affect the second generation of vigorous
channel flows significantly.  Thus, the breakup of these channels and
the values of the corresponding maximum Maxwell stress do not depend
strongly on the choice of the boundary condition.  On the other hand,
MRI termination (the termination of the initial exponential growth of
the MRI) does depend on whether velocity damping is applied or not,
and thus the ratio $M^{\mathrm{max}}_{\varpi\phi} /
M^{\mathrm{term}}_{\varpi\phi}$, too.

For boxes having a large toroidal aspect ratio, $L_{\phi} / L_z \ge
2$, we observe a quiet evolution during the non-linear saturation
phase of the MRI when varying the radial aspect ratio, $L_{\varpi} /
L_z$.  In particular, the fluctuations of $M_{\varpi\phi}$ are small
after MRI termination for models where both aspect ratios are large.
The models in the upper right corner of \figref{Fig:U3-aspect-ratio}
have values of $\langle M_{\varpi\phi} \rangle /
M_{\varpi\phi}^{\mathrm{term}}$ close to unity.

We may try to infer some consequences from these results for the MRI
in supernova cores.  Close to the core's equator the region, where the
MRI develops, can have to a small radial size, $L_{\varpi}$, ranging
from a few to a few hundred kilometers determined by the gradients of
$\Omega$ and $S$ in the core. The vertical extent of the unstable
region, $L_z$, can be expected to be of similar size.  The azimuthal
extent of the MRI unstable region, $L_{\phi}$, will be significantly
larger, leading to a non-violent evolution of the saturated state of
the MRI.  The geometry is different close to the pole.  However, we
cannot apply our results there without modifications as we have
considered only cases where the gradients of $\Omega$ and of all
thermodynamic quantities are aligned -- a situation which does not
apply near the poles.

\paragraph{Effects of resolution and initial magnetic fields in 3D vs 2D:}
After having discussed how the aspect ratios of the simulation box
determine the Maxwell stress at MRI termination and during the
subsequent saturation phase, we will now compare whether the behavior
of 3D models differs from that of 2D models.  The models discussed in
this paragraph are listed in \tabref{Tab:3d-models-1} and
\ref{Tab:3d-models-2}.

First, we note that in 3D, as in axisymmetry, the growth rate of the
instability is not affected by the choice of the grid provided the
fastest growing mode is resolved.

\begin{figure}
  \centering
  \includegraphics[width=7cm]{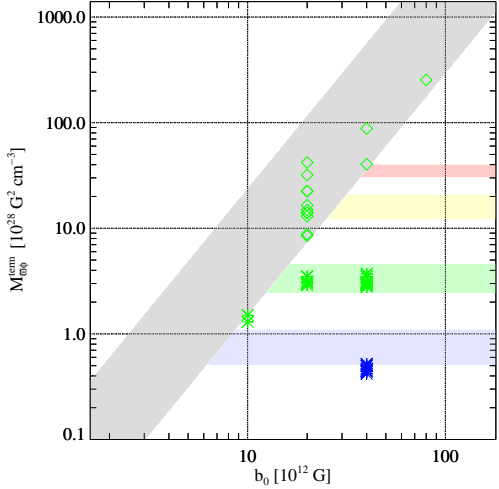}
  \caption{Volume-averaged Maxwell stress component
    $M_{\varpi\phi}^{\mathrm{term}}$ at MRI termination as a function
    of the initial magnetic field strength, $b_0$, for axisymmetric
    models with uniform initial magnetic field in $z$-direction, a
    rotational profile $\Omega= 1900\,$s$^{-1}\, \varpi^{-1.25}$, and
    a vanishing entropy gradient for a set of 3D models computed with
    (asterisks) and without (diamonds) velocity damping at the radial
    boundaries.  Models computed in a box with a radial size of
    $1\,$km and $0.5\,$km are shown with green and blue symbols,
    respectively.  The colored bands are the same as in
    \figref{Fig:UZ2-R4S0--termination}.  }
  \label{Fig:UZ3-M_b0}
\end{figure}

\figref{Fig:UZ3-M_b0} demonstrates that in 3D the dependence of
$M_{\varpi\phi}^{\mathrm{term}}$ on the initial magnetic field
strength is well described by the same power law as in axisymmetry:
models without damping of the radial velocity line up along a band
$\propto (b_0^z)^{16/7}$, and models with velocity damping are
characterized by a roughly constant value of
$M_{\varpi\phi}^{\mathrm{term}}$, which depends on the size of the
radial box (compare with \figref{Fig:UZ2-R4S0--termination}). This
agreement is to be expected as the growth and the resistive disruption
of channel flows are essentially axisymmetric processes, which are,
thus, not significantly modified by three-dimensional effects.

After MRI termination the evolution of $M_{\varpi\phi}$ depends on the
aspect ratio of the computational box (see discussion above).  When
averaging the fluctuating Maxwell stresses over the saturation phase,
we find values for $ \langle M_{\varpi\phi} \rangle$ which differ
considerably from those of $M_{\varpi\phi}^{\mathrm{term}}$.  Lacking
a thorough understanding of the instabilities involved in the MRI
saturation process, and having only a imited set of 3D models at hand,
one is not yet in a position to formulate a better description of the
dependence of the evolution after MRI termination on the aspect ratio
of the box, and to provide a unified description of MRI saturation
amplitudes.

\subsubsection{Uniform $b^z$ field, entropy gradients}

\paragraph{Mixed regime:}

Let us first consider models from the mixed regime having an initial
rotation law given by $\Omega_0 = 1900 \, \mathrm{s}^{-1}$ and
$\alpha_\Omega = -1.25$, and an entropy distribution given by $S_0=
0.2$ and $\partial_{\varpi} S = -0.038$ (i.e, with a negative entropy
gradient). In axisymmetry the MRI grows in these models with a rate of
$\sigma_{\mathrm{MRI}} \approx 1.7 \, \mathrm{ms}^{-1}$, i.e., close
to the theoretical value of $\approx 2.0 \mathrm{ms}^{-1}$.  The 3D
models show the same growth rate provided the spatial grid resolution
is sufficiently high.

The long-term evolution (i.e., many rotational periods into the
non-linear phase) of the models depends strongly on the choice of the
radial boundary conditions.  If the entropy at the inner and outer
boundary is allowed to change (i.e., using reflective boundaries), a
flat entropy profile develops after a short time.  To reduce the
influence of boundary effects, one could employ a technique widely
used in simulations of convective layers: add a cooling layer on top
of and an overshoot layer below the convection zone.  However,
exploring this approach was beyond the scope of the present work.

For models having a negative entropy gradient the growth of the MRI is
not influenced by 3D effects, if the fastest growing mode is resolved.
Thus, their behavior is similar to that of models with no entropy
gradient.  The Maxwell stress at MRI termination also does not differ
significantly from that of the corresponding axisymmetric models, and
due to the boundary conditions applied in the models (velocity
damping) its value is set by recombination of field lines close to
the inner and outer radial boundary.

\begin{figure}
  \centering
  \includegraphics[width=7cm]{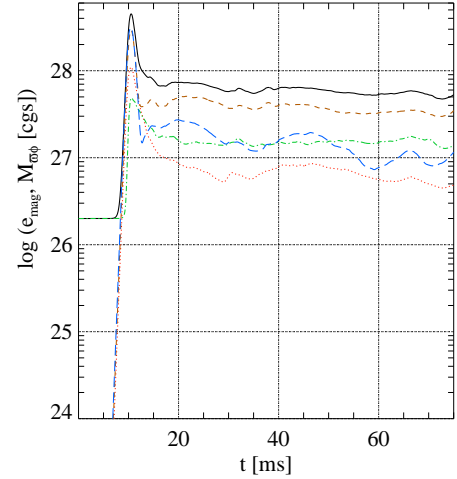}
  \caption{Evolution of the mean magnetic energy density
    $e^\mathrm{mag}$ (solid black line), the mean energy densities
    corresponding to the $\varpi$ (dotted red), $\phi$ (dashed brown),
    and $z$ (dash-dotted green) component of the magnetic field, and
    the absolute value of the mean Maxwell stress component
    $M_{\varpi,\phi}$ (dashed blue line) for a 3D model belonging to
    the mixed regime.  The model rotates differentially with $\Omega_0
    = 1900 \, \mathrm{s}^{-1}$ and $\alpha_\Omega = -1.25$. The
    initial entropy gradient is $\partial_{\varpi} S = -0.038\,
    \mathrm{km}^{-1}$, and the initial magnetic field strength is
    $b_0^z = 2\times 10^{13} \mathrm{G}$.  The model was simulated in
    a box of size $L_{\varpi} \times L_{\phi} \times L_z = 1 \times 2
    \times 1\, \mathrm{km}^{-3}$ and on a grid of $50 \times 100
    \times 50$ zones.  }
  \label{Fig:U3-R4Sm2-r150p2-R050p2-B13.2--tevo}
\end{figure}

\begin{figure}
  \centering
  \includegraphics[width=7cm]{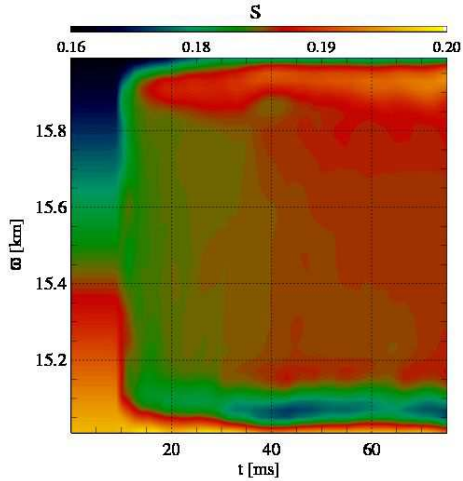}
  \caption{ Average radial entropy profile as a function of time for
    the model shown in
    \figref{Fig:U3-R4Sm2-r150p2-R050p2-B13.2--tevo}.  }
  \label{Fig:U3-R4Sm2-r150p2-R050p2-B13.2--Struktur}
\end{figure}

Contrary to axisymmetric models, the saturated state of the 3D models
does not show any sign of a late exponential growth phase
characterized by the re-appearance of channel modes, and the saturated
MRI stresses are smaller in magnitude than
$M_{\varpi\phi}^{\mathrm{term}}$, i.e., the maximum Maxwell stress is
reached at MRI termination (see
\figref{Fig:U3-R4Sm2-r150p2-R050p2-B13.2--tevo} and compare with
\figref{Fig:Res-2d--UZ2-4-tevo}). The evolution of the average radial
entropy profile profile, computed as the average of $S (\varpi, \phi,
z)$ at constant $\varpi$, is shown in
\figref{Fig:U3-R4Sm2-r150p2-R050p2-B13.2--Struktur}.  Until the
saturation of the instability (at $t \approx 11 \mathrm{ms}$), the
initial linear profile $S(\varpi)$ is basically unchanged. However,
afterward the entropy profile flattens, $S$ becoming nearly constant
for $15.2 \mathrm{km} \leq \varpi \leq 15.8 \mathrm{km}$. Close to
both radial boundaries, the entropy profile develops extrema, which
are most likely an artifact of our boundary conditions.  The flat
entropy profile is stable and does not vary strongly with time.  The
$\Omega$ profile flattens after the initial growth phase, too.  The
velocity field in the saturated state is dominated by a rich
small-scale structure, while the magnetic field is organized in a
multitude of flux tubes.

\paragraph{MBI regime:}
Next we consider a few models that, in axisymmetry, belong to the MBI
regime.  Initially, the models rotate rigidly with $\Omega = 1900\,
\mathrm{s}^{-1}$ and possess an entropy gradient $\partial_{\varpi} S
= -0.10 \, \mathrm{km}^{-1}$ (i.e., $\mathcal{C} = -3.6$).  We
computed models for $b_0^z = 10 \, \mathrm{G}$, $10^{10} \mathrm{G}$,
and $2\times 10^{13} \mathrm{G}$, respectively.  All models are
simulated in a box of size $L_{\varpi} \times L_{\phi} \times L_z = 1
\times 2 \times \times 1\, \mathrm{km}^{-3}$ and on a a grid of $50
\times 100 \times 50$ zones.

\begin{figure}
  \centering
  \includegraphics[width=7cm]{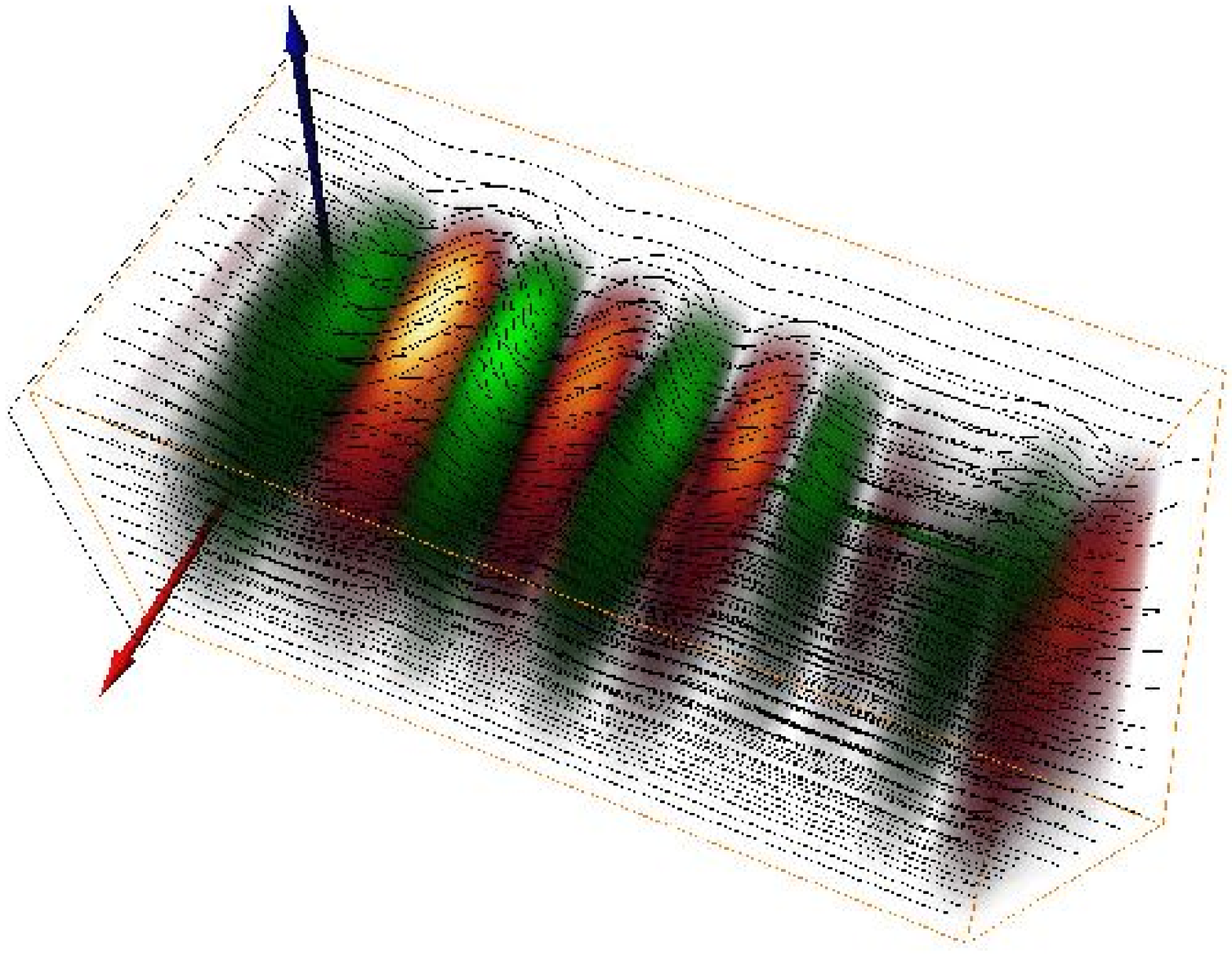}
  \includegraphics[width=7cm]{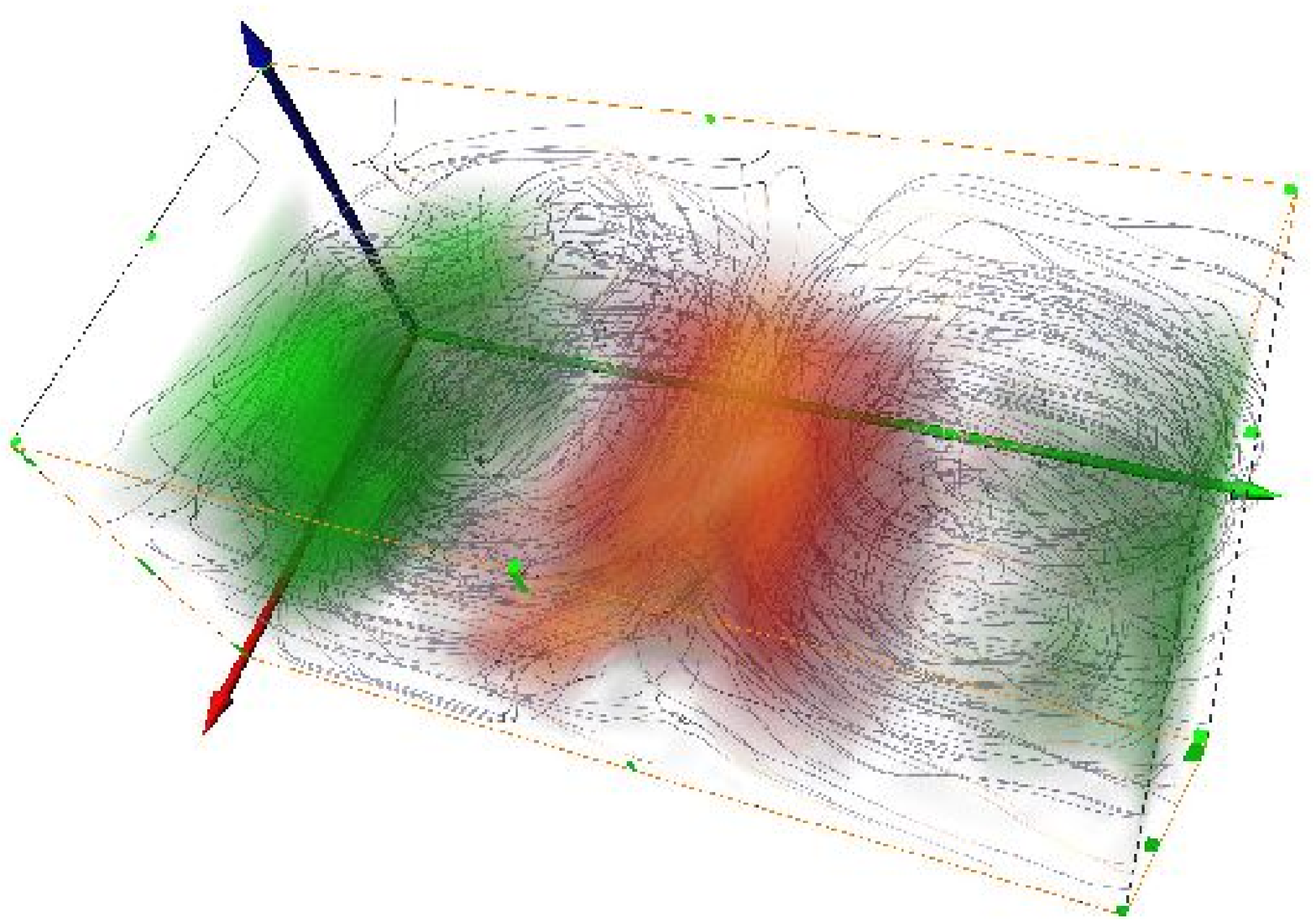}
  \caption{Flow structure of a MBI model with $b_0^z = 10 \mathrm{G}$
    at $t = 4.86 \mathrm{ms}$ (upper panel), and $t = 14.4
    \mathrm{ms}$ (lower panel), respectively.  The solid black lines
    are stream lines of the velocity field computed in a frame
    co-rotating with the mean angular velocity, and regions of
    positive and negative radial velocity are colored in red and
    green, respectively.  The colored arrows point into the same
    coordinate directions as in
    \figref{Fig:UZ3-R4S0-r150p1-R050-B13.2--z0x}.  }
  \label{Fig:U3-R0-19Sm3-r150p2-R050p2-B01-z026}
\end{figure}

Contrary to their axisymmetric counterparts (see
\secref{sSek:Res--2d}), these models develop convective modes even
when no magnetic field is present.  As described, e.g., by
\cite{Tassoul_Book__Theo_rotating_stars}, rotation can stabilize
axisymmetric modes in a convectively stable environment, but
non-axisymmetric modes can nevertheless grow in that situation.  The
model with the weakest initial field ($b_0^z = 10 \mathrm{G}$) shows a
growth of non-axisymmetric MBI modes, but these modes cannot be
resolved due to the extremely weak initial field.  Thus, the model
behaves essentially similar to an unmagnetized one, but can serve as a
reference model for initially more strongly magnetized models where we
can resolve $\lambda_\mathrm{MRI}$.  The growing convective modes
eventually extend over the entire domain in radial and $z$-direction,
while having a small wavelength in $\phi$ direction (see
\figref{Fig:U3-R0-19Sm3-r150p2-R050p2-B01-z026}, upper panel).  The
exponential growth of the convective instability saturates at $t
\approx 7 \mathrm{ms}$.  During this growth phase the mean magnetic
energy increases at the same rate as does the kinetic energy.  After
MRI termination the structure of the model is characterized by two
large, roughly cubic convective cells with a size of about $1\,
\mathrm{km}^{3}$ instead of a multitude of elongated structures (see
(\figref{Fig:U3-R0-19Sm3-r150p2-R050p2-B01-z026}, lower panel), and an
essentially flat entropy profile. The magnetic field is subject to
kinematic amplification at a smaller growth rate than before MRI
termination due to stretching in the convective vortices.  At much
later epochs the typical size of structures in the velocity field
decreases again, leading to more turbulent fields.  The magnetic
field, which is too weak to affect the dynamics, is passively advected
with the flow.

\begin{figure}[htbp]
  \centering
  \includegraphics[width=8cm]{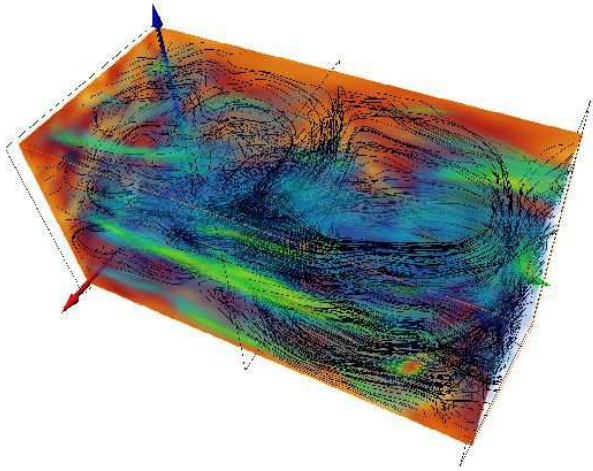}
  \caption{Structure of a rigidly rotating model with $b_0^z = 2\times
    10^{13} \mathrm{G}$ at $t = 19.3 \mathrm{ms}$.  The figure shows a
    volume rendering of the magnetic field strength (blue -- green),
    the value of the angular velocity, $\Omega$, at three slices
    parallel to the coordinate axes (red -- yellow), and stream lines
    of the velocity field in a co-rotating frame (black lines).  In
    the right half of the figure, one can identify one large
    convective cell, whereas the features of, e.g., the velocity field
    in the left half are of considerably smaller scale.  The magnetic
    field is comparably weak inside the convective cell, and strongest
    in the flux tubes both at the boundaries of the convective cell
    and in the left part.  The colored arrows point into the same
    coordinate directions as in
    \figref{Fig:UZ3-R4S0-r150p1-R050-B13.2--z0x}.  }
  \label{Fig:U3-R0-19Sm3-r150p2-R050p2-B13.2--z0800}
\end{figure}

For the model having the strongest initial magnetic field $b_0^z =
2\times 10 ^{13} \, \mathrm{G}$ we can resolve $\lambda_\mathrm{MRI}$.
The axisymmetric version of this model showed a MBI growth a rate
close to the theoretical one ($\sigma_\mathrm{MRI} \sim 1.7 \,
\mathrm{ms}^{-1}$).  For the 3D model we find $\sigma_\mathrm{conv}
\sim 2.6 \, \mathrm{ms}^{-1}$), i.e., its evolution is dominated by
convection (although we are able to resolve the MBI), and the MBI
growth rate is similar to that of a weakly magnetized model (see
previous paragraph).  MBI growth is mediated by non-axisymmetric modes
having the same elongated geometry as those in the essentially
unmagnetized model.  After saturation, a few large vortices of
approximately cubic shape form, which later decay into small-scale
structures again.  An intermediate stage of this decay process is
displayed in \figref{Fig:U3-R0-19Sm3-r150p2-R050p2-B13.2--z0800}, when
one large vortex is still present in the right half of the box, while
its left half is dominated by spatially less coherent fields.  At even
later times the vortex disappears and the structure of the whole model
is similar to that shown in the left half of the box.  The mean
magnetic energy and Maxwell stress are small compared to typical MSI
or mixed models.  Compared to a differentially rotating model
($\Omega_0 = 1900 \, \mathrm{s}^{-1}$, $\alpha_\Omega = -1.25$) with a
vanishing entropy gradient, the maximum magnetic fields are reduced by
a factor of $\lesssim 2$, and the mean magnetic energies and Maxwell
stresses by a factor of $\sim 10$, but we still find a slow growth at
the end of the simulation.

Finally, we add a few comments on a model where we cannot resolve
$\lambda_{\mathrm{MRI}}$ ($b_0^z = 10^{10} \mathrm{G}$), but where the
magnetic field saturates within $\sim 60 \mathrm{ms}$ after the onset
of convection.  The model evolves similarly to the essentially
unmagnetized one, but at $t \approx 60 \mathrm{ms}$ the energy of the
kinematically amplified magnetic field becomes almost as high as the
convective kinetic energy.  The amplification process ceases, and the
magnetic energy levels off.  Close to the end of the simulation
convective transport gives rise to a rotation law $\Omega \approx
\mathrm{const.}$.

\subsubsection{Magnetic fields with zero net flux}
Many results discussed above also apply analogously to models where
the initial magnetic field has a vanishing net flux through the
surface of the computational box.  We simulated models with $\Omega =
1900 \, \mathrm{s}^{-1}$ and $\alpha_\Omega = -1.25$, and find a
basically axisymmetric growth of channel flows, which decay due to
resistive instabilities at a level consistent with axisymmetric
models. The late evolution of the models is dominated by turbulent
fields.  Contrary to uniform-field models, there is no second phase of
channel activity.  Hence, the mean magnetic energies and Maxwell
stresses do not fluctuate violently in the saturated state, and - even
when no velocity damping is applied - the maximum values of, e.g.,
$\left| M_{\varpi\phi} \right|$, are reached at MRI termination. Later
on $\left| M_{\varpi\phi} \right|$ decreases by a factor of a few, and
stays roughly constant subsequently.

Performing similar simulations
\cite{Lesur_Ogilvie__2008__ApJ__zero_flux_MRI_dynamo} proposed a
non-linear dynamo that balances the dissipation of the magnetic energy
in MRI models with zero net flux.  In the turbulent saturated state,
they identified large-scale spatially (over a sizable fraction of the
box) and temporally (over several rotational periods) coherent
patterns of the toroidal magnetic field.  To study this process, we
looked for similar patterns in our models.

\begin{figure}
  \centering
  \includegraphics[width=7cm]{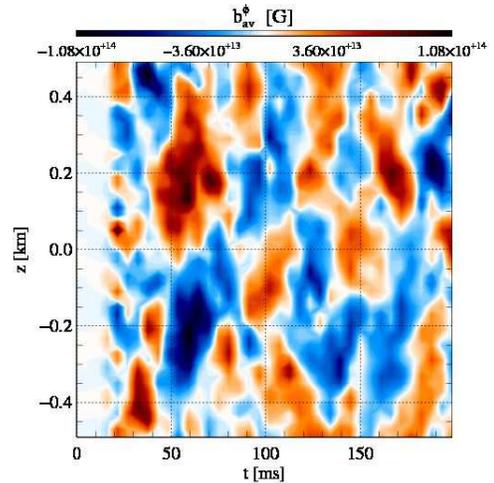}
  \caption{Evolution of large-scale coherent patterns exhibited by the
    mean (i.e., averaged over a plane at constant $z$-coordinate)
    toroidal field component of a 3D model having an initial magnetic
    field of zero flux and strength $b_0^z = 2\times 10^{13}
    \mathrm{G}$. The model was simulated in a box of $1 \mathrm{km}^3$
    covered by $50^3$ zones with no velocity damping applied.  This
    figure is similar to Fig.\,1 of
    \cite{Lesur_Ogilvie__2008__ApJ__zero_flux_MRI_dynamo}.
}
  \label{Fig:Z3-R4S0-r150p1-R050p1-B13.2-p--by-fluctuations}
\end{figure}

An example is shown in
\figref{Fig:Z3-R4S0-r150p1-R050p1-B13.2-p--by-fluctuations} for a
model with $b_0^z = 2\times 10^{13} \, \mathrm{G}$ simulated in a box
of $1 \mathrm{km}^{3}$ on a grid of $50^{3}$ zones applying no
velocity damping. The figure shows the mean value of the toroidal
field component $b^{\phi}_{\mathrm{av}}$ (i.e. $b^{\phi}$ averaged
over $\varpi$ and $\phi$) as a function of $z$ and time.  For $t
\lesssim 14 \mathrm{ms}$ the early channels flows can be identified in
which the magnetic field grows.  At $t \approx 14 \mathrm{ms}$, the
channels are disrupted, and the growth of the field seizes.  In the
saturated state that follows, the mean (vertical) size of the
structures is larger: at any time, we find only a few (typically two)
regions of opposite field polarity (blue and red), which remain stable
for a few rotational periods ($\Omega^{-1} \approx 0.53 \mathrm{ms}$).
Thus, we make similar observations as
\cite{Lesur_Ogilvie__2008__ApJ__zero_flux_MRI_dynamo} do for their
models.

\begin{figure}
  \centering
  \includegraphics[width=7cm]{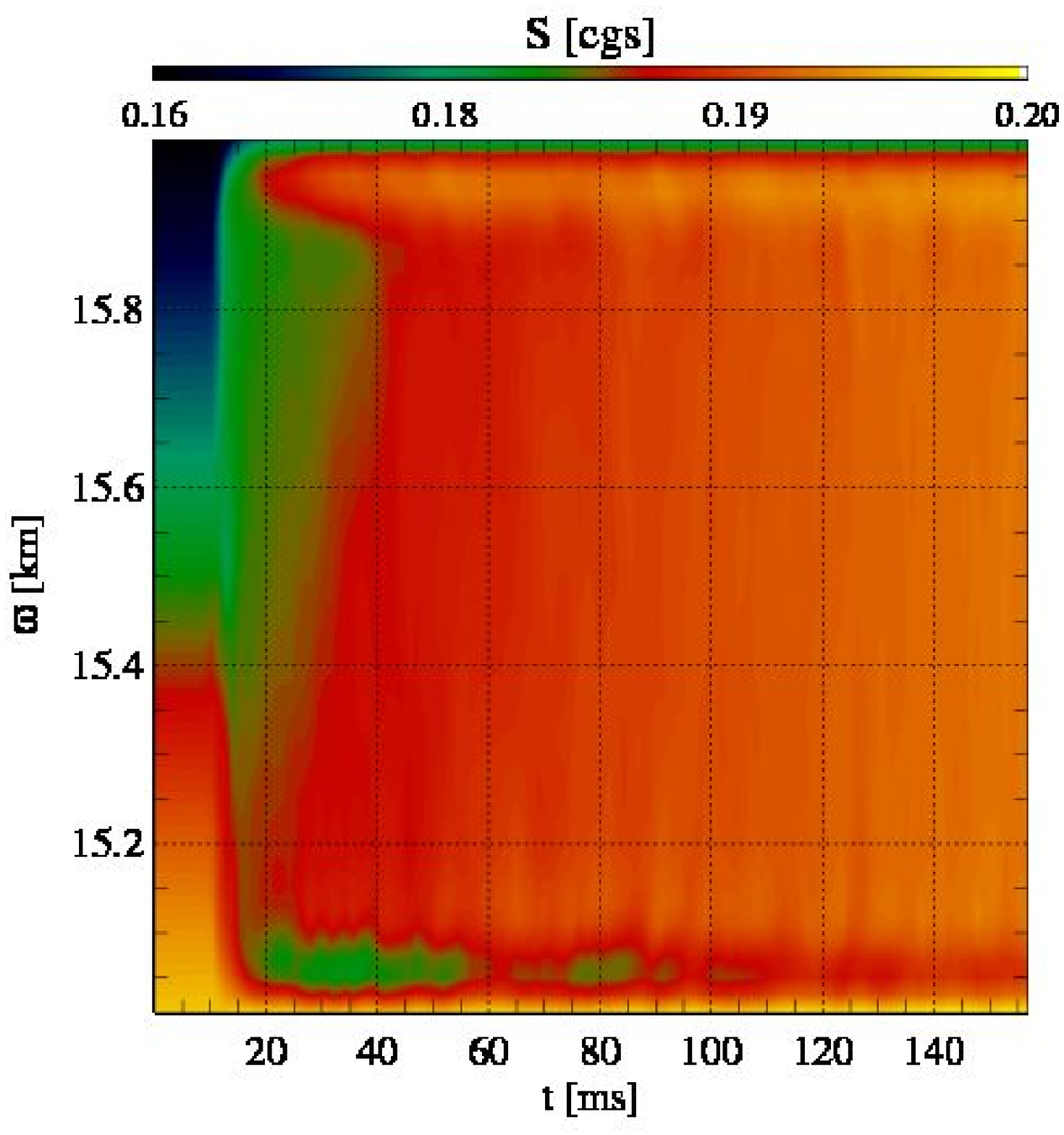}
  \includegraphics[width=7cm]{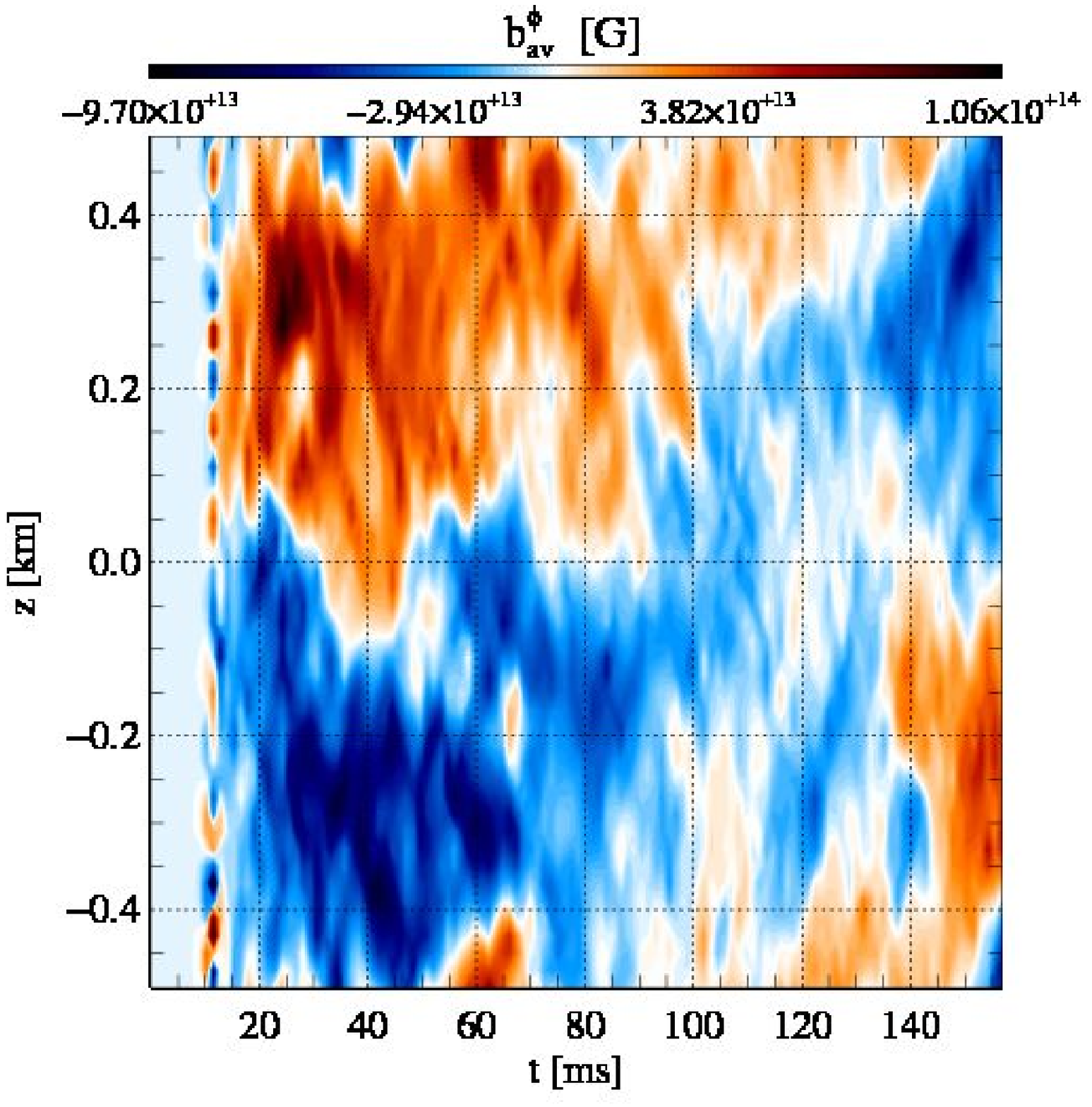}
  \caption{Evolution of a mixed-regime 3D model with zero net flux
    ($b_0^z = 2\times 10^{13} \mathrm{G}$), an entropy gradient
    $\partial_{\varpi} S = -0.038 \mathrm{km}^{-1}$, and a rotation
    law given by $\Omega = 1900 \, \mathrm{s}^{-1}$ and $\alpha_\Omega
    = -1.25$.  The upper panel shows the entropy of the model averaged
    over planes of constant $\varpi$ as a function of radius and time.
    The lower panel shows, similarly to
    \figref{Fig:Z3-R4S0-r150p1-R050p1-B13.2-p--by-fluctuations}, the
    average of $b^{\phi}$ over planes of constant $z$ as a function of
    $z$ and time.
}
  \label{Fig:Z3-R4Sm2-r150p2-R050p2-B13.2}
\end{figure}

We also simulated a few 3D models with zero-flux fields in the mixed
regime. The results are analogous to those obtained if the initial
fields are uniform(see, e.g.,
\figref{Fig:Z3-R4Sm2-r150p2-R050p2-B13.2}).  The MRI growth rates are
similar to those of the corresponding 2D models, and the mean Maxwell
stress in the saturated state is somewhat smaller. In the saturated
state, both the entropy (upper panel) and the angular velocity
profiles become flat rapidly.  The spatially and temporally coherent
large-scale structures in the magnetic field are even more pronounced
than in MSI models (compare the lower panel of
\figref{Fig:Z3-R4Sm2-r150p2-R050p2-B13.2} with
\figref{Fig:Z3-R4S0-r150p1-R050p1-B13.2-p--by-fluctuations}).  They
consist of two large regions characterized by an opposite sign of
$b^{\phi}$, which persist for the entire simulation of the saturated
state ($\approx 140 \mathrm{ms}$), subject only to a slow drift in
vertical direction.  The implications of this behavior for the
presence and properties of a nonlinear dynamo of the type proposed by
\cite{Lesur_Ogilvie__2008__ApJ__zero_flux_MRI_dynamo} remain to be
explored.

\section{Summary and conclusions}
\label{Sec:Sum}
We have studied the possible amplification of seed perturbations in
supernova cores by the magneto-rotational instability.  If the MRI
grows on dynamically relevant time scales (a few tens of
milliseconds), it can lead to MHD turbulence and efficient transport
of angular momentum.  Because the growth of the magnetic field and the
associated Maxwell stresses is exponential in time, the MRI is one of
the most promising mechanisms to amplify the -- most likely weak --
magnetic field of the supernova progenitor up to dynamically relevant
strengths.

As pointed out by \cite{Akiyama_etal__2003__ApJ__MRI_SN}, the
conditions for the instability are fulfilled in typical post-collapse
supernova cores.  Under the assumption that the MRI converts most of
the energy contained in differential rotation into magnetic energy,
these authors predicted saturation fields of approximately $10^{15}
\mathrm{G}$. This prediction derived from a semi-analytic analysis and
1D simulations can only be confirmed by detailed multi-dimensional
numerical simulations.  The reliability of global simulations of the
entire core, however, is limited due to the necessity to resolve
accurately small length scales (a few meters, at most) leading to
impracticable computational costs.

Traditionally, the MRI is studied in great detail in accretion discs,
i.e., in systems dominated by Keplerian rotation.  Because typical
post-collapse supernova cores differ from these systems in many
respects, e.g., by the importance of the thermal stratification and
the sub-Keplerian rotation, we investigated the instability under more
general physical conditions.  Analyzing the MRI dispersion relation of
\cite{Balbus__1995__ApJ__stratified_MRI,Urpin_1996_MNRAS_MHD-stability},
we identified the regimes of the instability relevant to supernova
cores.

We distinguish between \emph{Alfv\'en} and \emph{buoyant} modes of the
MRI. The former ones are generalizations of the standard MRI modes,
and the latter ones resemble standard convective modes.  Buoyant modes
are unstable only in systems dominated by a negative entropy gradient,
whereas Alfv\'en modes prevail if differential rotation is the main
agent of the instability.  Whereas Alfv\'en modes are rapidly
amplified only for a small range of wave numbers, buoyant modes grow
at essentially the same rate for a wide range of wave numbers, $k \leq
k_{\mathrm{max}}$.

We have identified six regimes of the MRI depending on the ratio of
the entropy and angular velocity gradient.  These MRI regimes and
their properties can be summarized as follows:
\begin{enumerate}
\item Sufficiently large positive gradients of the angular velocity
  or of the entropy define the \emph{stable regime} with oscillatory
  rather than growing modes.
\item For sufficiently strong differential rotation and small
  entropy gradients (or small buoyancy frequencies), we find the
  \emph{shear regime}, corresponding to the hydrodynamic shear
  instability.
\item If negative entropy gradients dominate the system, it is
  located in the \emph{convective regime}, which resembles ordinary
  hydrodynamic convection potentially modified in the non-linear
  phase by the presence of a magnetic field.
\item A small degree of differential rotation (e.g., Keplerian) and
  a small entropy gradient (if present at all) are the conditions
  for the \emph{magneto-shear (MSI) regime}, well studied for
  accretion discs.
\item When fast (nearly) rigid rotation suppresses convection, its
  stabilizing effect can be overridden by a weak magnetic field,
  giving rise to \emph{magneto-buoyant (MBI)} modes.  This regime is
  only encountered in axisymmetric flows as rotation can stabilize
  only axisymmetric modes of convection, i.e., in three dimensions
  convection may grow faster than the MBI.
\item Finally, a \emph{mixed regime} exists which shares properties
  of all unstable regimes listed above.
\end{enumerate}

To substantiate our stability analysis, we performed a set of more
than 200 models of semi-global high-resolution simulations of the MRI
in simplified models of post-bounce cores.  Our novel semi-global
simulations combine elements of both global and local simulations by
taking into account the presence of global background gradients and by
providing high local spatial resolution. In particular, we employed
the shearing-disc boundary conditions proposed by
\cite{Klahr_Bodenheimer__2003__ApJ__Global-baroclinic-inst-disc},
which allow for the treatment of global gradients of, e.g., density or
entropy, and studied the influence of a thermal stratification on the
MRI assuming various (radial) entropy profiles.  The presence of
gradients constitutes an important difference of our setting from that
of accretion discs.  We used a newly developed Eulerian
high-resolution MHD code to evolve the flow in a computational box
having an edge length of a few kilometers. The box was located in the
equatorial plane of the core at a distance of $15\,$km.  The initial
data were computed assuming hydrostatic equilibrium of differentially
rotating matter described by a simplified equation of state.  The gas
in the box was endowed with a weak initial magnetic field of different
topology and strength.  In most of the simulations, the magnetic field
of the progenitor had a strength of approximately $10^{10}
\mathrm{G}$.  We neglected the effects of neutrino radiation and
assumed an ideal MHD flow.

Computed under the assumption of rotational equilibrium, our
background models differ from realistic supernova cores in many
respects.  One of the most important shortcomings is the neglect of
the large-scale accretion flow through the post-shock region onto the
proto-neutron star.  From a physical point of view, such flows, whose
influence on the MRI has not been studied previously, add to the
already considerable complexity of the problem considered here,
possibly allowing for additional regimes and modifying growth and
saturation of the instability.  Technically, the proper treatment of
the boundary conditions required for a correct modeling of such flows
in non-global simulations is rather involved.  Thus, we have decided
to postpone the study of the MRI in the presence of a large-scale
accretion flow and focus on systems without overall radial motions.

The main results of our simulations are agree well with both our mode
analysis and with local simulations of the MRI in accretion
discs. They also confirm the estimates of
\cite{Akiyama_etal__2003__ApJ__MRI_SN}, and they are consistent with
the results of global MHD simulations of core collapse
\citep[e.g.,][]{Obergaulinger_Aloy_Mueller__2006__AA__MR_collapse,
  Obergaulinger_et_al__2006__AA__MR_collapse_TOV}.  We summarize our
results as follows:
\begin{enumerate}
\item Under the physical conditions considered in this study, i.e., in
  a background rotating in hydrostatic equilibrium, the MRI can act in
  supernova cores amplifying an initial magnetic field strongly.  The
  growth times are approximately equal to the rotational period of the
  core, which for rapidly rotating cores is sufficiently fast to
  influence the dynamics.
\item Due to our relatively fine numerical grids, we were able to
  resolve the fastest growing MRI modes for initial field strengths
  higher than a few $10^{12} \mathrm{G}$.  This threshold is
  considerably lower than the one of previous global simulations
  \citep[e.g.,][]{Obergaulinger_Aloy_Mueller__2006__AA__MR_collapse,
    Obergaulinger_et_al__2006__AA__MR_collapse_TOV}, enabling us to
  probe the MRI in a parameter range inaccessible to global
  simulations.
\item The growth of the instability is accompanied by the development
  of channel flows, predominantly radial flows of alternating
  direction stacked up in the $z$-direction.  This flow pattern is
  characteristic of both axisymmetric and three-dimensional
  simulations. The width of the channels is set by the wave length of
  the fastest growing MRI modes.
\item At MRI termination (i.e., at the end of the first exponential
  growth phase of the instability) the channels dissolve into a
  turbulent flow having a complex magnetic field topology.  During the
  subsequent evolution secondary generations of channel flows can
  (re-) appear which characterize secondary phases of exponential
  growth.
\item We identified the mechanism responsible for the breakup of the
  channel flows and MRI termination in our simulations.  Despite the
  absence of a physical resistivity in our model equations, we find
  that resistive MRI instabilities of the tearing-mode type develop
  due to the finite numerical resistivity of our MHD code.  A main
  characteristic of the channel flows is the presence of prominent
  current sheets immersed between layers of opposite magnetic
  polarity, which are unstable against current-driven instabilities.
  Using a simplified model of this kind of flows, we investigated the
  growth rates, $\sigma_{\mathrm{r}}$, of the resistive instabilities,
  and derived an approximate law for the scaling of
  $\sigma_{\mathrm{r}}$ with the magnetic field strength present in
  the channels, the channel width, and the grid resolution.  Comparing
  these growth rates with the MRI ones, we find that the MRI ceases to
  grow once the tearing modes grow faster than the MRI.  Using this
  criterion, we are able to explain the dependence of the conditions
  (i.e., field strength, Maxwell stresses) at MRI termination on,
  e.g., the initial field strength, the grid resolution, and the
  initial rotation profile.

  Strictly speaking, there should be no reconnection without physical
  resistivity, and the behavior of a magnetized ideal fluid subject to
  numerical resistivity may be quite different from that of a fluid
  having a large but finite conductivity.  Consequently, our results
  and their implications cannot replace a rigorous treatment of MRI
  growth in supernova cores with a non-ideal MHD model.  In
  particular, we have to be careful when drawing conclusions for the
  MRI in non-ideal plasmas.  Nevertheless, our results provide some
  qualitative insight into the basic processes of MRI saturation,
  highlighting the importance of tearing-mode-like instabilities.
  Quantitative conclusions, as e.g., the scaling laws for the field
  strengths and the Maxwell stresses at MRI termination as a function
  of the initial magnetic field strength, should be taken with a grain
  of salt.  These depend on the dissipative properties of the
  numerical scheme employed, which are likely to change when physical
  resistivity is considerd.
\item The saturation phase of the MRI differs considerably between
  axisymmetric and unrestricted 3D models, and between models having a
  different initial field configuration. In 2D the flow does not break
  down into small-scale turbulence, instead the channel flows merge
  until they form one pair of large-scale coherent up- and down-flows.
  When the strength of the magnetic field exceeds $10^{15}
  \mathrm{G}$, the rotational profile is modified within a few tens of
  milliseconds.
\item Axisymmetric models having an initial magnetic field with a
  vanishing net flux through the computational box become turbulent
  after a growth phase dominated by channel flows.  The saturation
  fields are considerably smaller than $10^{15} \mathrm{G}$.
\item The previous finding also holds for 3D models. Turbulence
  develops, but a spontaneous reorganization of the flow may lead to a
  re-appearance of channel modes, resulting in Maxwell stresses
  comparable to those found for axisymmetric models.  In models which
  do not develop late-stage channel flows, field strengths up to
  several $10^{14} \mathrm{G}$ are encountered.  The field is
  predominantely toroidal.  The extent of the late-time channel
  activity depends on the development of secondary (parasitic)
  instabilities, both flow-driven (e.g., Kelvin-Helmholtz) and
  current-driven (e.g., tearing modes), which feed off the channel
  flows.  The presence of these instabilities is determined to a large
  degree by the the aspect ratio of the computational box, i.e., we
  observe a strong dependence of the saturated state on the aspect
  ratio. For magneto-rotational core collapse, our results suggest
  that secondary instabilities are fairly efficient in suppressing
  coherent channel flows during saturation.
\item For models having an initial entropy gradient, we find an
  important influence of convective stabilization or destabilization
  on the evolution of the MRI.  We confirm the instability regimes
  predicted by our linear analysis with numerical simulations, the
  numerical growth rates being in accordance with the theoretical
  ones.  The MRI is suppressed in convectively stable regions, the
  growth rates are reduced, and the geometry of the flow changes
  favoring radially less extended patterns.  In the mixed regime,
  convectively unstable regions with comparably large entropy
  gradients are dominated by flows similar to volume-filling
  hydrodynamic convection.  The magnetic field is expelled from
  convective cells and accumulates near the box boundaries.  We note
  that the entropy gradients required for these effects are fairly
  shallow $\sim 0.1 \mathrm{km}^{-1}$.  We confirm the existence of
  the MBI regime for axisymmetric models, whereas the same models,
  computed in 3D, experience the growth of non-axisymmetric modes.
\item In 3D models having a zero net magnetic flux we observe the
  development of large-scale coherent field patterns similar those
  seen by \cite{Lesur_Ogilvie__2008__ApJ__zero_flux_MRI_dynamo},
  despite the turbulent nature of the velocity and magnetic fields. We
  also find differences in the flow patterns between MSI and mixed
  regime models.  The tentative connection to a non-linear dynamo
  operating in the models remains to investigated further.
\end{enumerate}

These results allow us to draw a few conclusions.  Firstly, the MRI
has the potential to play an important role for the dynamics of
supernova explosions, at least for relatively fast rotating
progenitors.  The details of the evolution of the MRI depend crucially
on the properties of the core, in particular its thermal
stratification.  This makes the study of the MRI in supernovae a
subject of its own, related to the MRI in accretion discs but also
quite different from it. Hence, there is a need for more
investigations focusing on MRI properties specific to core collapse
supernovae.

While local (or semi-global) simulations can yield interesting results
regarding the physics of the MRI, several important aspects can only
be addressed by global modeling.  The detailed dependence of the
geometry of the magnetic field at saturation, e.g., may depend
strongly on the global dynamics and on the position of the box inside
the core.  Our simulations did not account for any of these factors:
the background was in hydrostatic equilibrium, and we simulated models
only in the equatorial region.  Since the field geometry is of crucial
importance for the global dynamics, e.g., for the generation of
jet-like outflows in collapsars, conclusions on the dynamic influence
of the MRI based on local simulations cannot be drawn easily.  They
require global models.

Additional investigations are also required to study the interplay of
large-scale flows, e.g., the accretion of matter onto the
proto-neutron star and the influence of neutrino radiation.  Apart
from modifying the regimes of the instability discussed here, the
former process may allow for new instability regimes; in particular
the combination of the MRI and the standing accretion shock
instability may have interesting consequences for the dynamics of the
explosion.  While neutrino heating and cooling may, due to a net
transport of entropy, lead to a stratification qualitatively similar
to that of some models in this study with non-vanishing entropy
gradients, it is too early to speculate on how neutrino radiation
affects the MRI.

The inclusion of the MRI and its effects in global simulations
requires a considerably careful treatment.  The currently used
approach of artificially enhancing the initial field strength by a
constant factor is questionable.  On the other hand, finding a better
prescription relies on unraveling the dependence of saturated MRI
driven turbulence on the different parameters of the system, as e.g.,
the rotation law, the thermodynamic conditions, and probably also the
neutrino transport.

With our current simulations we are unfortunately not yet able to go
beyond the stage of a qualitative proof of principle and to address
important open questions of the MRI in core collapse supernovae.  This
would require additional 3D high-resolution non-ideal MHD simulations
covering a large parameter space of possible rotation profiles,
thermal stratifications, and magnetic field geometries. We are
planning to address these issues in future work.

\begin{acknowledgements}
  This research has been supported by the Spanish {\it Ministerio de
    Educaci\'on y Ciencia} (grants AYA2007-67626-C03-01,
  CSD2007-00050), and by the Collaborative Research Center on {\it
    Gravitational Wave Astronomy} of the Deutsche
  Forschungsgesellschaft (DFG SFB/Transregio 7).  MAA is a Ram\'on y
  Cajal fellow of the {\em Ministerio de Educaci\'on y Ciencia}.  Most
  of the simulations were performed at the Rechenzentrum Garching
  (RZG) of the Max-Planck-Society. We are also thankful for the
  computer resources, the technical expertise, and the assistance
  provided by the Barcelona Supercomputing Center - Centro Nacional de
  Supercomputaci\'on. Finally, we would like to acknowledge
  enlightening discussions with Tom Abel and Fen Zhao about the MRI.
\end{acknowledgements}


\appendix

\section{Growth rates of resistive instabilities}
\label{Sec:App-Resi}
Our simulations of the MRI indicate that the termination of the growth
of the instability is determined, at least partially, by resistive
instabilities of the tearing-mode type.  Although there exist detailed
investigations of this kind of resistive instabilities \citep[see,
  e.g.,][]{Biskamp__2000__Buch__Reconnection}, the application of the
results to our study is hampered by the completely different type of
dissipative effects we are facing here: all previous results hold for
instabilities due to \emph{physical} resistivity, whereas our
\emph{ideal MHD} simulations are affected by \emph{numerical}
resistivity, only.  Hence, we had to determine the growth rates of
resistive instabilities from numerical experiments without referring
to analytic results -- although, as we will see, there exist certain
similarities.

We simulated the evolution of two-dimensional current-sheet models on
a Cartesian grid $[x_0,x_1] \times [y_0, y_1]$ with $m_x \times m_y$
zones imposing periodic boundary conditions.  The fluid is described
by an ideal-gas equation of state with an adiabatic index $\Gamma =
4/3$.  We used initial data mimicking MRI-generated channel flows: the
initial magnetic field varies sinusoidally in a gas of constant
density $\rho_0$ and pressure $P_0$,
\begin{eqnarray}
    \label{Gl:App-Resi-init-bx}
    b^x
    & 
    = 
    & 
    b^x_0 
    \sin 
    \frac{2 \pi (y - y_0)}{a}
    ,
    \\
    \label{Gl:App-Resi-init-bz}
    b^z
    & 
    = 
    & 
    - b^x_0
    ,
  \end{eqnarray}
and having a velocity in $x$-direction given by
  \begin{equation}
    \label{Gl:App-Resi-init-v}
    v^x
    = 
    v^x_0 
    \sin 
    \frac{2 \pi (y - y_0) - \pi/2}{a}
    .
  \end{equation}
Here, $a$ denotes the initial width of the flux sheet, and $v^x_0$ is
equal to one half of the Alfv\'en velocity $c_\mathrm{A}^x$
corresponding to $b^x_0$. The presence of the (shear-free!) initial
$x$-velocity is not essential, as it changes the growth rates of the
instability only little. However, as we observe this kind of a
velocity in channel flows, we have included it in our simulations.

We perturbed the sheet by a small random $y$-velocity ($10^{-2}\times
c_\mathrm{A}^x$).  The parameters of the model are chosen to mimic the
situations encountered in MRI simulations (see
\tabref{Tab:App-Resi-Init}).  To isolate the dependence of the growth
rates on the physical and numerical parameters, we varied the initial
magnetic field strength, $b^x_0$, the initial density, $\rho_0$, the
width of the current sheet, $a$, and the grid resolution, $\delta x =
(x_1 - x_0 ) / m_x$.  We chose the grid resolution such that the
current sheet is covered by 12 to 100 zones.  The initial pressure is
$P_0 = \kappa \rho_0 ^ \Gamma$.

\begin{table} 
  \centering
  \caption{Parameters of the 2D current sheets simulated to determine
    the growth rates of resistive instabilities. The columns give from
    left to right the edge length of the square simulation box, the
    number of grid zones per dimension, the initial density (in units
    of $10^{13}\,$g/cm$^{-3}$), the sound speed (in units of
    $10^{9}\,$cm/s), the magnetic field (in units of $10^{14}\,$G),
    the wavelength of the initial field, and the growth rate,
    respectively.
  }
  \label{Tab:App-Resi-Init}
  \begin{tabular}{cc|ccc|c|c}
    \hline\hline
    $L$ 
    & $m_x$ 
    & $\rho_{0;13}$
    & $c_\mathrm{S;9}$
    & $b^x_{0;14}$
    & $a$
    & $\sigma$
    \\
    $\left[\mathrm{km}\right]$
    & 
    & $\left[ \mathrm{g~cm^{-3}}\right]$
    & $\left[ \mathrm{cm~s^{-1}} \right]$
    & $\left[ \mathrm{G} \right]$
    & $\left[\mathrm{m}\right]$
    & $\left[ \mathrm{ms^{-1}} \right]$
    \\

    \hline
    0.25
    & 50
    & 2.5
    & 3.1
    & 8
    & 125
    & 0.45
    \\
    0.25
    & 50
    & 2.5
    & 3.1
    & 16
    & 125
    & 1.6
    \\
    0.25
    & 50
    & 2.5
    & 3.1
    & 32
    & 125
    & 6.0
    \\
    0.25
    & 50
    & 2.5
    & 3.1
    & 64
    & 125
    & 32 
    \\

    \hline
    0.25
    & 50
    & 2.5
    & 3.1
    & 16
    & 62.5
    & 10 
    \\
    0.25
    & 50
    & 2.5
    & 3.1
    & 32
    & 62.5
    & 23 
    \\

    \hline
    0.25
    & 100
    & 2.5
    & 3.1
    & 16
    & 125
    & 0.60
    \\
    0.25
    & 100
    & 2.5
    & 3.1
    & 32
    & 125
    & 3.1 
    \\
    0.25
    & 100
    & 2.5
    & 3.1
    & 64
    & 125
    & 12 
    \\
    0.25
    & 100
    & 2.5
    & 3.1
    & 128
    & 125
    & 43
    \\

    \hline
    0.25
    & 100
    & 2.5
    & 3.1
    & 16
    & 62.5
    & 3.7
    \\
    0.25
    & 100
    & 2.5
    & 3.1
    & 32
    & 62.5
    & 14
    \\
    0.25
    & 100
    & 2.5
    & 3.1
    & 64
    & 62.5
    & 50
    \\
    0.25
    & 100
    & 2.5
    & 3.1
    & 128
    & 62.5
    & 110
    \\

    \hline
    0.25
    & 100
    & 0.025
    & 1.4
    & 3.2
    & 62.5
    & 25
    \\
    0.25
    & 100
    & 0.025
    & 3.1
    & 3.2
    & 62.5
    & 15
    \\
    0.25
    & 100
    & 0.025
    & 6.1
    & 3.2
    & 62.5
    & 8.0
    \\
    0.25
    & 100
    & 2.5
    & 1.4
    & 32
    & 62.5
    & 25
    \\
    0.25
    & 100
    & 2.5
    & 3.1
    & 32
    & 62.5
    & 14
    \\
    0.25
    & 100
    & 2.5
    & 1.4
    & 6.1
    & 62.5
    & 7.8
    \\
    0.25
    & 100
    & 250
    & 6.6
    & 320
    & 62.5
    & 7.4
    \\
    0.25
    & 100
    & 250
    & 3.1
    & 320
    & 62.5
    & 15
    \\
    0.25
    & 100
    & 2.5
    & 12.4
    & 64
    & 62.5
    & 16
    \\

    \hline
    0.25
    & 100
    & 2.5
    & 3.1
    & 4
    & 31.25
    & 1.2
    \\
    0.25
    & 100
    & 2.5
    & 3.1
    & 8
    & 31.25
    & 5.9
    \\
    0.25
    & 100
    & 2.5
    & 3.1
    & 16
    & 31.25
    & 20
    \\
    0.25
    & 100
    & 2.5
    & 3.1
    & 32
    & 31.25
    & 53
    \\
    0.25
    & 100
    & 2.5
    & 3.1
    & 64
    & 31.25
    & 110
    \\

    \hline
    0.25
    & 200
    & 2.5
    & 3.1
    & 32
    & 125
    & 0.9
    \\
    0.25
    & 200
    & 2.5
    & 3.1
    & 64
    & 125
    & 4.6
    \\
    0.25
    & 200
    & 2.5
    & 3.1
    & 128
    & 125
    & 19
    \\
    0.25
    & 200
    & 2.5
    & 3.1
    & 32
    & 62.5
    & 5.4
    \\
    0.25
    & 200
    & 2.5
    & 3.1
    & 64
    & 62.5
    & 24
    \\
    0.25
    & 200
    & 2.5
    & 3.1
    & 128
    & 62.5
    & 76
    \\

    \hline
    0.25
    & 200
    & 2.5
    & 3.1
    & 16
    & 31.25
    & 7.9
    \\
    0.25
    & 200
    & 2.5
    & 3.1
    & 32
    & 31.25
    & 28
    \\
    0.25
    & 200
    & 2.5
    & 3.1
    & 64
    & 31.25
    & 95
    \\
    0.25
    & 200
    & 2.5
    & 3.1
    & 128
    & 31.25
    & 250
    \\

    \hline
    0.25
    & 200
    & 2.5
    & 3.1
    & 4
    & 15.625
    & 2.5
    \\
    0.25
    & 200
    & 2.5
    & 3.1
    & 16
    & 15.625
    & 40
    \\
    0.25
    & 200
    & 2.5
    & 3.1
    & 64
    & 15.625
    & 260
    \\

    \hline
    0.5
    & 100
    & 2.5
    & 3.1
    & 32
    & 250
    & 1.2
    \\
    0.5
    & 100
    & 2.5
    & 3.1
    & 64
    & 250
    & 5.6
    \\
    0.5
    & 100
    & 2.5
    & 3.1
    & 128
    & 250
    & 20
    \\

    \hline
    0.5
    & 100
    & 2.5
    & 3.1
    & 16
    & 125
    & 1.9
    \\
    0.5
    & 100
    & 2.5
    & 3.1
    & 32
    & 125
    & 7.5
    \\
    0.5
    & 100
    & 2.5
    & 3.1
    & 64
    & 125
    & 26
    \\

    \hline
    0.5
    & 100
    & 2.5
    & 3.1
    & 8
    & 62.5
    & 3.0
    \\
    0.5
    & 100
    & 2.5
    & 3.1
    & 16
    & 62.5
    & 10
    \\
    0.5
    & 100
    & 2.5
    & 3.1
    & 32
    & 62.5
    & 24
    \\
    0.5
    & 100
    & 2.5
    & 3.1
    & 64
    & 62.5
    & 61
    \\
    \hline\hline
  \end{tabular}
\end{table}

\begin{figure}
  \centering
  \includegraphics[width=7cm]{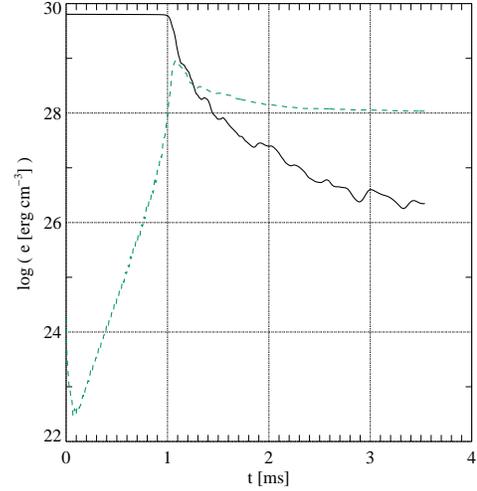}
  \caption{Evolution of the magnetic energy density of a current-sheet
    model simulated on a square computational grid (edge length
    0.25\,km; $200\times 200$ zones). The model has an initial density
    $\rho_0 = 2.5\times 10^{13} \mathrm{g\,cm^{-3}}$, and an initial
    field strength $b^x_0 = 1.6\times 10^{15} \mathrm{G}$. The
    wavelength of the initially sinusoidally varying magnetic field is
    $a = 31.25 \mathrm{m}$.  The black solid line and the green dashed
    line show the magnetic energy density corresponding to the $x$ and
    the $y$ component of the magnetic field, respectively.
  }
  \label{Fig:App-Resi-tevo}
\end{figure}

We show one typical result for the evolution of our models in
\figref{Fig:App-Resi-tevo}.  After a short initial phase, the
transverse magnetic energy density (green line) grows roughly
exponentially as tearing modes develop.  Initially the growth rate is
approximately constant, but it increases by a factor of $\sim 4$
towards saturation.  Simultaneously, the $x$-component of the magnetic
field decreases strongly until it is of similar strength as the
$y$-component.  At this point, the coherent current sheets are
completely disrupted by the resistive instability.

We determined estimates of the growth rates of the resistive
instability using the time derivative of the transverse magnetic energy
density, i.e., the time derivative of the magnetic energy density
corresponding to the $y$-component of the magnetic field.  To find a
scaling relation of the form
  \begin{equation}
    \label{Gl:App-Resi-sigma}
    \sigma 
    \propto
    \left( c_\mathrm{A}^x \right)^{\gamma_\mathrm{A}}
    \left( c_\mathrm{S} \right)^{\gamma_\mathrm{S}}
    \left( \delta x \right)^{\gamma_\delta}
    \left( a \right)^{\gamma_a}
  \end{equation}
we define the following functions of the growth rate $\sigma$,
  \begin{eqnarray}
    \label{Gl:App-Resi-fca}
    f_{cA} 
    & =
    & 
    \sigma 
    \left( c_\mathrm{A}^x \right)^{-\gamma_\mathrm{A}}
    \left( c_\mathrm{S} \right)^{-\gamma_\mathrm{S}}
    \left( \delta x \right)^{-\gamma_\delta}
    ,
    \\
    \label{Gl:App-Resi-faa}
    f_{a} 
    & =
    & 
    \sigma 
    \left( c_\mathrm{S} \right)^{-\gamma_\mathrm{S}}
    \left( a \right)^{-\gamma_a}
    \left( \delta x \right)^{-\gamma_\delta}
    ,
  \end{eqnarray}
and adjust the exponents $\gamma_\mathrm{A}$, $\gamma_\mathrm{S}$,
$\gamma_a$, and $\gamma_\delta$ to determine the scaling of $\sigma$
with the respective parameters.  Our preferred set of scaling
exponents is
  \begin{eqnarray}
    \label{Gl:App-Resi-ga-A}
    \gamma_\mathrm{A}  & = & 1.75,
    \\
    \label{Gl:App-Resi-ga-S}
    \gamma_\mathrm{S}  & = & - 0.75,
    \\
    \label{Gl:App-Resi-ga-a}
    \gamma_{a}  & = & - 2,
    \\
    \label{Gl:App-Resi-ga-d}
    \gamma_{\delta}  & = & 1,
  \end{eqnarray}
which implies the following scaling law:
  \begin{equation}
    \label{Gl:App-Resi-sigma-result}
    \sigma 
    \propto
    \left( 
      \frac{c_\mathrm{A}}{c_\mathrm{S}}
    \right)^{0.75}
    \left( 
      \frac{c_\mathrm{A} \delta x}{a^2}
    \right)
    .
  \end{equation}

\begin{figure*}
  \centering
  \includegraphics[width=18cm]{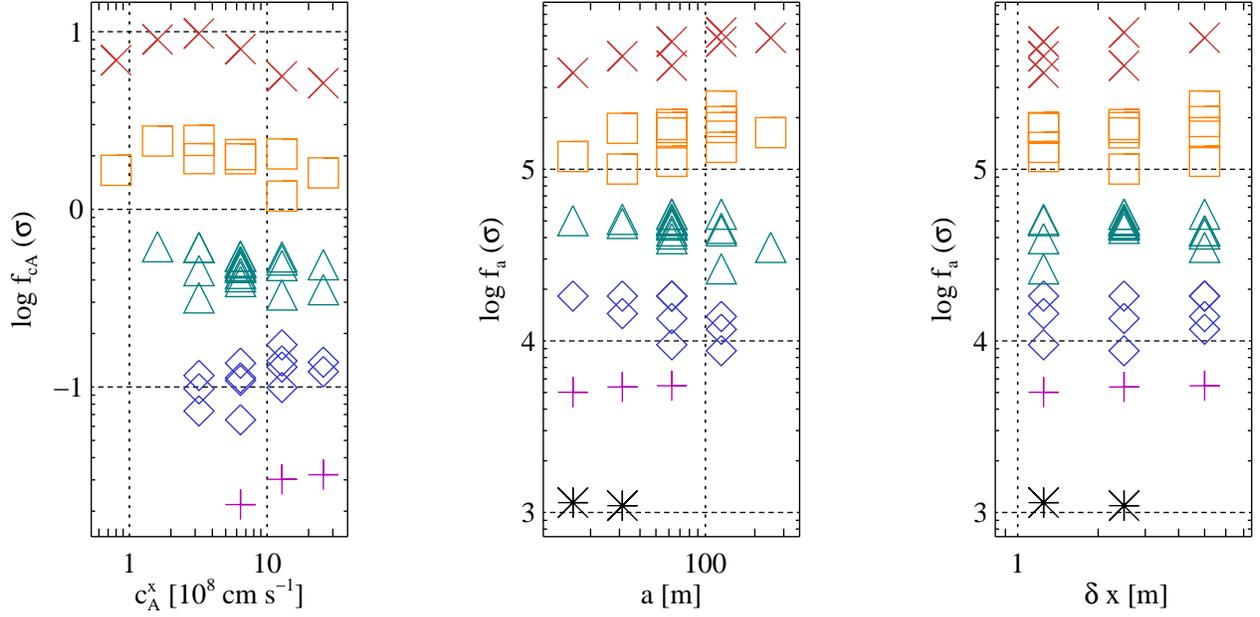}
  \caption{Dependence of the growth rate of resistive instabilities,
    $\sigma$, on various parameters of 2D current sheet models. The
    panels show $f_{cA}$ as a function of the initial Alfv\'en
    velocity ($c_\mathrm{A}^x$; left panel), and $f_a$ as a function
    of the width of the current sheet ($a$; middle panel) and of the
    grid resolution ($\delta x$; right panel).  The left panel shows
    groups of models with different $a$: 250\,m (pink plus sign),
    125\,m (blue diamond), 62.5\,m (green triangle), 32.25\,m (orange
    square), and 15.625\,m (red cross). The other two panels show
    groups of models with different Alfv\'en velocity (in units of
    $10^8 \mathrm{cm~s^{-1}}$): 0.8 (black asterisk), 1.6 (pink plus
    sign), 3.2 (blue diamond), 6.4 (green triangle), 12.8 (orange
    square), and 25.6 (red cross) respectively.
  }
  \label{Fig:App--Resi-sigma}
\end{figure*}

We demonstrate the quality of the fit parameters in
\figref{Fig:App--Resi-sigma} showing $f_{cA}$ as a function of the
initial Alfv\'en velocity (left panel), and $f_a$ as a function of the
width of the current sheet (middle panel) and of the grid resolution
(right panel).  None of the groups of models representing the
variation of one parameter (distinguished by a different color and
symbol in the figure) exhibits a strong trend with $c_A$, $a$ or
$\delta x$, i.e., our scaling exponents provide an adequate fit to the
data.

Due to a relatively large scatter in the growth rates, the scaling
relation, \eqref{Gl:App-Resi-sigma}, should not be taken too
literally.  We note, however, that our computed rates are compatible
with those of the resistive instabilities (for fields of similar
strength) in MRI models, i.e., approximately one millisecond for
$b_0^x \sim 10^{15} \mathrm{G}$.  An additional dependence of $\sigma$
on the domain size, $L$, cannot be excluded, but we did not examine
this possibility any further.  For small $a$ and coarse grids our
scaling formula tends to overestimate the growth rate for large
initial Alfv\'en velocities, and for sound speeds much larger than the
Alfv\'en velocity the rates depend more strongly on the sound speed
than predicted by our formula.  Because both situations do not apply
to our MRI models, we did not pursue these issues any further. Our
scaling law loses its validity, if the Alfv\'en velocity exceeds the
sound speed.  Thus, we excluded two respective models in the
derivation of our scaling relation.

Bearing in mind the uncertainties regarding the physical meaning of a
purely numerical resistivity and the precise values of the scaling we
may try to interpret our result summarized in
\eqref{Gl:App-Resi-sigma-result}.  As the product of the Alfv\'en
velocity and the grid spacing, $c_\mathrm{A} \delta x$, defines an
effective resistivity, we may conclude that the growth time of the
instability is set by the time scale for resistive diffusion across
the width of a current sheet, $\tau_\mathrm{r} = a^2 / ( c_\mathrm{A}
\delta x )$, modified by the ratio of the sound speed and the Alfv\'en
velocity. This interpretation has the nice property that it is
consistent with the fact that the magnetic Reynolds number is
proportional to the grid resolution.

\section{List of models}
\label{Sek:App-Tables}

In this appendix we provide a list of the models:
\begin{description}

\item[\tabref{Tab:U2-entr-MSI}]: This table contains a list of 2D
  models having a positive entropy gradient.  Their initial rotation
  profile is given by $\Omega_0 = 1900 \, \mathrm{s}^{-1}$ and
  $\alpha_\Omega = -1.25$, and their initial magnetic field is
  uniform.

  \item[\tabref{Tab:U2-entr-convection}]: This table contains a list
    of 2D models having a negative entropy gradient. The models rotate
    initially rigidly or differentially. The initial magnetic field is
    uniform.  

  \item[\tabref{Tab:3d-models-1} and \tabref{Tab:3d-models-2}]: This
    table contains a list of 3D models having different initial
    magnetic field strengths, entropy gradients, initial rotation
    laws, and simulated in computational boxes of various size with
    both types (with and without velocity damping) of radial boundary
    conditions.

\end{description}

\begin{table*}
  \centering
  \caption{List of 2D models having a positive entropy gradient and an
    initial rotation profile given by $\Omega_0 = 1900 \,
    \mathrm{s}^{-1}$ and $\alpha_\Omega = -1.25$. The columns give
    (from left to right) the number of grid zones $m_{\varpi} \times
    m_z$ (the box has an edge length $L_{\varpi} = L_z = 1
    \mathrm{km}$), the type of boundary condition which was applied
    (d: velocity damping; p: periodic), the adiabatic index of the
    gas, $\Gamma$ (models computed with an ideal-gas equation of state
    instead of the hybrid one are marked ``id''), the initial entropy,
    $S_0$, and the radial entropy gradient, $\partial_{\varpi} S$. The
    next three columns give the parameter $\mathcal{C}$ determining
    the instability regime (see \eqref{Gl:MRIds--C-param}), the
    theoretical growth rate, $\sigma_{\mathrm{th}}$, and the strength
    of the initially uniform magnetic field, $b_0^z$. The last two
    columns list the numerical growth rate, $\sigma$, and the value of
    the Maxwell stress component $M^{\mathrm{term}}_{\varpi\phi}$ at
    MRI termination.
  }
  \label{Tab:U2-entr-MSI}

  \begin{tabular}{c|c|cccccc|cc}
  \\
    \hline\hline
    $m_{\varpi} \times m_{z}$
    & BC
    & $\Gamma$
    & $S_0$ & $\partial_{\varpi} S$
    & $\mathcal{C}$
    & $\sigma_{\mathrm{th}}$
    & $b_0^z$
    & $\sigma$ 
    & $M_{\varpi\phi}^{\mathrm{term}}$
    \\
    & & 
    & & $[\mathrm{km}^{-1}]$
    & 
    & $\left[\mathrm{ms}^{-1}\right]$
    & $\left[10^{12}\,\mathrm{G}\right]$
    & $\left[\mathrm{ms}^{-1}\right]$
    & $\left[10^{28} \, \frac{\mathrm{G^2}}{\mathrm{cm^{-3}}} \right]$
    \\

    \hline\hline
    $ 100 \times 100 $
    & d
    & 1.31
    & 0.20 & 0.020
    & -1.7 & 0.73
    & 10
    & 0.72
    & 1.1
    \\
    $ 100 \times 100 $
    & d
    & 1.31
    & 0.20 & 0.020
    & -1.7 & 0.73
    & 20
    & 0.71
    & 1.3
    \\
    $ 100 \times 100 $
    & d
    & 1.31
    & 0.20 & 0.020
    & -1.7 & 0.73
    & 40
    & 0.70
    & 1.6
    \\
    $ 100 \times 100 $
    & d
    & 1.31
    & 0.20 & 0.020
    & -1.7 & 0.73
    & 80
    & 0.60
    & 1.8
    \\

    \hline
    $ 200 \times 200 $
    & d
    & 1.31
    & 0.20 & 0.020
    & -1.7 & 0.73
    & 10
    & 0.71
    & 1.3
    \\

    \hline
    $ 400 \times 400 $
    & d
    & 1.31
    & 0.20 & 0.020
    & -1.7 & 0.73
    & 10
    & 0.67
    & 1.2
    \\
    $ 400 \times 400 $
    & d
    & 1.31
    & 0.20 & 0.020
    & -1.7 & 0.73
    & 20
    & 0.69
    & 1.5
    \\

    \hline\hline
    $ 50 \times 50 $
    & d
    & $5/3$
    & 0.20 & 0.040
    & -1.1 & 0.49
    & 10
    & 0.46
    & 0.97
    \\
    $ 50 \times 50 $
    & d
    & $5/3$
    & 0.20 & 0.040
    & -1.1 & 0.49
    & 20
    & 0.61
    & 1.7
    \\
    $ 50 \times 50 $
    & d
    & $5/3$
    & 0.20 & 0.040
    & -1.1 & 0.49
    & 40
    & 0.48
    & 0.87
    \\
    $ 50 \times 50 $
    & d
    & $5/3$
    & 0.20 & 0.040
    & -1.1 & 0.49
    & 80
    & 0.39
    & 0.93
    \\

    \hline\hline
    $ 50 \times 50 $
    & d
    & 1.31, id
    & 0.20 & 0.040
    & -0.69 & 0.32
    & 20
    & 0.50
    & 0.53
    \\
    $ 50 \times 50 $
    & d
    & 1.31, id
    & 0.20 & 0.040
    & -0.69 & 0.32
    & 40
    & 0.30
    & 0.27
    \\
    $ 50 \times 50 $
    & d
    & 1.31, id
    & 0.20 & 0.040
    & -0.69 & 0.32
    & 80
    & 0.28
    & 0.22
    \\

    \hline\hline
    $ 50 \times 50 $
    & d
    & 1.31
    & 0.20 & 0.040
    & -0.69 & 0.32
    & 4
    & 0.25
    & 0.21
    \\
    $ 50 \times 50 $
    & d
    & 1.31
    & 0.20 & 0.040
    & -0.69 & 0.32
    & 20 
    & 0.49
    & 0.50
    \\
    $ 50 \times 50 $
    & d
    & 1.31
    & 0.20 & 0.040
    & -0.69 & 0.32
    & 40 
    & 0.28
    & 0.24
    \\
    $ 50 \times 50 $
    & d
    & 1.31
    & 0.20 & 0.040
    & -0.69 & 0.32
    & 80 
    & 0.28
    & 0.22
    \\

    \hline
    $ 100 \times 100 $
    & d
    & 1.31
    & 0.20 & 0.040
    & -0.69 & 0.32
    & 10 
    & 0.49
    & 0.55
    \\
    $ 100 \times 100 $
    & d
    & 1.31
    & 0.20 & 0.040
    & -0.69 & 0.32
    & 20 
    & 0.30
    & 0.29
    \\
    $ 100 \times 100 $
    & d
    & 1.31
    & 0.20 & 0.040
    & -0.69 & 0.32
    & 40 
    & 0.30
    & 0.29
    \\
    $ 100 \times 100 $
    & d
    & 1.31
    & 0.20 & 0.040
    & -0.69 & 0.32
    & 80 
    & 0.28
    & 0.22
    \\

    \hline
    $ 200 \times 200 $
    & d
    & 1.31
    & 0.20 & 0.040
    & -0.69 & 0.32
    & 4
    & 0.53
    & 0.40
    \\
    $ 200 \times 200 $
    & d
    & 1.31
    & 0.20 & 0.040
    & -0.69 & 0.32
    & 10 
    & 0.30
    & 0.20
    \\
    $ 200 \times 200 $
    & d
    & 1.31
    & 0.20 & 0.040
    & -0.69 & 0.32
    & 20 
    & 0.30
    & 0.22
    \\
    $ 200 \times 200 $
    & d
    & 1.31
    & 0.20 & 0.040
    & -0.69 & 0.32
    & 40 
    & 0.30
    & 0.30
    \\
    $ 200 \times 200 $
    & d
    & 1.31
    & 0.20 & 0.040
    & -0.69 & 0.32
    & 80 
    & 0.26
    & 0.18
    \\

    \hline\hline
    $ 100 \times 100 $
    & d
    & 1.31
    & 0.20 & 0.080
    & 1.1 & 0
    & 20
    & $<0.01$
    & 
    \\

    \hline
    $ 200 \times 200 $
    & d
    & 1.31
    & 0.20 & 0.080
    & 1.1 & 0
    & 10
    & $<0.05$
    & 
    \\

    \hline\hline
    $ 100 \times 100 $
    & p
    & $5/3$
    & 0.20 & 0.020
    & -1.8 & 0.82
    & 20
    & 0.79
    & 4.9
    \\

    \hline\hline
    $ 100 \times 100 $
    & p
    & 1.31
    & 0.20 & 0.020
    & -1.7 & 0.73
    & 10
    & 0.73
    & 1.9
    \\
    $ 100 \times 100 $
    & p
    & 1.31
    & 0.20 & 0.020
    & -1.7 & 0.73
    & 20
    & 0.71
    & 33
    \\
    $ 100 \times 100 $
    & p
    & 1.31
    & 0.20 & 0.020
    & -1.7 & 0.73
    & 40
    & 0.72
    & 202
    \\
    $ 100 \times 100 $
    & p
    & 1.31
    & 0.20 & 0.020
    & -1.7 & 0.73
    & 80
    & 0.62
    & 411
    \\

    \hline\hline
    $ 50 \times 50 $
    & p
    & 1.31
    & 0.20 & 0.040
    & -0.69 & 0.32
    & 20
    & 0.51
    & 1.7
    \\
    $ 50 \times 50 $
    & p
    & 1.31
    & 0.20 & 0.040
    & -0.69 & 0.32
    & 40
    & 0.31
    & 90
    \\
    $ 50 \times 50 $
    & p
    & 1.31
    & 0.20 & 0.040
    & -0.69 & 0.32
    & 80
    & 0.32
    & 159
    \\

    \hline
    $ 100 \times 100 $
    & p
    & 1.31
    & 0.20 & 0.040
    & -0.69 & 0.32
    & 10 
    & 0.53
    & 0.64
    \\
    $ 100 \times 100 $
    & p
    & 1.31
    & 0.20 & 0.040
    & -0.69 & 0.32
    & 20 
    & 0.30
    & 6.8
    \\
    $ 100 \times 100 $
    & p
    & 1.31
    & 0.20 & 0.040
    & -0.69 & 0.32
    & 40 
    & 0.30
    & 98
    \\
    $ 100 \times 100 $
    & p
    & 1.31
    & 0.20 & 0.040
    & -0.69 & 0.32
    & 80 
    & 0.31
    & 201
    \\

    \hline\hline
    $ 50 \times 50 $
    & p
    & $5/3$
    & 0.20 & 0.040
    & -1.1 & 0.49
    & 10
    & 0.45
    & 0.55
    \\
    $ 50 \times 50 $
    & p
    & $5/3$
    & 0.20 & 0.040
    & -1.1 & 0.49
    & 20
    & 0.59
    & 3.1
    \\
    $ 50 \times 50 $
    & p
    & $5/3$
    & 0.20 & 0.040
    & -1.1 & 0.49
    & 40
    & 0.47
    & 201
    \\
    $ 50 \times 50 $
    & p
    & $5/3$
    & 0.20 & 0.040
    & -1.1 & 0.49
    & 80
    & 0.38
    & 139
    \\

    \hline
    $ 100 \times 100 $
    & p
    & $5/3$
    & 0.20 & 0.040
    & -1.1 & 0.49
    & 4
    & 0.40
    & 0.43
    \\
    $ 100 \times 100 $
    & p
    & $5/3$
    & 0.20 & 0.040
    & -1.1 & 0.49
    & 10
    & 0.63
    & 0.83
    \\
    $ 100 \times 100 $
    & p
    & $5/3$
    & 0.20 & 0.040
    & -1.1 & 0.49
    & 20
    & 0.49
    & 5.0
    \\
    $ 100 \times 100 $
    & p
    & $5/3$
    & 0.20 & 0.040
    & -1.1 & 0.49
    & 40
    & 0.47
    & 262
    \\
    $ 100 \times 100 $
    & p
    & $5/3$
    & 0.20 & 0.040
    & -1.1 & 0.49
    & 80
    & 0.45
    & 978
    \\

    \hline\hline

  \end{tabular}
%
\end{table*}

\begin{table*}[htbp]
  \centering

  \caption{List of 2D models having a negative initial entropy
    gradient. The columns give (from left to right) the number of grid
    zones $m_{\varpi} \times m_z$ (the box has an edge length
    $L_{\varpi} = L_z = 1 \mathrm{km}$), the type of boundary
    condition which was applied (d: velocity damping; p: periodic),
    the rotation law (``d'': differential rotation with $\Omega_0 =
    1900 \, \mathrm{s}^{-1}$ and $\alpha_\Omega=-1.25$;
    ``$\mathrm{r}^{\Omega}$'' rigid rotation with an angular velocity
    of $\Omega$), the adiabatic index of the gas, $\Gamma$, the
    initial entropy, $S_0$, the radial entropy gradient,
    $\partial_{\varpi} S$, and the strength of the initially uniform
    magnetic field, $b_0^z$.  The next columns give the three
    quantities $\mathcal{R}_\varpi / \Omega_0^2$, $N^2/\Omega_0^2$ and
    $\mathcal{C}$ determining the instability regime (see
    \eqref{Gl:MRIds--C-param}), followed by the theoretical growth
    rate, $\sigma_{\mathrm{th}}$, and the type of the instability. The
    last two columns list the numerical growth rate, $\sigma$, and the
    value of the Maxwell stress component
    $M^{\mathrm{term}}_{\varpi\phi}$ at MRI termination.
  }
  \label{Tab:U2-entr-convection}
  \begin{tabular}{c|c|ccccc|ccccc|cc}

    \hline\hline
    $m_{\varpi} \times m_{z}$
    & BC
    & Rot
    & $\Gamma$
    & $S_0$ & $\partial_{\varpi} S$
    & $b_0^z$ ~ ~~
    & $\mathcal{R}_{\varpi} / \Omega_0^2$
    & $N^2 / \Omega_0^2$
    & $\mathcal{C}$
    & $\sigma_{\mathrm{th}}$
    & regime
    & $\sigma$ 
    & $M_{\varpi\phi}^{\mathrm{term}}$
    \\
    & & &
    & & $[\mathrm{km}^{-1}]$
    & $\left[10^{12}\,\mathrm{G}\right]$
    & & & 
    & $\left[\mathrm{ms}^{-1}\right]$
    &
    & $\left[\mathrm{ms}^{-1}\right]$
    & $\left[10^{28} \, \frac{\mathrm{G^2}}{\mathrm{cm^{-3}}} \right]$
    \\

    \hline\hline

    $100 \times 100$
    & d
    & d
    & 1.31
    & 0.20
    & -0.019
    & 0
    & -2.5
    & -0.90
    & -3.4
    & 1.6
    & mix
    & 0
    & 0
    \\
    $100 \times 100$
    & d
    & d
    & 1.31
    & 0.20
    & -0.019
    & 10
    & -2.5
    & -0.90
    & -3.4
    & 1.6
    & mix
    & 1.2
    & 2.6
    \\
    $100 \times 100$
    & d
    & d
    & 1.31
    & 0.20
    & -0.019
    & 20
    & -2.5
    & -0.90
    & -3.4
    & 1.6
    & mix
    & 1.3
    & 3.4
    \\
    $100 \times 100$
    & d
    & d
    & 1.31
    & 0.20
    & -0.019
    & 40
    & -2.5
    & -0.90
    & -3.4
    & 1.6
    & mix
    & 1.4
    & 5.7
    \\
    $100 \times 100$
    & d
    & d
    & 1.31
    & 0.20
    & -0.019
    & 80
    & -2.5
    & -0.90
    & -3.4
    & 1.6
    & mix
    & 1.3
    & 6.0
    \\

    \hline
    $50 \times 50$
    & d
    & d
    & 1.31
    & 0.20
    & -0.038
    & 20
    & -2.5
    & -1.8
    & -4.3
    & 2.0
    & mix
    & 1.6
    & 4.2
    \\

    \hline
    $100 \times 100$
    & d
    & d
    & 1.31
    & 0.20
    & -0.038
    & 0
    & -2.5
    & -1.8
    & -4.3
    & 2.0
    & mix
    & 0
    & 0
    \\
    $100 \times 100$
    & d
    & d
    & 1.31
    & 0.20
    & -0.038
    & 10
    & -2.5
    & -1.8
    & -4.3
    & 2.0
    & mix
    & 1.5
    & 3.3
    \\
    $100 \times 100$
    & d
    & d
    & 1.31
    & 0.20
    & -0.038
    & 20
    & -2.5
    & -1.8
    & -4.3
    & 2.0
    & mix
    & 1.7
    & 5.3
    \\
    $100 \times 100$
    & d
    & d
    & 1.31
    & 0.20
    & -0.038
    & 40
    & -2.5
    & -1.8
    & -4.3
    & 2.0
    & mix
    & 1.7
    & 7.8
    \\
    $100 \times 100$
    & d
    & d
    & 1.31
    & 0.20
    & -0.038
    & 80
    & -2.5
    & -1.8
    & -4.3
    & 2.0
    & mix
    & 1.7
    & 9.5
    \\

    \hline
    $100 \times 100$
    & d
    & d
    & 1.31
    & 0.20
    & -0.075
    & 0
    & -2.5
    & -3.6
    & -6.1
    & 2.8
    & mix
    & 0
    & 0
    \\
    $100 \times 100$
    & d
    & d
    & 1.31
    & 0.20
    & -0.075
    & 10
    & -2.5
    & -3.6
    & -6.1
    & 2.8
    & mix
    & 2.4
    & 1.3
    \\
    $100 \times 100$
    & d
    & d
    & 1.31
    & 0.20
    & -0.075
    & 20
    & -2.5
    & -3.6
    & -6.1
    & 2.8
    & mix
    & 2.6
    & 4.8
    \\
    $100 \times 100$
    & d
    & d
    & 1.31
    & 0.20
    & -0.075
    & 40
    & -2.5
    & -3.6
    & -6.1
    & 2.8
    & mix
    & 2.5
    & 9.7
    \\

    \hline
    $100 \times 100$
    & d
    & d
    & 1.31
    & 0.20
    & -0.15
    & 0
    & -2.5
    & -7.2
    & -9.7
    & 4.4
    & conv
    & 3.0
    & 0
    \\
    $100 \times 100$
    & d
    & d
    & 1.31
    & 0.20
    & -0.15
    & 20
    & -2.5
    & -7.2
    & -9.7
    & 4.4
    & conv
    & 4.0
    & 1.2
    \\
    $100 \times 100$
    & d
    & d
    & 1.31
    & 0.20
    & -0.15
    & 40
    & -2.5
    & -7.2
    & -9.7
    & 4.4
    & conv
    & 3.8
    & 3.7
    \\

    \hline\hline
    $100 \times 100$
    & p
    & d
    & 5/3
    & 0.20
    & -0.019
    & 20
    & -2.5
    & -0.70
    & -3.2
    & 1.5
    & mix
    & 1.3
    & 6.0
    \\
    $100 \times 100$
    & p
    & d
    & 5/3
    & 0.20
    & -0.019
    & 40
    & -2.5
    & -0.70
    & -3.2
    & 1.5
    & mix
    & 1.4
    & 50
    \\

    \hline\hline
    $50 \times 50$
    & p
    & d
    & 1.31
    & 0.20
    & -0.038
    & 20
    & -2.5
    & -1.8
    & -4.3
    & 2.0
    & mix
    & 1.6
    & 12
    \\
    $50 \times 50$
    & p
    & d
    & 1.31
    & 0.20
    & -0.038
    & 40
    & -2.5
    & -1.8
    & -4.3
    & 2.0
    & mix
    & 1.7
    & 19.5
    \\

    \hline
    $100 \times 100$
    & p
    & d
    & 1.31
    & 0.20
    & -0.038
    & 20
    & -2.5
    & -1.8
    & -4.3
    & 2.0
    & mix
    & 1.7
    & 27
    \\
    $100 \times 100$
    & p
    & d
    & 1.31
    & 0.20
    & -0.038
    & 40
    & -2.5
    & -1.8
    & -4.3
    & 2.0
    & mix
    & 1.7
    & 22
    \\

    \hline\hline

    $100 \times 100$
    & d
    & $\mathrm{r}^{1000}$
    & 1.31
    & 0.20
    & -0.075
    & 0
    & 0
    & -14
    & -14
    & 3.4
    & conv
    & 2.7
    & 
    \\
    $100 \times 100$
    & d
    & $\mathrm{r}^{1500}$
    & 1.31
    & 0.20
    & -0.075
    & 0
    & 0
    & -5.7
    & -5.7
    & 2.1
    & conv
    & 1.6
    & 
    \\
    $100 \times 100$
    & d
    & $\mathrm{r}^{1900}$
    & 1.31
    & 0.20
    & -0.075
    & 0
    & 0
    & -3.2
    & -3.2
    & 1.5
    & MBI
    & 0
    & 
    \\

    \hline
    $100 \times 100$
    & d
    & $\mathrm{r}^{1500}$
    & 1.31
    & 0.20
    & -0.075
    & $10^{-8}$
    & 0
    & -5.7
    & -5.7
    & 2.1
    & conv
    & 1.5
    & $1.4 \times 10^{-18}$
    \\
    $100 \times 100$
    & d
    & $\mathrm{r}^{1500}$
    & 1.31
    & 0.20
    & -0.075
    & 1
    & 0
    & -5.7
    & -5.7
    & 2.1
    & conv
    & 1.5
    & 0.013
    \\
    $100 \times 100$
    & d
    & $\mathrm{r}^{1500}$
    & 1.31
    & 0.20
    & -0.075
    & 10
    & 0
    & -5.7
    & -5.7
    & 2.1
    & conv
    & 1.8
    & 1.7
    \\

    \hline
    $100 \times 100$
    & d
    & $\mathrm{r}^{1900}$
    & 1.31
    & 0.40
    & -0.10
    & 4
    & 0
    & -3.6
    & -3.6
    & 1.7
    & MBI
    & 1.1
    & 0.22
    \\
    $100 \times 100$
    & d
    & $\mathrm{r}^{1900}$
    & 1.31
    & 0.40
    & -0.10
    & 8
    & 0
    & -3.6
    & -3.6
    & 1.7
    & MBI
    & 1.2
    & 1.6
    \\
    $100 \times 100$
    & d
    & $\mathrm{r}^{1900}$
    & 1.31
    & 0.40
    & -0.10
    & 10
    & 0
    & -3.6
    & -3.6
    & 1.7
    & MBI
    & 1.1
    & 1.6
    \\
    $100 \times 100$
    & d
    & $\mathrm{r}^{1900}$
    & 1.31
    & 0.40
    & -0.10
    & 20
    & 0
    & -3.6
    & -3.6
    & 1.7
    & MBI
    & 1.4
    & 5.2
    \\
    $100 \times 100$
    & d
    & $\mathrm{r}^{1900}$
    & 1.31
    & 0.40
    & -0.10
    & 40
    & 0
    & -3.6
    & -3.6
    & 1.7
    & MBI
    & 
    & 
    \\

    \hline
    $200 \times 200$
    & d
    & $\mathrm{r}^{1900}$
    & 1.31
    & 0.40
    & -0.10
    & 0
    & 0
    & -3.6
    & -3.6
    & 1.7
    & MBI
    & 0
    & 0
    \\
    $200 \times 200$
    & d
    & $\mathrm{r}^{1900}$
    & 1.31
    & 0.40
    & -0.10
    & 0.01
    & 0
    & -3.6
    & -3.6
    & 1.7
    & MBI
    & $\lesssim 0.006$
    & 
    \\
    $200 \times 200$
    & d
    & $\mathrm{r}^{1900}$
    & 1.31
    & 0.40
    & -0.10
    & 4
    & 0
    & -3.6
    & -3.6
    & 1.7
    & MBI
    & 1.0
    & 0.99
    \\
    $200 \times 200$
    & d
    & $\mathrm{r}^{1900}$
    & 1.31
    & 0.40
    & -0.10
    & 8
    & 0
    & -3.6
    & -3.6
    & 1.7
    & MBI
    & 1.4
    & 1.4
    \\
    $200 \times 200$
    & d
    & $\mathrm{r}^{1900}$
    & 1.31
    & 0.40
    & -0.10
    & 10
    & 0
    & -3.6
    & -3.6
    & 1.7
    & MBI
    & 1.4
    & 2.1
    \\
    $200 \times 200$
    & d
    & $\mathrm{r}^{1900}$
    & 1.31
    & 0.40
    & -0.10
    & 20
    & 0
    & -3.6
    & -3.6
    & 1.7
    & MBI
    & 1.5
    & 5.6
    \\

    \hline\hline

  \end{tabular}

\end{table*}

\begin{table*}[htbp]
  \centering

  \caption{List of 3D models.  The first column (from left to right)
    gives the geometry of the initial field (``U'': uniform; ``V''
    zero-flux).  The next two columns show the box size ($L_{\varpi}
    \times L_{\phi} \times L_z$) and the number of grid zones
    ($m_{\varpi} \times m_{\phi} \times m_z$). The next four columns
    list the rotation law (``d'': differential rotation with $\Omega_0
    = 1900 \, \mathrm{s}^{-1}$ and $\alpha_\Omega=-1.25$;
    ``$\mathrm{r}^{\Omega}$'' rigid rotation with an angular velocity
    of $\Omega$), the entropy gradient, $\partial_{\varpi} $
    ($s_0=0.2$), the strength of the initial magnetic field, $b_0^z$,
    and the type of boundary condition which was applied (d: velocity
    damping; p: periodic).  The remaining four columns give the growth
    rate of the MRI, $\sigma$, and the Maxwell stress component
    $M^{\mathrm{term}}_{\varpi\phi}$ at MRI termination, its maximum
    value, and its time averaged value, respectively.
  }
  \label{Tab:3d-models-1}

  \begin{tabular}{l|cc|cccc|ccccc}

    \hline\hline
    field
    & grid size
    & resolution
    & Rot
    & $\partial_{\varpi} S$
    & $b^z_0$
    & BC
    & $\sigma$ 
    & $M_{\varpi \phi}^\mathrm{term}$
    & $M_{\varpi \phi}^\mathrm{max}$
    & $\left\langle M_{\varpi \phi} \right\rangle$
    \\
    & $[\textrm{km}^3]$
    &
    & 
    & $\left[\mathrm{km}^{-1}\right]$
    & $\left[10^{12}\,\mathrm{G}\right]$
    &
    & $\left[\mathrm{ms}^{-1}\right]$
    & \multicolumn{3}{c}{$\left[10^{28}~\mathrm{G}^2 ~ \mathrm{cm}^{-3}\right]$}
    \\

    \hline\hline
    U3
    & $ 0.5 \times 0.25 \times 0.5 $
    & $ 26 \times 12 \times 26 $
    & d & 0 & 40 & d
    & 0.98
    & 0.47
    & 298
    & 47
    \\
    & $ 0.5 \times 0.5 \times 0.5 $
    & $ 26 \times 26 \times 26 $
    & d & 0 & 40 & d
    & 0.96
    & 0.45
    & 211
    & 36
    \\
    & $ 0.5 \times 1 \times 0.5 $
    & $ 26 \times 50 \times 26 $
    & d & 0 & 40 & d
    & 0.96
    & 0.45
    & 16
    & 2.9
    \\
    & $ 0.5 \times 2 \times 0.5 $
    & $ 26 \times 100 \times 26 $
    & d & 0 & 40 & d
    & 0.92
    & 0.44
    & 5.0
    & 1.5
    \\

    \hline\hline
    & $1 \times 0.5 \times 1$
    & $50 \times 26 \times 50$
    & d & 0
    & 20
    & d
    & 1.05
    & 3.1
    & 924
    & 69
    \\
    & $1 \times 0.5 \times 1$
    & $50 \times 26 \times 50$
    & d & 0
    & 40
    & d
    & 1.10
    & 3.7
    & 553
    & 68
    \\

    \hline
    & $1 \times 1 \times 1$
    & $50 \times 50 \times 50$
    & d & 0
    & 10
    & d
    & 0.76
    & 1.5
    & 2.7
    & 1.1
    \\
    & $1 \times 1 \times 1$
    & $50 \times 50 \times 50$
    & d & 0
    & 20
    & d
    & 1.03
    & 3.1
    & 347
    & 44
    \\
    & $ 1 \times 1 \times 1 $
    & $ 50 \times 50 \times 50 $
    & d & 0 & 40 & d
    & 1.09
    & 3.1
    & 482
    & 63
    \\

    \hline
    & $1 \times 2 \times 1$
    & $50 \times 50 \times 50$
    & d & 0
    & 10
    & d
    & 0.75
    & 1.3
    & 1.3
    & 0.47
    \\
    & $1 \times 2 \times 1$
    & $50 \times 50 \times 50$
    & d & 0
    & 20
    & d
    & 1.03
    & 3.0
    & 6.5
    & 2.9
    \\
    & $1 \times 2 \times 1$
    & $50 \times 50 \times 50$
    & d & 0
    & 40
    & d 
    & 1.09
    & 2.9
    & 22
    & 5.8
    \\
    & $1 \times 2 \times 1$
    & $50 \times 100 \times 50$
    & d & 0
    & 20
    & d
    & 1.02
    & 3.1
    & 9.5
    & 3.4
    \\
    & $1 \times 2 \times 1$
    & $50 \times 100 \times 50$
    & d & 0
    & 40
    & d
    & 1.09
    & 2.8
    & 60
    & 9.2
    \\

    \hline
    & $1 \times 4 \times 1$
    & $50 \times 200 \times 50$
    & d & 0
    & 20
    & d
    & 1.03
    & 2.9
    & 4.4
    & 2.3
    \\
    & $1 \times 4 \times 1$
    & $50 \times 200 \times 50$
    & d & 0
    & 40
    & d
    & 1.08
    & 3.0
    & 10.5
    & 3.4
    \\
    & $1 \times 4 \times 1$
    & $50 \times 200 \times 50$
    & d & 0
    & 80
    & d
    & 1.01
    & 2.7
    & 45
    & 9.0
    \\

    \hline
    & $1 \times 4 \times 1$
    & $100 \times 400 \times 100$
    & d & 0
    & 20
    & d
    & 1.05
    & 3.5
    & 6.5
    & 2.5
    \\
    & $1 \times 4 \times 1$
    & $100 \times 400 \times 100$
    & d & 0
    & 40
    & d
    & 1.09
    & 3.5
    & 28
    & 8.2
    \\

    \hline\hline
    & $ 0.5 \times 0.25 \times 1 $
    & $ 26 \times 12 \times 50 $
    & d & 0 & 40 & d
    & 1.04
    & 0.44
    & 325
    & 64
    \\
    & $ 0.5 \times 0.5 \times 1 $
    & $ 26 \times 26 \times 50 $
    & d & 0 & 40 & d
    & 1.03
    & 0.52
    & 227
    & 40
    \\
    & $ 0.5 \times 1 \times 1 $
    & $ 26 \times 50 \times 50 $
    & d & 0 & 40 & d
    & 1.00
    & 0.42
    & 298
    & 33
    \\
    & $ 0.5 \times 2 \times 1 $
    & $ 26 \times 100 \times 50 $
    & d & 0 & 40 & d
    & 0.96
    & 0.50
    & 289
    & 38
    \\

    \hline\hline
    & $ 1 \times 0.25 \times 0.5 $
    & $ 50 \times 12 \times 26 $
    & d & 0 & 40 & d
    & 1.08
    & 2.9
    & 310
    & 59
    \\
    & $ 1 \times 0.38 \times 0.5 $
    & $ 50 \times 18 \times 26 $
    & d & 0 & 40 & d
    & 1.08
    & 3.0
    & 234
    & 45
    \\
    & $ 1 \times 0.5 \times 0.5 $
    & $ 50 \times 26 \times 26 $
    & d & 0 & 40 & d
    & 1.07
    & 2.9
    & 16
    & 6.2
    \\
    & $ 1 \times 1 \times 0.5 $
    & $ 50 \times 50 \times 26 $
    & d & 0 & 40 & d
    & 1.07
    & 3.1
    & 9.0
    & 2.3
    \\
    & $ 1 \times 2 \times 0.5 $
    & $ 50 \times 100 \times 26 $
    & d & 0 & 40 & d
    & 1.06
    & 3.0
    & 3.4
    & 1.8
    \\

    \hline\hline
    & $ 0.5 \times 0.25 \times 0.5 $
    & $ 26 \times 12 \times 26 $
    & d & 0 & 20 & p
    & 1.07
    & 97
    & 2940
    & 145
    \\
    & $ 0.5 \times 0.5 \times 0.5 $
    & $ 26 \times 26 \times 26 $
    & d & 0 & 20 & p
    & 0.99
    & 64
    & 4196
    & 335
    \\
    & $ 0.5 \times 1 \times 0.5 $
    & $ 26 \times 50 \times 26 $
    & d & 0 & 20 & p
    & 1.06
    & 60
    & 535
    & 19
    \\
    & $ 0.5 \times 2 \times 0.5 $
    & $ 26 \times 100 \times 26 $
    & d & 0 & 20 & p
    & 1.08
    & 66
    & 66
    & 8.7
    \\

    \hline\hline
    & $ 1 \times 0.25 \times 1 $
    & $ 50 \times 12 \times 50 $
    & d & 0 & 20 & p
    & 1.02
    & 13
    & 1316
    & 110
    \\
    & $ 1 \times 0.5 \times 1 $
    & $ 50 \times 26 \times 50 $
    & d & 0 & 20 & p
    & 1.02
    & 8.8
    & 1424
    & 60
    \\
    & $ 1 \times 1 \times 1 $
    & $ 50 \times 50 \times 50 $
    & d & 0 & 20 & p
    & 0.98
    & 42
    & 2570
    & 128
    \\
    & $ 1 \times 2 \times 1 $
    & $ 50 \times 100 \times 50 $
    & d & 0 & 20 & p
    & 1.04
    & 16.5
    & 736
    & 79
    \\

    \hline
    & $1 \times 4 \times 1$
    & $50 \times 200 \times 50$
    & d & 0
    & 20
    & p
    & 1.05
    & 22.5
    & 59
    & 6.5
    \\
    & $1 \times 4 \times 1$
    & $50 \times 200 \times 50$
    & d & 0
    & 40
    & p
    & 1.09
    & 40.4
    & 361
    & 18
    \\
    & $1 \times 4 \times 1$
    & $50 \times 200 \times 50$
    & d & 0
    & 80
    & p
    & 1.09
    & 254
    & 254
    & 5.5
    \\

    \hline
    & $1 \times 4 \times 1$
    & $100 \times 400 \times 100$
    & d & 0
    & 20
    & p
    & 1.08
    & 14.9
    & 27.6
    & 6.4
    \\
    & $1 \times 4 \times 1$
    & $100 \times 400 \times 100$
    & d & 0
    & 40
    & p
    & 1.11
    & 88
    & 88
    & 11
    \\

    \hline\hline
    & $ 0.5 \times 0.25 \times 1 $
    & $ 26 \times 12 \times 50 $
    & d & 0 & 20 & p
    & 1.04
    & 64
    & 2182
    & 146
    \\
    & $ 0.5 \times 0.25 \times 1 $
    & $ 26 \times 26 \times 50 $
    & d & 0 & 20 & p
    & 1.03
    & 31
    & 1540
    & 170
    \\
    & $ 0.5 \times 1 \times 1 $
    & $ 26 \times 50 \times 50 $
    & d & 0 & 20 & p
    & 1.01
    & 8.9
    & 1735
    & 102
    \\
    & $ 0.5 \times 2 \times 1 $
    & $ 26 \times 100 \times 50 $
    & d & 0 & 20 & p
    & 1.05
    & 30
    & 825
    & 72
    \\
    & $ 0.5 \times 4 \times 1 $
    & $ 26 \times 200 \times 50 $
    & d & 0 & 20 & p
    & 1.04
    & 27
    & 103
    & 15
    \\

    \hline\hline
    & $ 1 \times 0.25 \times 0.5 $
    & $ 50 \times 12 \times 26 $
    & d & 0 & 20 & p
    & 1.07
    & 32
    & 1654
    & 171
    \\
    & $ 1 \times 0.5 \times 0.5 $
    & $ 50 \times 26 \times 26 $
    & d & 0 & 20 & p
    & 1.06
    & 8.5
    & 2902
    & 170
    \\
    & $ 1 \times 1 \times 0.5 $
    & $ 50 \times 50 \times 26 $
    & d & 0 & 20 & p
    & 1.05
    & 14
    & 682
    & 13
    \\
    & $ 1 \times 2 \times 0.5 $
    & $ 50 \times 100 \times 26 $
    & d & 0 & 20 & p
    & 1.06
    & 14
    & 14
    & 4.0
    \\

    \hline\hline

  \end{tabular}
\end{table*}

\begin{table*}[htbp]
  \centering
  \caption{Continuation of \tabref{Tab:3d-models-1}.  For the MBI
    model with $b_0^z = 10 \mathrm{G}$, we do not provide values for
    the Maxwell stress, because the magnetic field behaves similar to
    a purely passive field.
  }
  \label{Tab:3d-models-2}

  \begin{tabular}{l|cc|cccc|ccccc}

    \hline\hline
    field
    & grid size
    & resolution
    & Rot
    & $\partial_{\varpi} S$
    & $b^z_0$
    & BC
    & $\sigma$ 
    & $M_{\varpi \phi}^\mathrm{term}$
    & $M_{\varpi \phi}^\mathrm{max}$
    & $\left\langle M_{\varpi \phi} \right\rangle$
    \\
    & $[\textrm{km}^3]$
    &
    & 
    & $\left[\mathrm{km}^{-1}\right]$
    & $\left[10^{12}\,\mathrm{G}\right]$
    &
    & $\left[\mathrm{ms}^{-1}\right]$
    & \multicolumn{3}{c}{$\left[10^{28}~\mathrm{G}^2 ~ \mathrm{cm}^{-3}\right]$}
    \\

    \hline\hline
    V3
    & $1 \times 1 \times 1$
    & $50 \times 50 \times 50$
    & d & 0
    & 10
    & d
    & 0.66
    & 0.13
    & 0.24
    & 0.16
    \\
    & $1 \times 1 \times 1$
    & $50 \times 50 \times 50$
    & d & 0
    & 20
    & d
    & 0.96
    & 0.53
    & 0.53
    & 0.21
    \\
    & $1 \times 1 \times 1$
    & $50 \times 50 \times 50$
    & d & 0
    & 40
    & d
    & 1.03
    & 1.5
    & 1.5
    & 0.27
    \\

    \hline
    & $1 \times 4 \times 1$
    & $50 \times 200 \times 50$
    & d & 0
    & 10
    & d
    & 0.66
    & 0.13
    & 0.52
    & 0.24
    \\
    & $1 \times 4 \times 1$
    & $50 \times 200 \times 50$
    & d & 0
    & 20
    & d
    & 0.90
    & 0.36
    & 0.80
    & 0.43
    \\

    \hline
    & $1 \times 4 \times 1$
    & $100 \times 400 \times 100$
    & d & 0
    & 20
    & d
    & 1.08
    & 1.1
    & 1.1
    & 0.20
    \\

    \hline
    & $1 \times 4 \times 1$
    & $100 \times 400 \times 100$
    & d & 0
    & 20
    & p
    & 1.08
    & 1.5
    & 1.5
    & 0.21
    \\

    \hline\hline

    U3
    & $1 \times 2 \times 1$
    & $26 \times 50 \times 26$
    & d & $-0.038$
    & 20
    & d
    & 1.5
    & 1.9
    & 5.3
    & 0.83
    \\
    & $1 \times 2 \times 1$
    & $26 \times 50 \times 26$
    & d & $-0.038$
    & 40
    & d
    & 1.7
    & 4.1
    & 7.2
    & 1.1
    \\

    \hline
    & $1 \times 2 \times 1$
    & $50 \times 100 \times 50$
    & d & $-0.038$
    & 20
    & d
    & 1.7
    & 3.1
    & 3.1
    & 0.16
    \\
    & $1 \times 2 \times 1$
    & $50 \times 100 \times 50$
    & d & $-0.038$
    & 40
    & d
    & 1.9
    & 4.4
    & 4.4
    & 0.23
    \\

    \hline\hline
    & $1 \times 2 \times 1$
    & $50 \times 100 \times 50$
    & r${}^{1900}$ & $-0.10$
    & $10^{-11}$ & d
    & 2.6
    \\
    & $1 \times 2 \times 1$
    & $50 \times 100 \times 50$
    & r${}^{1900}$ & $-0.10$
    & 0.01 & d
    & 2.6
    & $2.3 \times 10^{-8}$
    & 0.11
    & 0.040
    \\
    & $1 \times 2 \times 1$
    & $50 \times 100 \times 50$
    & r${}^{1900}$ & $-0.10$
    & 20 & d
    & 2.6
    & 0.060
    & 0.87
    & 0.30
    \\

    \hline\hline
    V3
    & $1 \times 2 \times 1$
    & $50 \times 100 \times 50$
    & d & $-0.038$
    & 20
    & d
    & 1.6
    & 0.54
    & 0.54
    & 0.011
    \\

    \hline\hline

  \end{tabular}
\end{table*}

\bibliographystyle{aa}
\bibliography{1323.bib}

\end{document}